\documentclass[reqno,11pt]{amsart}
\usepackage{graphicx}
\usepackage{amscd}
\usepackage{slashed}
\usepackage{amssymb}
\usepackage{ushort}
\usepackage{mathtools}
\usepackage{esint}
\usepackage[usenames,dvipsnames]{pstricks}
\usepackage{pst-grad} 
\usepackage{pst-plot} 
\usepackage[mathscr]{eucal}
\numberwithin{equation}{section}
\allowdisplaybreaks[1]

\title[Spin-Weighted Spheroidal Wave Operators]{A Spectral Representation
for Spin-Weighted Spheroidal Wave Operators with Complex Aspherical Parameter}

\author[F.\ Finster]{Felix Finster}
\thanks{F.F.\ is supported in part by the Deutsche Forschungsgemeinschaft.}
\address{Fakult\"at f\"ur Mathematik \\ Universit\"at Regensburg \\ D-93040 Regensburg \\ Germany}
\email{Felix.Finster@mathematik.uni-regensburg.de}

\author[J.\ Smoller]{Joel Smoller \\ \\ July 2015}
\thanks{J.S.\ is supported in part by the National Science Foundation,
Grant No.\ DMS-1105189.}
\address{Mathematics Department \\ The University of Michigan \\ Ann Arbor, MI 48109, USA}
\email{smoller@umich.edu}

\newtheorem{Def}{Def.}[section]
\newtheorem{Thm}[Def]{Theorem}
\newtheorem{Prp}[Def]{Proposition}
\newtheorem{Lemma}[Def]{Lemma}
\newtheorem{Remark}[Def]{Remark}
\newtheorem{Corollary}[Def]{Corollary}

\newcommand{\Thanks}{\vspace*{.5em} \noindent \thanks}
\newcommand{\Proof}{\begin{proof}}
\newcommand{\QED}{\end{proof} \noindent}
\newcommand{\QEDrem}{\ \hfill $\Diamond$}

\newcommand{\C}{\mathbb{C}}
\newcommand{\R}{\mathbb{R}}
\newcommand{\1}{\mbox{\rm 1 \hspace{-1.05 em} 1}}
\newcommand{\Z}{\mathbb{Z}}

\newcommand{\N}{\mathbb{N}}

\newcommand{\beq}{\begin{equation}}
\newcommand{\eeq}{\end{equation}}

\newcommand{\umax}{{u_{\mbox{\tiny{\rm{max}}}}}}

\newcommand{\WKB}{{\text{\tiny{\rm{WKB}}}}}

\newcommand{\A}{\mathcal{A}}
\renewcommand{\O}{\mathscr{O}}

\newcommand{\phiD}{\phi^\text{\tiny{\rm{D}}}}
\newcommand{\rhoD}{\rho^\text{\tiny{\rm{D}}}}
\newcommand{\uu}{\underbracket[0.8pt][0pt]{u}}
\newcommand{\p}{\mathfrak{p}}
\newcommand{\piot}{{\textstyle{\frac{\pi}{2}}}}
\renewcommand{\H}{\mathscr{H}}
\newcommand{\la}{\langle}
\newcommand{\ra}{\rangle}
\newcommand{\Lin}{\text{\rm{L}}}
\newcommand{\ul}{u_\ell}
\newcommand{\ur}{u_r}
\newcommand{\Const}{\mathscr{C}}
\newcommand{\const}{\mathfrak{c}}
\newcommand{\D}{\mathscr{D}}

\DeclareMathOperator{\re}{Re}
\DeclareMathOperator{\im}{Im}

\begin{document}
\maketitle

\begin{abstract}
A family of spectral decompositions of the spin-weighted spheroidal wave operator
is constructed for complex aspherical parameters with bounded imaginary part.
As the operator is not symmetric, its spectrum is complex and Jordan chains may appear.
We prove uniform upper bounds for the length of the Jordan chains and the norms of the
idempotent operators mapping onto the invariant subspaces.
The completeness of the spectral decomposition is proven.
\end{abstract}

\tableofcontents

\section{Introduction and Statement of Results}
The spin-weighted spheroidal wave equation arises in the study of electromagnetic, 
gravitational and neutrino-field perturbations of rotating black holes 
when separating variables in the so-called Teukolsky master equation
(see~\cite{chandra, teukolsky} or the survey paper~\cite{bull}).
In the spin-weighted spheroidal wave equation,
the {\em{spin}} of the wave enters as a parameter~$s \in \{0, \frac{1}{2}, 1,
\frac{3}{2}, 2, \ldots\}$.
We are mainly interested in the cases~$s=1$ of an electromagnetic and~$s=2$ of a gravitational field.
If~$s$ is an integer, the spin-weighted wave equation is the eigenvalue equation
\beq \label{eigen}
\A \,\Psi = \lambda \Psi\:,
\eeq
where the spin-weighted spheroidal wave operator~$\A$ is an elliptic operator
with smooth coefficients on the unit sphere~$S^2$. More specifically, choosing polar
coordinates~$\vartheta \in (0, \pi)$ and~$\varphi \in [0, 2 \pi)$, we have (see for
example~\cite{whiting})
\[ \A = - \frac{\partial}{\partial \cos \vartheta} \,\sin^2 \vartheta \,\frac{\partial}
{\partial \cos \vartheta}
+ \frac{1}{\sin^2 \vartheta} \Big( \Omega\, \sin^2 \vartheta + i \frac{\partial}{\partial \varphi}
- s \cos \vartheta \Big)^2 . \]
Here~$\Omega \in \C$ is the {\em{aspherical parameter}}. In the special case~$\Omega=0$,
we obtain the spin-weighted Laplacian on the sphere, whose eigenvalues and eigenfunctions can be given
explicitly~\cite{goldberg}. In the case~$s=0$ and~$\Omega \neq 0$, one gets
the spheroidal wave operator (\cite{flammer, angular}). Setting~$\Omega=0$ and~$s=0$,
one simply obtains the Laplacian on the sphere.
We consider~$\A$ on the Hilbert space~$\H = L^2(S^2)$ with domain of
definition~$\D(\A) = C^\infty(S^2)$.
We remark that~$\A$ clearly is an elliptic operator on the sphere.
However, even in the case~$s=0$ and for real~$\Omega$, in general there is no Riemannian metric
on the sphere which realizes the spheroidal wave operator as the Laplace-Beltrami operator.
Thus the spheroidal wave operator cannot be 
be identified with the Laplace-Beltrami operator on a Riemannian manifold.
For general spin, this means in particular that the methods used for spin-weighted spherical
harmonics in~\cite[Section~4.15]{penrose+rindler}
do not seem to generalize to the spheroidal situation.

As the spin-weighted spheroidal wave operator is axisymmetric, we can separate out the
$\varphi$-dependence with a plane wave ansatz,
\[ \Psi(\vartheta, \varphi) = e^{-i k \varphi}\: \Theta(\vartheta) \qquad \text{with~$k \in \Z$}\:. \]
Then $\A$ becomes the ordinary differential operator
\beq \label{swswo}
\A_k :=  - \frac{\partial}{\partial \cos \vartheta} \,\sin^2 \vartheta \,\frac{\partial}
{\partial \cos \vartheta}
+ \frac{1}{\sin^2 \vartheta} \left( \Omega\, \sin^2 \vartheta + k - s \cos \vartheta \right)^2  \:.
\eeq
This operator acts on the vectors in~$\H$ with the prescribed~$\phi$-dependence, which
we denote by~$\H_k$,
\[ \H_k := L^2(S^2) \cap \{ e^{-i k \varphi}\: \Theta(\vartheta) \:|\: \Theta : (0, \pi) \rightarrow \C \} \:. \]
The domain of definition reduces to
\[ \D(\A_k) = C^\infty(S^2) \cap \H_k \:. \]

The Hilbert space~$\H_k$ can be identified with
\[ \H_k = L^2((0, \pi), \sin \vartheta \,d \vartheta) \:. \]
Also, one can consider~$\A_k$ as an ordinary differential operator on this Hilbert space,
for example with the domain of definition~$C^\infty((0,\pi)) \cap L^2((0, \pi)$.
However, when doing so, one still needs to specify boundary conditions at~$\vartheta = 0, \pi$.
As will be explained in detail in Section~\ref{secSL} below, the correct boundary conditions are
that the limits
\beq \label{Thetalim}
\lim_{\vartheta \rightarrow 0, \pi} \Theta(\vartheta) \quad \text{must exist}\:.
\eeq
In this formulation as a pure ODE problem, the spheroidal wave equation~\eqref{swswo}
can also be used in the case of half-integer spin (to describe neutrino or Rarita-Schwinger fields),
if~$k$ is chosen to be a half-integer. Thus in what follows, we fix the parameters~$s$ and~$k$ such that
\beq \label{skrange}
2s \in \N_0 \qquad \text{and} \qquad k-s \in \Z\:.
\eeq

We are interested in the case that~$\Omega$ is {\em{complex}}. Then the potential in~\eqref{swswo}
is complex, so that the operator~$\A_k$ is not a symmetric operator on~$\H_k$.
As a consequence, the spectral theorem in Hilbert spaces does not apply.
The spectrum will in general be complex.
Moreover, the operator need not be diagonalizable, because Jordan chains may form.
The main task of the present paper is to control the length of these Jordan chains
to obtain a decomposition of the Hilbert space into invariant subspaces of~$\A_k$.
This is our main result:

\begin{Thm} \label{thmmain}
For any~$s$ and~$k$ in the range~\eqref{skrange} and any~$c>0$, we 
let~$U \subset \C$ be the strip
\beq \label{ocond}
|{\mbox{\rm{Im}}}\, \Omega| < c \:.
\eeq
Then there is a positive integer~$N$ and a family of bounded
linear operators~$Q_n(\Omega)$ on~$\H_k$
defined for all~$n \in \N \cup \{0\}$ and $\Omega \in U$
with the following properties:
\begin{itemize}
\item[(i)] The image of the operator~$Q_0$ is an $N$-dimensional invariant
subspace of $\A_k$. 
\item[(ii)] For every $n\geq1$, the image of the operator~$Q_n$ is an at most
two-dimensional invariant subspace of $\A_k$.
\item[(iii)] The $Q_n$ are uniformly bounded in~$\Lin(\H_k)$,
i.e. for all $n \in \N \cup \{0\}$ and~$\Omega \in U$,
\beq \label{Qnb}
\|Q_n\| \leq c_2
\eeq
for a suitable constant $c_2=c_2(s,k,c)$ (here~$\| \cdot \|$ denotes the $\sup$-norm on~$\H_k$).
\item[(iv)] The~$Q_n$ are idempotent and mutually orthogonal in the sense that
\[ Q_n\, Q_{n'} = \delta_{n,n'}\: Q_n \qquad \text{for all~$n,n' \in \N \cup \{0\}$}\:. \]
\item[(v)] The $Q_n$ are complete in the sense that for every~$\Omega \in U$,
\beq \label{strongcomplete}
\sum_{n=0}^\infty Q_n = \1
\eeq
with strong convergence of the series.
\end{itemize}
\end{Thm} \noindent
Note that the operators~$Q_n$ are in general not symmetric (i.e.\ $Q_n^* \neq Q_n$).
This corresponds to the fact that for non-symmetric operators, the eigenvectors
corresponding to different eigenvalues are in general not orthogonal.

\section{Reformulation as a Sturm-Liouville Problem} \label{secSL}
We first bring the operator~\eqref{swswo} to the standard Sturm-Liouville form
(for more details see~\cite[Section~2]{tinvariant}). To this end, we first write the operator in the
variable~$u=\vartheta \in (0,\pi)$,
\[ \A_k =  -\frac{1}{\sin u}\: \frac{d}{d u}\: \sin u\: \frac{d}{d u}
+ \frac{1}{\sin^2 u} \left( \Omega\, \sin^2 u + k - s \cos u \right)^2  \:. \]
Introducing the function $\phi$ by
\beq \label{Ydef}
\phi = \sqrt{\sin u}\: \Theta \:,
\eeq
we get the eigenvalue equation
\beq \label{Schrodinger}
H \phi = \lambda \phi \;,
\eeq
where~$H$ has the form of a one-dimensional Hamiltonian
\beq \label{Hamilton}
H = -\frac{d^2}{d u^2} + W
\eeq
where~$W$ is the complex potential
\begin{align}
W &= -\frac{1}{4}\: \frac{\cos^2 u}{\sin^2 u} - \frac{1}{2} +\frac{1}{\sin^2 u}(\Omega \sin^2 u + k - s \cos u)^{2}
\label{Wdef} \\
&= \Omega^2\: \sin^2 u + \left(k^2 + s^2 - \frac{1}{4} \right) \frac{1}{\sin^2 u}
\:+\: 2 \Omega k - s^2 - \frac{1}{4} \label{Wsymm} \\
&\quad - 2 s \Omega \cos u - 2 s k\, \frac{\cos u}{\sin^2 u} \:. \label{Wasy}
\end{align}
For what follows, it is usually most convenient to write~\eqref{Schrodinger} as the the Sturm-Liouville equation
\begin{equation} \label{5ode}
\left( -\frac{d^2}{du^2} + V \right) \phi = 0 \:,
\end{equation}
where~$V$ is the potential
\begin{equation} \label{Vdef}
V = \Omega^2\: \sin^2 u + \left(k^2 + s^2 - \frac{1}{4} \right) \frac{1}{\sin^2 u}
- 2 s \Omega \cos u - 2 s k\, \frac{\cos u}{\sin^2 u} \:-\: \mu\:,
\end{equation}
and~$\mu$ is the constant
\begin{equation}
\mu = \lambda - 2 \Omega k + s^2 + \frac{1}{4} . \label{mudef}
\end{equation}
The transformation~\eqref{Ydef} from~$\Theta$ to~$Y$ becomes
a unitary transformation if the integration measure in the
corresponding Hilbert spaces is transformed from~$\sin u\:du$ to~$du$.
Hence the eigenvalue equation~\eqref{eigen} on~$\H_k$ is equivalent
to~\eqref{5ode} on the Hilbert space $L^2((0,\pi), du)$.

If the potential~$V$ were continuous on the interval~$[0, \pi]$, we would get
a well-defined boundary problem by imposing Dirichlet or Neumann or more general
mixed boundary values at~$0$ and~$\pi$ (for details see~\cite[Chapter~12]{coddington}).
In our situation, there is the complication that the potential~\eqref{Vdef} has poles at the boundary points.
As a consequence, the fundamental solutions will also have singularities, so that it is no longer
obvious how to introduce suitable boundary conditions. In the case when~$s$ is an integer,
the correct boundary values can be determined by going back to the eigenfunctions
on the sphere~\eqref{eigen}, as we now explain.
Due to elliptic regularity theory, the eigenfunctions~$\Psi$ of the angular operator~\eqref{eigen}
are smooth functions on the sphere. Therefore, we obtain~\eqref{Thetalim} as a necessary condition.
In view of the transformation~\eqref{Ydef}, this implies that the limits
\beq \label{Dirichlet}
\lim_{u \searrow 0} u^{-\frac{1}{2}} \,\phi(u) \quad \text{and}
\quad \lim_{u \nearrow \pi} (\pi-u)^{-\frac{1}{2}}\, \phi(u) \qquad \text{must exist}\:.
\eeq
These boundary conditions can also be understood by looking at the asymptotics of the
solutions of~\eqref{5ode} near the boundary points. Namely, expanding the potential~\eqref{Vdef} near the boundary points, we obtain
\begin{align*}
V(u) &= \frac{1}{u^2} \left( (k-s)^2 - \frac{1}{4} \right) + \O \big( u^{-1} \big) \\
V(u) &= \frac{1}{(\pi-u)^2} \left( (k+s)^2 - \frac{1}{4} \right) + \O \big( (\pi-u)^{-1} \big)\:.
\end{align*}
If the factors~$k \pm s$ are non-zero, the solutions have the asymptotics
\beq \label{genasy}
\phi(u) \sim u^{\frac{1}{2} \pm |k-s|} \:\big(1+ \O(u) \big) \qquad \text{and} \qquad
\phi(u) \sim (\pi-u)^{\frac{1}{2} \pm |k+s|} \:\big(1+ \O(\pi-u) \big) \:.
\eeq
If on the other hand, the factors~$k \pm s$ are zero, the asymptotic solutions involve an additional logarithm
(for detail see~\cite[Sections~7 and~8]{tinvariant}),
\begin{align}
\phi(u) &= c_1 \,\sqrt{u} + c_2\, \sqrt{u}\, \log u + \O(u) && \text{if~$k=s$} \label{ex1} \\
\phi(u) &= c_1 \,\sqrt{\pi-u} + c_2\, \sqrt{\pi - u} \:\log (\pi - u) + \O(\pi-u) && \text{if~$k=-s$} \:. \label{ex2}
\end{align}
In each case, the boundary conditions~\eqref{Dirichlet} single out one of the two fundamental solutions.
In this way, the conditions~\eqref{Dirichlet} give mathematically reasonable boundary conditions.
We remark that in the case~\eqref{genasy}, our boundary conditions are equivalent to
Dirichlet boundary conditions. Alternatively, these boundary conditions could be implemented
simply by demanding that the eigenfunctions must be square integrable
(note that, in view of~\eqref{skrange}, the parameter $k-s$ is always an integer).
In the exceptional cases~\eqref{ex1} and~\eqref{ex2}, however, both fundamental solutions
satisfy Dirichlet boundary conditions and are square integrable. Thus in these cases, it is essential
to state the boundary conditions in the form~\eqref{Dirichlet}.

In order to bring the boundary conditions~\eqref{Dirichlet} into a more tractable form,
it is convenient to work with solutions of the corresponding Riccati equation:
For any solution~$\phi$ of the Sturm-Liouville equation~\eqref{5ode}, the function~$y := \phi'/\phi$
satisfies the corresponding Riccati equation
\begin{equation} \label{riccati}
y' = V-y^2\:.
\end{equation}
Using the results and methods in~\cite{invariant, tinvariant}, we can construct a solution~$y$
of the Riccati equation with rigorous error bounds. With this in mind, let us assume that
a solution~$y$ of the Riccati equation is known. Then a particular
solution of the corresponding Sturm-Liouville equation is obtained by integration,
\beq \label{phidef}
\phi(u) = \exp \Big( \int_{u_0}^u y \Big) .
\eeq
The general solution can be constructed by integrating the equation for the Wronskian. Namely, if~$\hat{\phi}$
is another solution of the Sturm-Liouville equation, the Wronskian
\[ w(\phi, \hat{\phi}) := \phi' \hat{\phi} - \phi \hat{\phi}' \]
is a constant, and thus
\beq \label{phi2}
\hat{\phi}(u) = \phi(u) \left( \frac{\hat{\phi}(u_0)}{\phi(u_0)} - \int_{u_0}^u
\frac{w(\phi, \hat{\phi})}{\phi^2} \right) \:.
\eeq

In particular, this relation can be used to construct solutions of the Sturm-Liouville equation~\eqref{5ode}
which satisfy the boundary conditions~\eqref{Dirichlet}.
We denote these solutions by~$\phiD_L$ and~$\phiD_R$ (where the subscript~D refers to
``Dirichlet'', and~$L\!/\!R$ to the left and right boundary points at~$u=0$ and~$u=\pi$, respectively).
To this end, we let~$\phi_L$ and~$\phi_R$ be generic solutions which do {\em{not}} satisfy
the boundary conditions~\eqref{Dirichlet}, i.e.\
\begin{align*}
\phi_L(u) &\sim \left\{ \begin{array}{cl} u^{\frac{1}{2} - |k-s|} \:\big(1+\O(u) \big) & \text{if~$k \neq s$} \\[0.2em]
\sqrt{u} \,\log u \:\big( 1+\O(u) \big) & \text{if~$k = s$} \end{array} \right. \\
\phi_R(u) &\sim \left\{ \begin{array}{cl} (\pi-u)^{\frac{1}{2} - |k+s|}
\:\big( 1+\O(\pi-u) \big) & \text{if~$k \neq -s$} \\[0.2em]
\sqrt{\pi-u} \,\log (\pi-u) \:\big(1+\O(\pi-u) \big)& \text{if~$k = -s \:.$} \end{array} \right.
\end{align*}
Then, using~\eqref{phi2}, the solutions which do satisfy~\eqref{Dirichlet}
are given (up to irrelevant prefactors) by
\beq \label{phiDdef}
\phiD_L(u) = \phi_L(u) \int_0^u \frac{1}{\phi_L^2} \qquad \text{and} \qquad
\phiD_R(u) = -\phi_R(u) \int_u^\pi \frac{1}{\phi_R^2} \:.
\eeq

If~$\phi$ is a solution of~\eqref{5ode} subject to the boundary conditions~\eqref{Dirichlet},
then this solution must be a multiple of both~$\phiD_L$ and~$\phiD_R$. Hence~$\phiD_L$
and~$\phiD_R$ are linearly dependent, and their Wronskian vanishes,
\beq \label{evalcond}
w\big( \phiD_L, \phiD_R \big) = 0 \:.
\eeq
In this way, we have reformulated the existence problem for solutions
satisfying the boundary conditions~\eqref{Dirichlet} in terms of the vanishing of the Wronskian~\eqref{evalcond}.
More generally, the Wronskian can be used to describe the spectrum of the Hamiltonian.
Namely, we just saw that if the Wronskian vanishes, then there is an eigensolution
which satisfies the boundary conditions~\eqref{5ode}. Conversely, if the Wronskian is non-zero,
we may introduce the {\em{Green's function}} by
\beq \label{sldef}
s_\lambda(u,u') = \frac{1}{w(\phiD_L, \phiD_R)} \times \left\{
\begin{aligned} \phiD_L(u)\: \phiD_R(u') &\quad&& \text{if~$u \leq u'$} \\
\phiD_L(u')\: \phiD_R(u) &&& \text{if~$u' < u$}\:.
\end{aligned}  \right.
\eeq
By direct computation one verifies that the Green's function satisfies the equation
\[ \left( H-\lambda \right) s_\lambda(u,u') = \delta(u-u') \:. \]
Thus taking~$s_\lambda(u,u')$ as the integral kernel of a corresponding operator~$s_\lambda$
on~$\H_k = L^2((0,\pi), du)$, this operator is a bounded inverse of the operator~$(H-\lambda)$.
Thus~$\lambda$ is in the {\em{resolvent set}}, and~$s_\lambda$ is the {\em{resolvent}}.
We conclude that the {\em{spectrum}} of~$H$, defined as the complement of the resolvent set,
is given as the set of all~$\lambda$ for which the Wronskian~\eqref{evalcond} vanishes
for non-trivial solutions~$\phiD_L$ and~$\phiD_R$ satisfying the boundary conditions~\eqref{Dirichlet}
at~$u=0$ and~$u=\pi$, respectively.

\section{The Qualitative Behavior of the Spectrum}
\begin{figure}
\includegraphics[width=6.5cm]{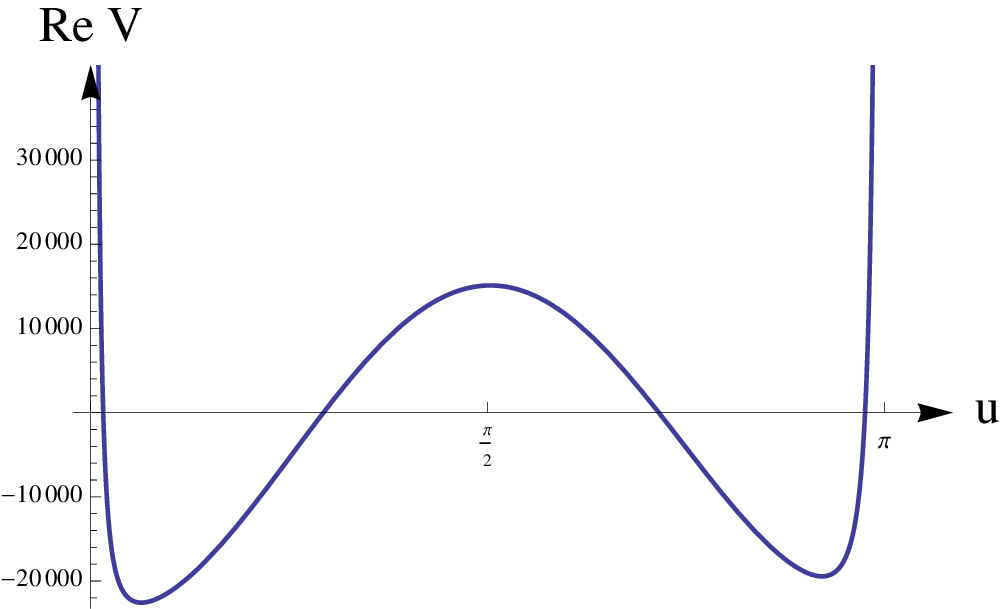} $\qquad$
\includegraphics[width=6.5cm]{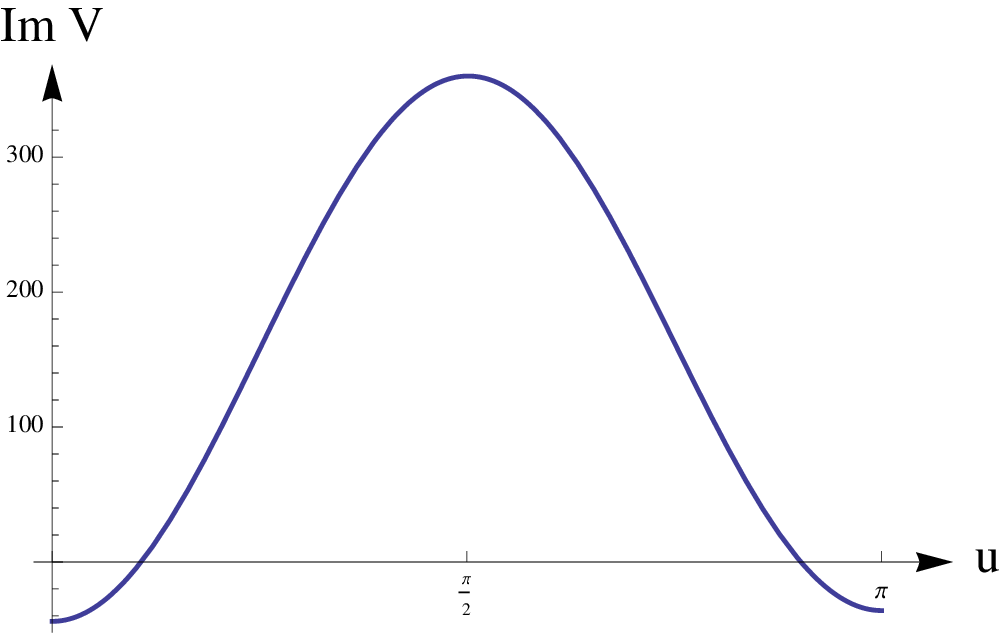} 
\caption{Typical plot of the potential~$V$.}
\label{figpot1}
\end{figure}

We now explain qualitatively how the spectrum of the angular operator looks and how
this qualitative behavior can be understood. This will also motivate and explain the statements
in Theorem~\ref{thmmain}.
Before discussing the effect of the imaginary part, we consider the situation
that~$\Omega$ and~$\lambda$ are real, so that~$V$ is real-valued.
Then the spectrum can be understood most easily by considering the
Sturm-Liouville equation~\eqref{Schrodinger} as a
one-dimensional Schr\"odinger equation with Hamiltonian~\eqref{Hamilton}.
As shown on the left of Figure~\ref{figpot1},
the potential looks typically like a {\em{double-well potential}}.
This potential is {\em{approximately symmetric}} (because the quadratic terms in~$\Omega$ are symmetric around~$\pi/2$ according to~\eqref{Wsymm}, but the terms~\eqref{Wasy} are anti-symmetric).
If instead of a double-well potential we had a single-well potential, the $n^\text{th}$ eigenvalue
could be computed approximately for large~$n$ by the {\em{Bohr-Sommerfeld-Wilson quantization condition}}
(see~\cite[\S48]{landau3} or~\cite[eq.~(2.5.51)]{sakurai})
\[ \oint p \,dq = 2 \pi n \:, \]
where one integrates momentum along a closed classical path of the particle. Thus, denoting
the zeros of~$V=W-\lambda$ by~$\ul$ and~$\ur$ (with~$\ul < \ur < \pi/2$), we obtain
\beq \label{bohr}
\int_{\ul}^{\ur} 2 \,\sqrt{\lambda_n - W} \, du = 2 \pi n\:.
\eeq
From this formula, the expected eigenvalue gaps can be computed by
\begin{align*}
\pi &\approx \int_{\ul(n+1)}^{\ur(n+1)} \sqrt{\lambda_{n+1} - W} \: du
- \int_{\ul(n)}^{\ur(n)} \sqrt{\lambda_n - W}\: du
\approx \int_{\ul}^{\ur} \frac{\lambda_{n+1} - \lambda_n}{2 \,\sqrt{\lambda_n - W}} du
\end{align*}
so that
\beq \label{gaps}
\lambda_{n+1}-\lambda_n \approx 2 \pi 
\left( \int_{\ul}^{\ur} \frac{1}{\sqrt{\lambda_n - W}} \, du \right)^{-1} \:.
\eeq
In particular, for the large eigenvalues we obtain Weyl's asymptotics
(see~\cite[Section~11.6]{straussPDE})
\beq \label{weyl}
\lambda_n \simeq n^2 \:, \qquad \lambda_{n+1} - \lambda_n \simeq n \qquad
\text{(as~$n \rightarrow \infty$)}\:.
\eeq
Another spectral region of interest is if~$\lambda$ lies near the minimum of the potential.
Approximating~$W$ by a quadratic potential and using that~$W'' \simeq \Omega^2$,
in this case we obtain the scaling
\beq \label{harmonic}
\lambda_n \simeq n \,|\Omega| \:,\qquad \lambda_{n+1} - \lambda_n \simeq |\Omega| \qquad
\text{(if~$1 \ll n \ll |\Omega|$)}\:.
\eeq

These formulas describe the behavior of the eigenvalues if we had a single-well potential.
The eigenvalues of the double-well potential can be understood by
considering two Hamiltonians with a single-well potential and by weakly coupling them
together via a potential barrier (see for example~\cite[Section~3.3]{schwabl1}).
If our double-well potential was symmetric about~$\pi/2$, the two
single-well Hamiltonians would have degenerate eigenvalues. Coupling them
together slightly removes the degeneracy, leading to the well-known eigenfunctions
with even and odd parity (similar as considered for example in~\cite[Sections~3.4 and~3.5]{schwabl1}).
In this way, we would end up with pairs of eigenvalues.
These pairs would be separated by spectral gaps having the behavior~\eqref{gaps}.
Since in our situation, the double-well potential is not symmetric about~$\pi/2$,
we do not know a-priori whether the eigenvalues of the two single-well Hamiltonians
are degenerate or not. But we can conclude that the eigenvalues of the
double-well Hamiltonian can appear at most in pairs, separated by gaps which again
scale according to~\eqref{gaps}. If~$\lambda$ is chosen much larger than the potential
barrier, the eigenfunctions no longer see a double-well potential. Therefore,
Weyl's asymptotics~\eqref{weyl} should again hold for the large eigenvalues.

These simple qualitative arguments already allow us to understand the statement of Theorem~\ref{thmmain}
in the special case of a real potential. Namely, the operator~$Q_0$ is the spectral projection
on all the small eigenvalues, for which the Born-Sommerfeld rule is not a good approximation.
The operators~$Q_1, Q_2, \ldots$ are spectral projection operators corresponding to
one or two eigenvalues (depending on whether there is a spectral pair or not).

\begin{figure}
\includegraphics[width=6.5cm]{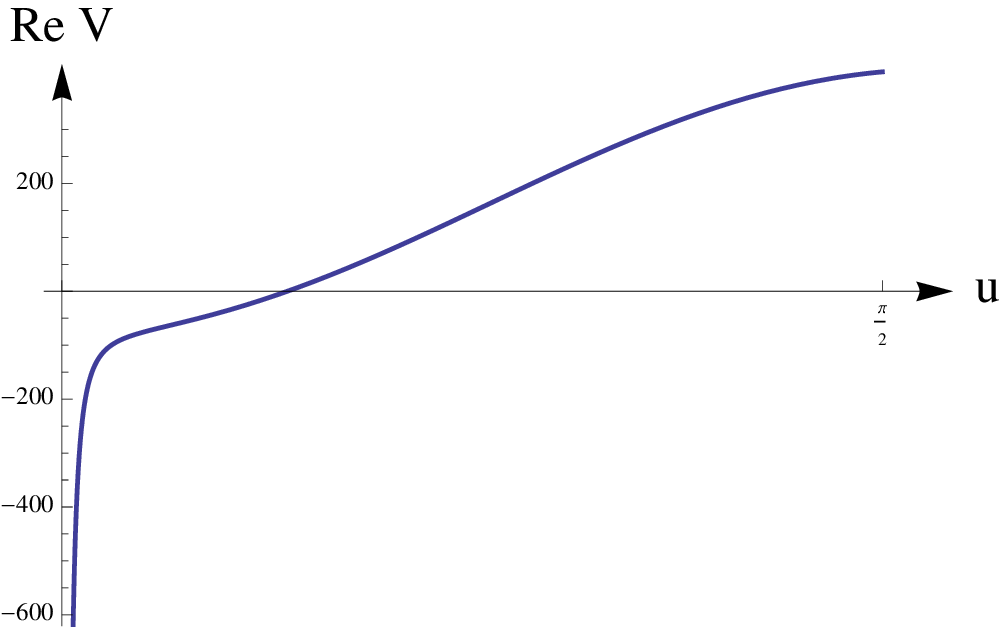} $\qquad$
\includegraphics[width=6.5cm]{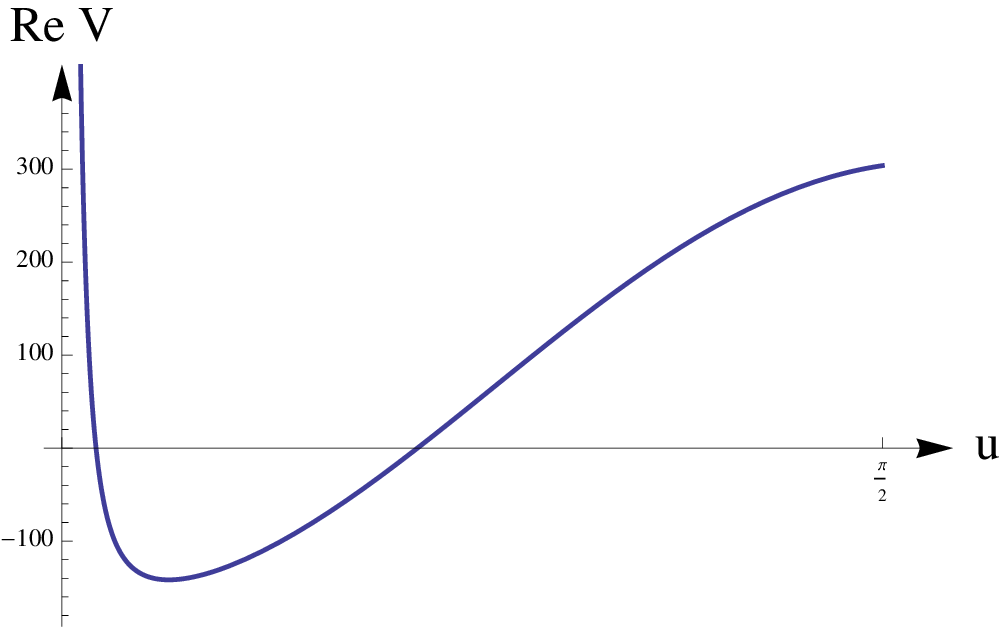} 
\caption{The poles of~$\re V$ in the case~$k=s$ (left) and~$k \neq s$ (right).}
\label{figpot2}
\end{figure}

Before moving on to the complex potential, we remark that in the case~$k = \pm s$, the
potential at the pole goes to minus infinity (for a typical example see Figure~\ref{figpot2}).
However, it turns out that, using the known asymptotics of the wave functions near the pole,
the above qualitative arguments still go through if we choose~$\ul \sim |\Omega|^{-\frac{1}{2}}$ close to the
inflection point of the potential.

We next discuss the situation for a complex potential.
One potential method is to treat the imaginary part of~$W$ as a {\em{slightly non-selfadjoint perturbation}}
(see~\cite[V.4.5]{kato} or as the application to the spheroidal wave operator in~\cite[Section~8]{angular}).
For this method to be applicable, the imaginary part of the potential must be small compared to the gaps, i.e.\
\beq \label{slightly}
|\im W| \leq \lambda_{n+1} - \lambda_n \:.
\eeq
For any fixed~$\Omega$, this condition will be satisfied for sufficiently large~$n$
in view of Weyl's asymptotics~\eqref{weyl}.
But the inequality~\eqref{slightly} cannot be satisfied uniformly in~$\Omega$, as the
following argument shows: Using~\eqref{ocond} in~\eqref{Wdef}, one sees that
\[ \sup_{(0, \pi)} |\im W| \gtrsim c\, |\Omega| \]
with a constant~$c$ which may be large. Therefore, the inequality~\eqref{slightly}
is in general violated if we are in the asymptotic regime~\eqref{harmonic}.
By choosing~$\Omega$ large, one can arrange that this asymptotic regime 
includes arbitrarily many eigenvalues. We conclude that $\im W$ {\em{cannot}} in general
be treated as a slightly non-selfadjoint perturbation.
This means qualitatively that the imaginary part of~$W$ shifts the eigenvalues considerably
on the scale of the gaps. The eigenvalues will typically move into the complex plane.
Moreover, degeneracies and Jordan chains may form.

In order to locate the spectrum in the complex plane, for a complex potential
whose real part has a single well one can again use the Bohr-Sommerfeld condition~\eqref{bohr},
which now makes a statement on both the real and imaginary parts of the integral on
the left~\eqref{bohr}. Treating the imaginary part of~$\lambda-W$ as a perturbation, we thus obtain
to first order
\begin{align}
\int_{\ul}^{\ur} \sqrt{\re \big( \lambda_n - W \big)} \, du &= \pi n \label{bohrc1} \\
\int_{\ul}^{\ur} \frac{\im (\lambda_n - W)}{\sqrt{\re ( \lambda_n - W )}} \, du &= 0 \:. \label{bohrc2}
\end{align}
We will prove that these relations really make it possible to locate the spectrum in the complex plane.
Applying these relations naively, we find for the Hamiltonian with the single-well potential
that the real part of the eigenvalues behaves just as discussed for the real potential.
The imaginary part of the potential, however, must be adjusted such that~\eqref{bohrc2} holds.
In particular, we again find that the spectral points form at most pairs, separated by
spectral gaps which scale similar to~\eqref{gaps}.
If the two spectral points of the pair coincide, a Jordan chain of length at most two may form.
In this way, one can understand all statements of Theorem~\ref{thmmain}.

\section{Overview of the Proof of the Main Theorem}
Making the above qualitative arguments precise requires an intricate combination
of different mathematical methods.
In order to facilitate reading, we now give a short overview of the proof of Theorem~\ref{thmmain}.
In Section~\ref{secgenFA}, we collect general statements on Sturm-Liouville operators
with a complex potential. We show that the spectrum is purely discrete, and that
the Hilbert space can be decomposed into a direct sum of invariant subspaces.
Moreover, idempotent operators mapping onto these invariant subspaces can be
constructed using contour integral methods.
In Section~\ref{secosc} we introduce a useful method for analyzing the oscillatory behavior
of solutions of the Sturm-Liouville equation of the form~\eqref{phiDdef}.
These estimates are essential for locating the spectrum and for making the
Bohr-Sommerfeld condition~\eqref{bohrc1} precise.

Our proof involves a deformation argument where we continuously deform
a real potential to our complex potential (Section~\ref{secdeform}).
Moreover, in our proof we will sometimes be able to treat the imaginary part of the potential
as a perturbation (cf.\ Section~\ref{secslightly}).
The starting point of these methods is to have detailed information on the spectrum and the
spectral gaps for a real potential. These estimates are worked out in Section~\ref{seclower}.

In Section~\ref{secslightly} we employ the method of slightly non-selfadjoint perturbations
to obtain the desired spectral representation provided that~$\Omega$ lies in
bounded set (see Proposition~\ref{prpbounded}). Therefore, all the subsequent sections are
devoted to the problem of getting estimates for large~$|\Omega|$, uniformly in the spectral parameter~$\lambda$.

In Section~\ref{secimVes} we derive an a-priori estimate for the imaginary parts of all eigenvalues.
The method is to evaluate an expectation value (see~\eqref{imVexpect}) giving an equation which makes the
Bohr-Sommerfeld condition~\eqref{bohrc2} precise.
This a-priori estimate is needed in order to distinguish the different cases and regions in Section~\ref{secoverview}.
In Section~\ref{secimVes2} we shall return to the method and refine it considerably.

For the remaining estimates we shall construct approximate solutions of the
Sturm-Liouville equation by glueing together WKB, Airy and parabolic cylinder functions
as well as asymptotic solutions near the poles at~$u=0$ and~$u=\pi$. 
Moreover, we derive rigorous error bounds.
In Section~\ref{secoverview} we give an overview of the different cases and regions
and explain how to locate the spectrum.
The detailed estimates are worked out in Section~\ref{secesregion}.

Section~\ref{secimVes2} gives refined integral estimates of the imaginary part of the potential
(see Propositions~\ref{prpimVsqrtV} and~\ref{prpdelta}). These estimates make use of the
specific form of our potential and will be needed several times in the subsequent sections.

In order to show that the Jordan chains have length at most two, our method is to show that if~$\lambda_0$
is an eigenvalue, then there is an annular region around~$\lambda_0$ which contains at
most one other eigenvalue (see Figure~\ref{figannulus}). In order to construct this annular region, we
differentiate the equations with respect to~$\lambda$ and use an implicit function argument.
The $\lambda$-derivatives are computed and estimated in Section~\ref{seclambda}.
The construction of the annular regions is given in Section~\ref{secannular}.

Section~\ref{secdwp} is devoted to estimates of the Green's function.
Here the main task is to estimate the Wronskian~$w(\phiD_L, \phiD_R)$ of the
fundamental solutions in~\eqref{phiDdef}.

In Section~\ref{secdeform} we continuously deform the potential
from a real potential to our complex potential. Combining all the results from the previous sections,
we can track the eigenvalues and control the spectral gaps. We also derive uniform norm estimates for
the operators~$Q_n$ and show that their sum converges strongly to the identity.

\section{General Functional Analytic Results} \label{secgenFA}
In order to get into the standard functional analytic framework, we consider
the Sturm-Liouville operator~\eqref{Hamilton} as an operator on the Hilbert space $L^2((0, \pi))$.
As the dense domain of definition we choose those function in~$C^2((0, \pi)) \cap L^2((0, \pi))$
which satisfy the boundary conditions~\eqref{Dirichlet}.

\begin{Lemma} \label{lemmadiscrete}
The spectrum of the Hamiltonian~\eqref{Hamilton} is discrete and has no limit points.
\end{Lemma}
\Proof For Sturm-Liouville equations with a continuous potential, this is proved in~\cite[Chapter~12]{coddington}.
Since the potential~\eqref{Vdef} has poles at~$u=0$ and~$u=\pi$, we give the proof in detail.
For any~$\lambda \in \C$, we choose non-trivial solutions~$\phi_L$ and~$\phi_R$ with the generic
asymptotic behavior~\eqref{phiDdef}. 
These solutions can be chosen to depend locally holomorphically on~$\lambda$ in the sense that
every~$\lambda_0 \in \C$ has an open neighborhood~$U$ such that the functions~$\phi_L$
and~$\phi_R$ are holomorphic in~$\lambda \in U$
(these holomorphic families can be constructed for example by taking the solutions of the
Sturm-Liouville equation for variable~$\lambda$ but fixed boundary values at some~$u \in (0, \pi)$).
Then the functions~$\phiD_L$ and~$\phiD_R$ defined by~\eqref{phiDdef} as well as
their Wronskian in~\eqref{evalcond} are also holomorphic in~$\lambda \in U$.

Let us show that the function~$w(\phiD_L, \phiD_R)$ does not vanish identically.
If this were the case, by analytic continuation we would conclude that~$w(\phiD_L, \phiD_R)$
vanishes identically for all~$\lambda \in \C$. Thus for every~$\lambda \in \C$ there would
exist a non-trivial solution~$\phi$ satisfying the boundary conditions~\eqref{Dirichlet}.
On the other hand, the computation
\[ \frac{d^2}{du^2} |\phi|^2 = 2 \,\big| \phi' \big|^2 + 2 |\phi|^2\, \re (W-\lambda) \]
shows that if~$\lambda$ is large and negative, then the absolute square of~$\phi$
is convex away from small neighborhoods of the poles at~$0, \pi$.
But this convexity is incompatible with the asymptotics near the poles in~\eqref{Dirichlet},
a contradiction.

The result follows because holomorphic functions which do not vanish identically have isolated zeros.
\QED

For a self-adjoint operator, one can construct the spectral projection operators
by integrating the resolvent along a closed contour. In our non-selfadjoint setting,
where the operator need not be diagonalizable, we cannot expect to obtain a
spectral decomposition. But we can detect invariant subspaces:
\begin{Lemma} \label{lemmacontourproduct}
Let~$\Gamma$ be a closed contour which lies entirely in the resolvent set and encloses
points in the spectrum with winding number one.
Then the contour integral
\beq \label{QGamma}
Q_\Gamma := -\frac{1}{2 \pi i} \ointctrclockwise_{\Gamma} s_\lambda\: d\lambda
\eeq
defines a bounded linear operator whose image is the invariant subspace corresponding to
the spectral points enclosed by~$\Gamma$.

The operator~$Q_\Gamma$ is idempotent. Moreover, the product of two operators~$Q_\Gamma$
and~$Q_{\Gamma'}$ is given by
\beq \label{Qprod}
Q_\Gamma \,Q_{\Gamma'} = Q_{\tilde{\Gamma}} \:,
\eeq
where~$\tilde{\Gamma}$ is any contour which encloses precisely all the spectral points enclosed by~$\Gamma$
and~$\Gamma'$, all with winding number one.
\end{Lemma}
\Proof We first show that~$Q_\Gamma$ is idempotent. Multiplying the identity
\[ (H-\lambda) - (H-\lambda') = \lambda'-\lambda \]
for~$\lambda \neq \lambda'$ from the left by~$s_\lambda$ and from the right by~$s_{\lambda'}$,
one obtains the resolvent identity (see for example~\cite[Theorem~VI.5]{reed+simon})
\[ s_\lambda \:s_{\lambda'} = \frac{1}{\lambda-\lambda'} \left(
s_\lambda - s_{\lambda'} \right) . \]
We let~$\Gamma'$ be a contour obtained by continuously deforming the contour~$\Gamma$
in the resolvent set such that every point of~$\Gamma$
is enclosed by~$\Gamma'$ with winding number one. Then, using the resolvent identity,
\begin{align}
Q_\Gamma \, Q_\Gamma &= Q_\Gamma \, Q_{\Gamma'}
= -\frac{1}{4 \pi^2} \ointctrclockwise_{\Gamma} d\lambda \ointctrclockwise_{\Gamma'} d\lambda'\:
\frac{1}{\lambda-\lambda'} \left( s_\lambda - s_{\lambda'} \right) \notag \\
&= -\frac{1}{4 \pi^2} \ointctrclockwise_{\Gamma}  
\bigg( \ointctrclockwise_{\Gamma'} \frac{d\lambda'}{\lambda-\lambda'} \bigg)\, s_\lambda\: d\lambda
+\frac{1}{4 \pi^2} \ointctrclockwise_{\Gamma'}
\bigg( \ointctrclockwise_{\Gamma} \frac{d\lambda}{\lambda-\lambda'} \bigg)\, s_{\lambda'}\: d\lambda' \:.
\label{contour1}
\end{align}
Carrying out the inner contour integrals with residues, the integral in the first summand
gives~$-2 \pi i$, whereas the integral in the second summand vanishes.
We conclude that~$Q_\Gamma \, Q_\Gamma = Q_\Gamma$.

In order to prove the more general formula~\eqref{Qprod}, it is convenient to deform the contours
and to decompose each contour integral into a finite sum of integrals where each contour encloses
only one spectral point. Then, in view of the idempotence of the~$Q_\lambda$,
it remains to prove~\eqref{Qprod} in the case that~$\Gamma$ and~$\Gamma'$
enclose different points of the spectrum. By continuously deforming~$\Gamma'$ without
crossing spectral points we can again arrange that
the contours~$\Gamma$ and~$\Gamma'$ do not intersect. As in~\eqref{contour1}, we obtain
\beq \label{contour2}
Q_\Gamma \, Q_{\Gamma'}
= -\frac{1}{4 \pi^2} \ointctrclockwise_{\Gamma}  
\bigg( \ointctrclockwise_{\Gamma'} \frac{d\lambda'}{\lambda-\lambda'} \bigg)\, s_\lambda\: d\lambda
+\frac{1}{4 \pi^2} \ointctrclockwise_{\Gamma'}
\bigg( \ointctrclockwise_{\Gamma} \frac{d\lambda}{\lambda-\lambda'} \bigg)\, s_{\lambda'}\: d\lambda' \:.
\eeq
Since the contours enclose different
points in the spectrum, no point of~$\Gamma$ is enclosed by~$\Gamma'$ and vice versa.
Hence the inner integrals in~\eqref{contour2} vanish, proving that~$Q_\Gamma \, Q_{\Gamma'}=0$.

It remains to show that the image of~$Q_\Gamma$ consists of the invariant subspaces
corresponding to all the spectral points enclosed in~$\Gamma$.
Let~$\Omega \subset \C$ be the open set enclosed by~$\Gamma$.
Since the spectral points are isolated, we may decompose~$\Omega$ into a finite number
of subsets such that the boundary of each subset is a closed contour enclosing only one
spectral point. Thus it suffices to consider the situation that~$\Gamma$ encloses exactly
one spectral point~$\lambda_0$. Since the Wronskian in~\eqref{sldef} is holomorphic in~$\lambda$,
the resolvent~$s_\lambda$ at~$\lambda_0$ has a pole of finite order~$n$. Iterating the identity
\[ H Q_\Gamma = -\frac{1}{2 \pi i} \ointctrclockwise_{\Gamma} (\lambda s_\lambda-\1)\: d\lambda
= -\frac{1}{2 \pi i} \ointctrclockwise_{\Gamma} \lambda \:s_\lambda\: d\lambda \:, \]
we obtain
\[ (H-\lambda_0)^n Q_\Gamma = -\frac{1}{2 \pi i} \ointctrclockwise_{\Gamma} (\lambda-\lambda_0)^n
\:s_\lambda\: d\lambda = 0 \:, \]
where in the last step we used that the integrand is holomorphic.
Hence every vector in the image of~$Q_\Gamma$ is contained in the invariant subspace
corresponding to~$\lambda_0$. Conversely, let~$\psi$ be a vector in this invariant subspace. Then
there is~$N \in \N$ such that~$(H-\lambda_0)^N \psi = 0$, and thus
\[ 0 = \big( (H-\lambda) + (\lambda-\lambda_0) \big)^N \psi = \sum_{k=0}^N
\begin{pmatrix} N \\ k \end{pmatrix} (\lambda-\lambda_0)^{N-k} \: (H-\lambda)^k \psi \:. \]
Multiplying by~$s_\lambda$ and solving for~$s_\lambda \psi$, we obtain
\[ s_\lambda \psi = -\sum_{k=1}^N
\begin{pmatrix} N \\ k \end{pmatrix} (\lambda-\lambda_0)^{-k}\, (H-\lambda)^{k-1} \,\psi \:. \]
Taking the contour integral, a computation with residues yields
\begin{align*}
Q_\Gamma \psi &= \frac{1}{2 \pi i} \sum_{k=1}^N
\begin{pmatrix} N \\ k \end{pmatrix} \ointctrclockwise_{\Gamma} \frac{(H-\lambda)^{k-1} \psi}
{(\lambda-\lambda_0)^k}\: d\lambda 
= \frac{1}{2 \pi i} \sum_{k=1}^N
\begin{pmatrix} N \\ k \end{pmatrix} \ointctrclockwise_{\Gamma} \frac{\psi}
{(\lambda-\lambda_0)}\: d\lambda \\
&= \sum_{k=1}^N \begin{pmatrix} N \\ k \end{pmatrix} (-1)^{k-1}\: \psi
= \psi - \sum_{k=0}^N \begin{pmatrix} N \\ k \end{pmatrix} (-1)^{k}\: \psi =
\psi - (1-1)^N \psi = \psi \:.
\end{align*}
Hence~$\psi$ really lies in the image of~$Q_\lambda$. This concludes the proof.
\QED
This lemma also shows that the dimension of the invariant subspace is at most the
order of the pole of the resolvent.

The next lemma bounds the resolvent away from the real axis.
\begin{Lemma} \label{lemmases}
If
\beq \label{imV}
\inf_{(0, \pi)} |\im V| >0 \:,
\eeq
then the Wronskian in~\eqref{evalcond} has no zeros. Moreover, the resolvent is bounded by
\[ \|s_\lambda \| \leq \left( \inf_{(0, \pi)} | \im V | \right)^{-1}\]
(where~$\| \cdot \|$ denotes the $\sup$-norm on~$\H_k$).
\end{Lemma}
\Proof Since~$\im V$ is continuous, the condition~\eqref{imV} implies that~$\im V$ is 
either always positive or always negative.
We only give the proof in the first case because the second case is similar.
If the Wronskian in~\eqref{evalcond} is zero, there is a non-trivial solution~$\phi$
of the Sturm-Liouville equation~\eqref{5ode} with the asymptotics~\eqref{Dirichlet}.
In the case~$k \pm s \neq 0$, differentiating the asymptotics~\eqref{genasy},
one sees that
\[ \phi(u) \sim u^{\frac{1}{2} + |k-s|} \:\big(1+ \O(u) \big) \qquad \text{and} \qquad
\phi'(u) \sim u^{-\frac{1}{2} + |k-s|} \:\big(1+ \O(u) \big) \:. \]
As a consequence, we do not get boundary terms when integrating by parts as follows,
\beq \label{PI}
0 = \la \phi \,|\, (-\partial_u^2+V) \phi \ra_{L^2} = \la \partial_u \phi \,|\, \partial_u \phi \ra_{L^2}
+ \la \phi \,|\, V \phi \ra_{L^2}\:.
\eeq
Taking the imaginary part, we conclude that
\beq \label{im0}
0 = \int_0^\pi \im V\: |\phi|^2 \:,
\eeq
in contradiction to~\eqref{imV}. In the case~$k=s$, the
situation is a bit more subtle because differentiating the asymptotics~\eqref{ex1}, one sees that
\[ \phi(u) = c_1 \sqrt{u} + \O(u) \qquad \text{and} \qquad
\phi'(u) = \frac{c_1}{2 \sqrt{u}} + \O \big(u^0 \big)\:. \]
This implies that integrating by parts in~\eqref{PI} we get real-valued boundary terms,
so that~\eqref{im0} again holds. The remaining case~$k=-s$ is treated similarly by
differentiating~\eqref{ex2}.

Next, setting~$\psi = s_\lambda \phi$ and again integrating by parts, we obtain
\[ \|\psi\|\, \|\phi\| \geq \big| \la \psi | \phi \ra_{L^2} \big| =  \big| \la \psi | (-\partial_u^2+V) \psi \ra_{L^2} \big|
\geq \big| \im \la \psi |  V \psi \ra_{L^2} \big| \geq \|\psi \|^2 \:\inf_{(0,\pi)} \im V \:, \]
implying that
\[ \|\phi\| \geq \|s_\lambda \phi \| \:\inf_{(0,\pi)} \im V  \:. \]
Since this inequality holds for all~$\phi \in \H_k$, the result follows.
\QED

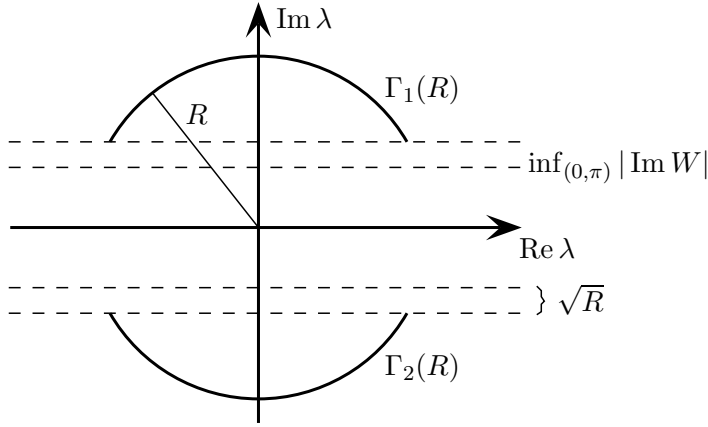
\begin{figure}
\scalebox{1} 
{
\begin{pspicture}(0,-2.82)(9.94,2.82)
\usefont{T1}{ptm}{m}{n}
\rput(7.16,-0.475){$\re \lambda$}
\psarc[linewidth=0.04](3.32,-0.2){2.28}{30.0}{150.0}
\psarc[linewidth=0.04](3.32,-0.2){2.28}{-150.0}{-30.0}
\psline[linewidth=0.04cm,arrowsize=0.02cm 8.0,arrowlength=1.4,arrowinset=0.4]{->}(0.02,-0.2)(6.82,-0.2)
\psline[linewidth=0.04cm,arrowsize=0.02cm 8.0,arrowlength=1.4,arrowinset=0.4]{->}(3.32,-2.8)(3.32,2.8)
\psline[linewidth=0.02cm](3.32,-0.2)(1.92,1.58)
\usefont{T1}{ptm}{m}{n}
\rput(2.5,1.305){$R$}
\psline[linewidth=0.02cm,linestyle=dashed,dash=0.16cm 0.16cm](0.0,0.6)(6.8,0.6)
\psline[linewidth=0.02cm,linestyle=dashed,dash=0.16cm 0.16cm](0.0,-1.0)(6.8,-1.0)
\psline[linewidth=0.02cm,linestyle=dashed,dash=0.16cm 0.16cm](0.0,0.94)(6.8,0.94)
\psline[linewidth=0.02cm,linestyle=dashed,dash=0.16cm 0.16cm](0.0,-1.34)(6.8,-1.34)
\usefont{T1}{ptm}{m}{n}
\rput(3.93,2.585){$\im \lambda$}
\usefont{T1}{ptm}{m}{n}
\rput(8.11,0.605){$\inf_{(0, \pi)} |\im W|$}
\usefont{T1}{ptm}{m}{n}
\rput(7.6,-1.155){$\sqrt{R}$}
\usefont{T1}{ptm}{m}{n}
\rput(5.5,1.625){$\Gamma_1(R)$}
\usefont{T1}{ptm}{m}{n}
\rput(5.5,-2.075){$\Gamma_2(R)$}
\psline[linewidth=0.02](7.02,-0.98)(7.08,-1.04)(7.08,-1.12)(7.12,-1.18)(7.08,-1.22)(7.08,-1.32)(7.02,-1.36)
\end{pspicture} 
}
\caption{The contour~$\Gamma$.}
\label{figcontour}
\end{figure}

We are now in the position to state a general completeness result. The method is based on an idea
in~\cite[proof of Theorem~2.12]{bachelot} and was used previously in~\cite{schdecay}.
First, we write the potential~\eqref{Vdef} and~\eqref{mudef} in the form
\[ V = W - \lambda \]
with~$W$ independent of~$\lambda$.
For given~$R>0$, we consider the two contours~$\Gamma_1$ and~$\Gamma_2$ in the complex
$\lambda$-plane defined by
\begin{align*}
\Gamma_1 &= \partial B_R(0) \cap \Big\{ {\mbox{Im}}\, \lambda < -\inf_{(0, \pi)} |\im W| - \sqrt{R} \Big\} \\
\Gamma_2 &= \partial B_R(0) \cap \Big\{ {\mbox{Im}}\, \lambda > \inf_{(0, \pi)} |\im W| + \sqrt{R} \Big\} \:.
\end{align*}
and set~$\Gamma(R) = \Gamma_1 \cup \Gamma_2$ (see Figure~\ref{figcontour}).

\begin{Thm} \label{thmcomplete} For any~$\phi \in C^\infty_0((0, \pi))$,
\beq \label{psirep}
\phi(u) = -\frac{1}{2 \pi i} \lim_{R \rightarrow \infty}
\int_{\Gamma(R)} (s_\lambda \phi)(u)\: d\lambda\:.
\eeq
\end{Thm}
\Proof Since the length of the contour~$S_1 \cup S_2 := \partial B_R(0) \setminus \Gamma(R)$
only grows like~$\sqrt{R}$,
\[ \left| \ointctrclockwise_{\partial B_R(0)} \frac{d\lambda}{\lambda} -
\int_{\Gamma(R)}  \frac{d\lambda}{\lambda} \right| \;\leq\; 
\frac{1}{R} \int_{S_1 \cup S_2} |d\lambda| 
\;\stackrel{R \rightarrow \infty}{\longrightarrow}\; 0\:. \]
As a consequence,
\beq \label{contour}
\frac{1}{2 \pi i} \lim_{R \rightarrow \infty} \int_{\Gamma(R)} \frac{d\lambda}{\lambda} = 1\:.
\eeq

Since our contours lie in the resolvent set, we know that for every~$\lambda \in \Gamma((R)$,
\[ \phi = s_\lambda\, (-\partial_u^2 + W - \lambda) \,\phi\:. \]
Dividing by~$\lambda$ and integrating over~$\Gamma(R)$, we can apply~\eqref{contour}
to obtain
\begin{align*}
\phi(u) &= \frac{1}{2 \pi i} \lim_{R \rightarrow \infty}
\int_{\Gamma(R)} \frac{d\lambda}{\lambda}\: \Big( s_\lambda\, \big(-\partial_u^2 + W - \lambda \big)
\,\phi \Big)(u) \\
&= -\frac{1}{2 \pi i} \lim_{R \rightarrow \infty}
\int_{\Gamma(R)} \left\{ \big( s_\lambda \phi \big)(u) \:-\:
\frac{1}{\lambda} \big( s_\lambda\, (-\partial_u^2 + W)\, \phi \big)(u) \right\} d\lambda\:.
\end{align*}
But the second term in the curly brackets vanishes in the limit,
because by Lemma~\ref{lemmases},
\[ \left| \int_{\Gamma(R)} \big( s_\lambda\, (-\partial_u^2 + W)\, \phi \big)(u) \:\frac{d\lambda}{\lambda} \right|
\leq \int_{\Gamma(R)} \frac{C}{\sqrt{R}}\: \frac{|d\lambda|}{|\lambda|} \;\leq\; \frac{2 \pi C}{\sqrt{R}} \:. \]
Thus~\eqref{psirep} holds.
\QED
This theorem shows that the operators~$Q_\Gamma$ defined by~\eqref{QGamma} converge
to the identity if~$\Gamma$ tends to a contour which encloses the whole spectrum.
The advantage of this method is that it does not require a functional analytic framework,
but only uses properties of the Green's function~$s_\lambda(u,u')$.
The drawback is that one obtains strong convergence only on a the dense subspace
of test functions. In order to prove strong convergence
on the whole Hilbert space, we will rely on the theory of slightly non-selfadjoint perturbations
(see Section~\ref{secslightly} and Section~\ref{seccomplete}).

\section{An Osculating Circle to the $\zeta$-Curve} \label{secosc}
In order to locate the spectrum, we need to find the zeros of the Wronskian in~\eqref{evalcond}.
The main difficulty is to understand the behavior of the integrals in~\eqref{phiDdef}.
To this goal, we now develop a method referred to as the ``osculating circle method.''
For ease in notation, we only consider the solution~$\phiD_L$ and omit the subscript~$L$.
We denote the integral in~\eqref{phiDdef} by
\beq \label{zetadef}
\zeta(u) := \int_0^u \frac{1}{\phi^2}\:.
\eeq
Then
\beq \label{phiDzeta}
\phiD = \zeta \,\phi \:, 
\eeq
making it possible to relate the behavior of~$\phiD$ to properties of the function~$\zeta(u)$.

\begin{figure}
\begin{picture}(0,0)%
\includegraphics{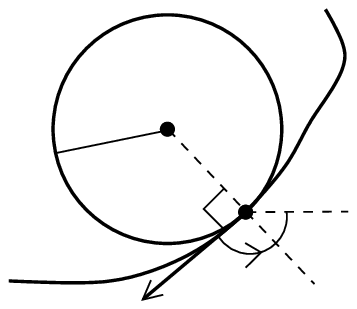}%
\end{picture}%
\setlength{\unitlength}{1989sp}%
\begingroup\makeatletter\ifx\SetFigFont\undefined%
\gdef\SetFigFont#1#2#3#4#5{%
  \reset@font\fontsize{#1}{#2pt}%
  \fontfamily{#3}\fontseries{#4}\fontshape{#5}%
  \selectfont}%
\fi\endgroup%
\begin{picture}(3975,3191)(-1557,-6579)
\put(1303,-5096){\makebox(0,0)[lb]{\smash{{\SetFigFont{11}{13.2}{\familydefault}{\mddefault}{\updefault}$\zeta$}}}}
\put(503,-4421){\makebox(0,0)[lb]{\smash{{\SetFigFont{11}{13.2}{\familydefault}{\mddefault}{\updefault}${\mathfrak{p}}$}}}}
\put(-17,-5026){\makebox(0,0)[lb]{\smash{{\SetFigFont{11}{13.2}{\familydefault}{\mddefault}{\updefault}$R$}}}}
\put(-1542,-6091){\makebox(0,0)[lb]{\smash{{\SetFigFont{11}{13.2}{\familydefault}{\mddefault}{\updefault}$\zeta(u)$}}}}
\put(1209,-6175){\makebox(0,0)[lb]{\smash{{\SetFigFont{11}{13.2}{\familydefault}{\mddefault}{\updefault}$2\vartheta$}}}}
\put(553,-6470){\makebox(0,0)[lb]{\smash{{\SetFigFont{11}{13.2}{\familydefault}{\mddefault}{\updefault}$\zeta'$}}}}
\end{picture}%
\caption{The osculating circle to the curve~$\zeta(u)$.}
\label{figosc}
\end{figure}

In order to clarify the evolution of the function~$\zeta(u)$ in the complex plane,
it is useful to consider the osculating circle to the curve~$\zeta$ at a point~$\zeta(u)$ (see Figure~\ref{figosc}).
The curvature~$K$ of the curve~$\zeta(u)$ and the radius~$R$ of the osculating circle are given by
(see for example~\cite[Theorem~5.1.6]{dubrovin+fomenko+novikov})
\beq \label{KRdef}
K = -\frac{\im(\zeta''\, \overline{\zeta'})}{|\zeta'|^3} \:,\qquad R = \frac{1}{|K|} \:.
\eeq
The center~$\p$ of the osculating circle is
\beq \label{Pdef}
\p = \zeta - \frac{i}{K}\: \frac{\zeta'}{|\zeta'|} \:.
\eeq
Moreover, we introduce the angle~$\vartheta$ as the argument of~$\phi$,
\beq \label{phiarg}
\phi = |\phi|\: e^{i \vartheta} \:.
\eeq
Then
\[ \frac{\zeta'}{|\zeta'|} = \frac{|\phi|^2}{\phi^2} = e^{-2 i \vartheta} \:, \] 
so that~\eqref{Pdef} becomes
\beq \label{zetarel}
\zeta = \p + \frac{i}{K}\: e^{-2 i \vartheta}\:.
\eeq
In order to simplify the computations, we always choose the phase and normalization of~$\phi$ such that
\beq \label{initial}
\phi(u_0)^2 \, \im y(u_0) = 1
\eeq
for some~$u_0$ which will be specified later. Moreover, we set
\beq \label{ybound}
y_0 = y(u_0) \:.
\eeq

Using the definition~\eqref{zetadef} of~$\zeta$ as well as the differential equation~\eqref{5ode},
we obtain useful formulas for~$\vartheta$, $K$, $\p$ and their derivatives.
\begin{Lemma} \label{lemmaosc}
\beq \label{thetaprel}
\vartheta' = \im y \:.
\eeq
Furthermore,
\begin{align}
K(u) &= 2\,  |\phi|^2\, \im y \:, & \hspace*{-1cm}
\p(u) &= \zeta - \frac{i}{2 \,\phi^2 \,\im y} \label{KPrel} \\
K'(u) &= 2\,|\phi|^2 \,\im V \:,&  \hspace*{-1cm}
\p'(u) &= \frac{i}{2 \,\phi^2 \,\im^2 y}\: \im V \:. \label{KPprel}
\end{align}
and
\beq \label{Rprel}
R'(u) = -\frac{\im V}{2\, |\phi|^2\, \im y\, |\im y|} \:.
\eeq
\end{Lemma} \noindent
Before coming to the proof, we point out that for a real potential, this lemma shows that~$K$, $\p$
and~$R$ are constant. Thus for a real potential, the curve~$\zeta(u)$ lies on a fixed circle with radius~$R$
centered at~$\p$. The position of~$\zeta$ on the circle is described by the angle~$\vartheta$,
and its evolution is described completely by~\eqref{thetaprel}. Moreover, we point out
that~$|\p'|=|R'|$, which can be understood from the fact that the point~$\zeta$ stays on the
circle as the osculating circles move.

\Proof[Proof of Lemma~\ref{lemmaosc}.] A direct computation yields
\begin{gather*}
\zeta' = \frac{1}{\phi^2} \:, \qquad \zeta'' = -2\, \frac{\phi'}{\phi^3} = -\frac{2 y}{\phi^2} \\
\vartheta' = -\frac{1}{2}\: \frac{d}{du}\: \im \log \zeta' = -\frac{1}{2}\: \im \frac{\zeta''}{\zeta'}
= \im \left( \phi^2 \,\frac{\phi'}{\phi^3} \right) = \im y \\
\im(\zeta''\, \overline{\zeta'}) = -2 \:\frac{\im y}{|\phi|^4} \:.
\end{gather*}
This gives~\eqref{thetaprel} as well as the formula for~$K$ in~\eqref{KPrel}. Using this formula in~\eqref{Pdef},
we obtain the second equation in~\eqref{KPrel}.

Next, we write out the real and imaginary parts of the Riccati equation~\eqref{riccati},
\begin{align}
\re y' &= \re V - \re^2 y + \im^2 y \label{rey} \\
\im y' &= \im V - 2 \re y \im y \:. \label{imy}
\end{align}
Using~\eqref{imy}, we obtain
\[ K' = 2  |\phi|^2 \im(y') +4 \re y  \:|\phi|^2\:\im y  = 2 \,\im V\, |\phi|^2 \:, \]
giving the formula for~$K'$ in~\eqref{KPprel}. Moreover,
\begin{align*}
\p' &= \zeta' + \frac{i}{2 \im^2 y}\:\im(y') \:\frac{1}{\phi^2}
+ \frac{i}{2 \im y}\: \frac{2}{\phi^3}\: \phi' \\
&= \frac{1}{\phi^2} + \frac{i}{2 \im^2 y} \Big( \im V - 2 \re y \im y \Big) \frac{1}{\phi^2}
+ \frac{i}{2 \im y}\: \frac{2}{\phi^2}\: y \\
&= \frac{1}{\phi^2} + \frac{i \im V}{2 \phi^2 \im^2 y} 
-\frac{i}{\phi^2 \im y} \:\re y + \frac{i}{\phi^2 \im y}\: y 
= \frac{i  \im V}{2 \phi^2 \im^2 y} \:.
\end{align*}
Finally, we differentiate~\eqref{KRdef},
\[ R' = -\frac{K K'}{|K|^\frac{3}{2}} \:, \]
and using~\eqref{KPrel} and~\eqref{KPprel} gives~\eqref{Rprel}.
\QED

\section{Estimates for a Real Potential} \label{seclower}
In Section~\ref{secdeform} we shall consider a homotopy of the potential
which joins the potential~$V$ with its real part.
In preparation for this analysis, we now derive eigenvalue estimates
for a Sturm-Liouville equation with a real potential. More precisely, we
replace the potential in the Sturm-Liouville equation~\eqref{5ode} by its real part,
\begin{equation} \label{5odere}
\Big( -\frac{d^2}{du^2} + \re V \Big) \phi = 0\:,
\end{equation}
where~$V$ is again given by~\eqref{Vdef} and~\eqref{mudef}.
We can assume
that~$\lambda$ is real, so that the equation can be written in the
Schr\"odinger form~\eqref{Schrodinger} with the Hamiltonian
\beq \label{Hamiltonreal}
H = -\frac{d^2}{d u^2} + \re W \:.
\eeq
This Hamiltonian has a unique self-adjoint extension, as the following
consideration shows: In the case~$k \neq s$, the asymptotics in~\eqref{genasy}
shows that one of the fundamental solutions
is square-integrable near~$u=0$, whereas the other fundamental solution is not.
Using Weyl's notion, the Sturm-Liouville operator
is in the limit point case at~$u=0$ (see~\cite[Sections~9.2, 9.3]{coddington}.
In the case~$k=s$, on the other hand, according to~\eqref{ex1}
both fundamental solutions are square integrable.
This is the so-called limiting circle case (see~\cite[Sections~9.4]{coddington}).
In all of these cases, our boundary conditions~\eqref{Dirichlet} give rise to
a unique self-adjoint extension (for details see~\cite[Sections~9.2, 9.3, 9.4]{coddington}
or~\cite[Chapter XIII.2]{dunford2}).

For ease in notation, we denote the selfadjoint extension of~\eqref{Hamiltonreal}
again by~$H$, and its domain of definition~$\D(H)$.
For the analysis of the spectrum, it is again useful to consider the
Riccati equation corresponding to~\eqref{5odere}, which we write as
\beq \label{Ricreal}
y' = \re V - y^2 \:,
\eeq
where we again set~$y:=\phi'/\phi$. We consider complex-valued solutions of this equation.
A direct computation (see also~\cite[eq.~(3.8)]{angular})
shows the product~$|\phi|^2\, \im y$ is a constant,
\beq \label{wrel}
w := |\phi|^2 \,\im y = \text{const}\:.
\eeq
This implies in particular that the function~$y$ cannot cross the real axis.

\subsection{A Node Theorem}
The classical node theorem (see for example~\cite[Theorem~14.10]{weidmann}) states that
the~$n^\text{th}$ eigenfunction of a Sturm-Liouville operator has exactly~$(n-1)$ zeros.
We now state and prove this node theorem in our setting. In the subsequent Sections~\ref{seclowbound}
and~\ref{secweyl}, we will apply the node theorem to obtain eigenvalue estimates and the Weyl asymptotics.
There are two reasons why we decided to give the proof of the node theorem in detail.
First, due to our singular boundary conditions, the proof given in most textbooks does not
apply to our problem. Second, our proof works with osculating circles and complex solutions of the corresponding
Riccati equation. It can be used as an introduction to the methods needed later in this paper.

\begin{Prp} \label{prpnode}
The spectrum of the Hamiltonian~\eqref{Hamiltonreal} is a discrete subset of~$\R$ which is bounded from below.
Numbering the eigenvalues in increasing order, $\lambda_0 \leq \lambda_1 \leq \ldots$,
the eigenfunction corresponding to~$\lambda_n$ has exactly~$n$ zeros on the open interval~$(0, \pi)$.
Moreover, choosing~$\lambda=\lambda_n$, any solution~$y$ of the Riccati equation~\eqref{Ricreal}
with~$\im y>0$ satisfies the relation
\beq \label{intn}
\int_0^\pi \im y = (n+1)\, \pi \:.
\eeq
\end{Prp}
\Proof We let~$\phi_1$ and~$\phi_2$ be two real-valued fundamental solutions of the ODE~\eqref{5odere}.
Since their Wronskian
\[ w := \phi'_1 \phi_2 - \phi_1 \phi'_2 \]
is a non-zero constant, the functions~$\phi_1$ and~$\phi_2$ cannot have common zeros.
Hence the complex solution~$\phi := \phi_1+i \phi_2$ has no zeros.
By choosing suitable fundamental solutions, we can arrange that the corresponding
solution of the Riccati equation~\eqref{Ricreal} satisfies~\eqref{wrel} with~$w=1$, so that
\beq \label{normalize}
|\phi|^2 \,\im y \equiv 1 \:.
\eeq

The relations~\eqref{phiDzeta} and~\eqref{zetadef} define a solution~$\phiD$ which
satisfies the boundary condition at~$u=0$. The boundary conditions at~$u=\pi$
are satisfied if and only if~$\zeta(\pi)=0$. Hence the condition for an eigenvalue can be stated as
\beq \label{evalcond4}
\zeta(\pi) = 0 \:.
\eeq
In order to control the behavior of the function~$\zeta$, we again use the
osculating circle method of Section~\ref{secosc}. For a real potential, the
relations~\eqref{KPprel} show that the center and the radius of the osculating circle are
fixed. Moreover, combining the first identity in~\eqref{KPrel} with~\eqref{normalize},
one sees that~$K=2$. Hence the formula~\eqref{zetarel} simplifies to
\beq \label{oscreal}
\zeta(u) = \p + \frac{i}{2}\: e^{-2 i \vartheta} \:,
\eeq
where~$\vartheta$ satisfies the differential equation~\eqref{thetaprel}.
As a consequence, the eigenvalue condition~\eqref{evalcond4} can be written as
\beq \label{econd}
\int_0^\pi \im y \in \pi \Z \:.
\eeq

The above formulas are valid if we let~$y$ be any solution of the Riccati equation in the upper half plane
and if we satisfy~\eqref{initial} (and consequently also~\eqref{normalize}) by letting
\beq \label{phifromy}
\phi(u) = \frac{1}{\im y(u_0)}\: \exp \Big( \int_{u_0}^u y \Big) \:.
\eeq
We now consider in particular a family of solutions~$y$ parametrized by~$\lambda \in \R$
such that
\beq \label{yllim}
\lim_{u \nearrow \pi} \phi^2(u)\: \partial_\lambda y(u) = 0 \qquad \text{for all~$\lambda \in \R$} \:.
\eeq
Such a family exists in view of the asymptotics near~$u=\pi$ as worked out in~\cite[Section~8]{tinvariant}
(namely, one chooses~$\phi$ with the asymptotics as in~\cite[Section~8]{tinvariant} with
coefficients adjusted such that the leading asymptotics is independent of~$\lambda$,
implying that~$\partial_\lambda y$ vanishes to leading order).

Differentiating~\eqref{riccati} and~\eqref{ybound} with respect to~$\lambda$ 
and using that~$\partial_\lambda V=-1$ gives
\beq \label{yleq}
y_\lambda' = -1 - 2 y y_\lambda \:,\qquad y_\lambda(u_0) = \partial_\lambda y_0\:.
\eeq
Solving this linear ODE by integration, we obtain
\[ \phi^2 y_\lambda = \phi^2(u_0)\: \partial_\lambda y_0 - \int_{u_0}^u \phi^2 \:, \]
so that
\beq \label{ylam}
y_\lambda(u) = \frac{\phi^2(u_0)}{\phi^2(u)}\: \partial_\lambda y_0 - \frac{1}{\phi^2(u)} \int_{u_0}^u \phi^2\:.
\eeq
Integrating this differential equation with respect to~$\lambda$ yields
\beq \label{ylform}
\int_0^\pi y_\lambda
= \phi^2(u_0)\: \partial_\lambda y_0 \: \zeta \big|_0^\pi
- \int_0^\pi \left( \frac{1}{\phi^2(u)} \int_{u_0}^u \phi^2 \right) du\:.
\eeq
In the last integral we perform the transformations
\begin{align*}
\int_{u_0}^\pi du \int_{u_0}^u dv \:\cdots
&= \int_{u_0}^\pi dv \int_v^\pi du \\
\int_0^{u_0} du \int_{u_0}^u dv \:\cdots
&=-\int_0^{u_0} du \int_u^{u_0} dv \:\cdots
= -\int_0^{u_0} dv \int_0^v du \:\cdots
\end{align*}
to obtain
\[ \int_0^\pi \left( \frac{1}{\phi^2(u)} \int_{u_0}^u \phi^2 \right) du
= \int_{u_0}^\pi \phi^2(v)\: \zeta \big|_v^\pi \:dv - 
\int_0^{u_0} \phi^2(v)\: \zeta \big|_0^v \:dv \:. \]
Taking the limit~$u_0 \nearrow \pi$ and using~\eqref{yllim}, the relation~\eqref{ylform} simplifies to
\[ \int_0^\pi \partial_\lambda y =\int_0^{u_0} \phi^2 \zeta = \int_0^{u_0} \phi \,\phiD \:. \]
The representation for~$\phi \phiD$ derived in Lemma~\ref{lemmasign} below shows that the
function~$\phi \phiD$ has a non-negative imaginary part, and that its imaginary part is
even strictly positive on a set of positive measure. Therefore,
\[ \int_0^\pi \partial_\lambda \im y > 0 \qquad \text{for all~$\lambda \in \R$}\:, \]
showing that for our family of functions~$y$, the integral~\eqref{econd} is indeed strictly increasing in~$\lambda$.

Combining this strict monotonicity of the integral~\eqref{econd} with the continuous dependence
on the parameter~$\lambda$, the intermediate value theorem gives rise to
eigenfunctions~$\lambda_n$ which are uniquely characterized by their number of zeros.
Since the integral~\eqref{econd} is strictly positive, converges to zero as~$\lambda \rightarrow -\infty$
and tends to infinity as~$\lambda \rightarrow \infty$ (using the WKB asymptotics),
we conclude that there is a sequence
of eigenvalues~$\lambda_0 < \lambda_1 < \ldots$ and that the eigenfunction corresponding
to~$\lambda_n$ has precisely $n$ zeros in the open interval~$(0, \pi)$.
Moreover, we conclude that~\eqref{intn} holds for the family of functions~$y$ satisfying~\eqref{yllim}.

In order to show that~\eqref{intn} holds for any smooth family of solutions~$y$ with~$\im y>0$,
we use the following continuity argument: For any fixed~$\lambda = \lambda_n$,
we denote the solution satisfying~\eqref{yllim}
by~$y_0$, and let~$y$ be any other solution with~$\im y>0$.
For any~$\tau \in [0,1]$, we let~$(y_\tau)$ be the family of solutions of~\eqref{Ricreal} with initial conditions
\[ y_\tau(\piot) = \tau \,y_0(\piot) + (1-\tau)\, y(\piot)\:. \]
Then the condition~\eqref{econd} is satisfied for any~$\tau$,
\[ \int_0^\pi \im y_\tau \in \pi \Z \:. \]
By continuity, this integral is independent of~$\tau$. We conclude that~\eqref{intn} holds
for any~$\tau \in [0,1]$, and in particular for~$\tau=1$.
\QED

\begin{Remark} {\em{
We remark that for a real potential, the eigenvalue condition~\eqref{econd} can also be
understood without going through the osculating circle estimates, as we now explain.
Since the real and imaginary parts of~$\phi$ form a fundamental system, the
solution~$\phiD$ can be represented as
\beq \label{phiDim}
\phiD = c\, \im(e^{-i \alpha} \phi)
\eeq
for a suitable phase~$\alpha$ and a complex prefactor~$c$. The zeros of~$\phiD$ are then
determined by the phase of~$\phi$,
\[ \phiD=0 \qquad \Longleftrightarrow \qquad \arg \phi \in \alpha + \pi \Z\:. \]
In particular, for~$\phiD$ to satisfy the Dirichlet boundary conditions, it follows that
\[ \arg \phi \big|_0^\pi \in \pi \Z \:. \]
Differentiating gives
\beq \label{argder}
\frac{d}{du} \arg \phi = \frac{d}{du} \im \log \phi = \im \: \frac{\phi'}{\phi} = \im y \:,
\eeq
and applying the fundamental theorem of calculus again gives~\eqref{econd}.
Moreover, one sees again that the integral in~\eqref{econd} gives $\pi$ times the number of 
zeros on~$(0, \pi)$ plus one.
Using the osculating circle method has the advantage that with~\eqref{phiDzeta} we have
an explicit formula for~$\phiD$, making it unnecessary to think about how the angle~$\alpha$
in~\eqref{phiDim} is to be chosen. }} \QEDrem
\end{Remark}

We append the lemma which shows that the integrand~$\im(\phi \phiD)$ has a definite sign.
\begin{Lemma} \label{lemmasign}
For every solution~$\phi$ satisfying the normalization condition~\eqref{initial}
(for any~$u_0 \in (0, \pi)$),
\[ \im \big( \phi(u) \, \phiD(u) \big) =
\frac{1}{2}\: |\phi(u)|^2 \left( 1 - \frac{\phi(u)^2}{|\phi(u)|^2} \: \lim_{v \searrow 0} \frac{|\phi(v)|^2}{\phi(v)^2}
\right) . \]
\end{Lemma}
\Proof Clearly, $\phiD$ is a linear combination of the fundamental solutions~$\phi$ and~$\overline{\phi}$, i.e.
\beq \label{phiDansatz}
\phiD = \alpha \,\phi + \beta \,\overline{\phi}
\eeq
for suitable coefficients~$\alpha, \beta \in \C$. In order to compute these coefficients, we compute the
Wronskians of~$\phiD$ with both~$\phi$ and~$\overline{\phi}$. First, using the ansatz~\eqref{phiDansatz},
we get
\[ w(\phi, \phiD) = \beta\, w(\phi, \overline{\phi}) \:,\qquad
w(\overline{\phi}, \phiD) = -\alpha \:w(\phi, \overline{\phi}) \:. \]
Next, using the representation~\eqref{phiDzeta} and~\eqref{zetadef}, we obtain
\begin{align*}
w(\phi, \phiD) &= \phi' \phi \zeta - \phi \:( \phi \zeta)' = -\phi^2 \,\zeta' = -1 \\
w(\overline{\phi}, \phiD) &= \overline{\phi}' \phi \zeta - \overline{\phi} \:\big( \phi \zeta)' =
-w(\phi, \overline{\phi})\, \zeta - |\phi|^2\, \zeta' = -w(\phi, \overline{\phi})\, \zeta - \frac{|\phi|^2}{\phi^2}\:.
\end{align*}
Comparing these formulas, we can compute~$\alpha$ and~$\beta$. We obtain that
for any~$v \in (0, \pi)$,
\[ \alpha = \zeta(v) + \frac{1}{w(\phi, \overline{\phi})}\: \frac{|\phi(v)|^2}{\phi(v)^2}
\:,\qquad \beta = -\frac{1}{w(\phi, \overline{\phi})}\:. \]
Finally, the normalization condition~\eqref{initial} implies that~$w(\phi, \phi')=2 i$.
Computing $\im \phi \phiD$ using the above relations,
taking the limit~$v \searrow 0$ and using that~$\lim_{v \searrow 0} \zeta(v)=0$,
we obtain the result.
\end{proof}

\subsection{Lower Bounds for Small Eigenvalues} \label{seclowbound}
In order to obtain eigenvalue estimates, we need to count the number of zeros of the
function~$\phiD$. Our method is to decompose the domain~$(0, \pi)$ into subintervals
on which the potential~$\re V$ has a definite sign.
On every interval where~$\re V$ is positive, the number of zeros is a-priori bounded:
\begin{Lemma} \label{lemmaatmostone}
If~$I$ is a closed interval with~$\re V|_I \geq 0$, then~$\phiD$ has at most one
zero on~$I$.
\end{Lemma}
\Proof Assume conversely that there is more than one zero on~$I$. We choose
two neighboring zeros~$u_1<u_2$. Then, possibly by flipping the sign of~$\phi$ we can arrange
that~$\phiD|_{(u_1, u_2)} > 0$. As a consequence, the function~$\phiD$ is convex on~$[u_1, u_2]$
(for details and other estimates using this convexity property see~\cite[Section~5]{angular}).
This implies that $\phiD$ is non-positive on~$(u_1, u_2)$, a contradiction.
\QED

\begin{Lemma} \label{lemmaZes}
Let~$y$ be any solution of the Riccati equation~\eqref{Ricreal} with~$\im y >0$.
Then the number~$Z$ of zeros of~$\phiD$ on an open interval~$I \subset (0, \pi)$ is bounded by
\[ -1 + \frac{1}{\pi} \int_I \im y \;\leq\; Z \;<\; 1+ \frac{1}{\pi} \int_I \im y \:. \]
\end{Lemma}
\Proof One method of proof is to consider the osculating circle for a real potential~\eqref{oscreal}
and to note that the change of the phase~$\vartheta$ is given by the differential equation~\eqref{thetaprel}.
Finally, the representation~\eqref{phiDzeta} shows that the zeros of~$\phiD$ coincide with the zeros of~$\zeta$.
An alternative method is to use the representation~\eqref{phiDim}
with~$\phi$ according to~\eqref{phifromy}, and to make use of the fact that~$\arg \phi$
satisfies the differential equation~\eqref{argder}.
\QED

Combining the node theorem of Proposition~\ref{prpnode} with the
last two lemmas, we obtain the following corollary.
\begin{Corollary} \label{corlower}
Let~$I_1, \ldots, I_k \subset (0, \pi)$ be open intervals such that~$\re V$ is non-negative
on the complement of~$I_1 \cup \cdots \cup I_k$. On the~$I_\ell$ we choose any solutions~$y_\ell$
of the Riccati equation~\eqref{Ricreal} with~$\im y_\ell >0$.
Then for the $N^\text{th}$ eigenvalue $\lambda_N$,
\[ \pi \left( N-2 k-1 \right) \:<\: \sum_{\ell=1}^k \int_{I_\ell} \im y_\ell \:\leq\: \pi \left( N+k \right)  \:. \]
\end{Corollary}

We now apply this corollary to the spheroidal wave operator. We restrict
attention to lower bounds for the eigenvalues, but remark that upper bounds could be
derived with similar methods.
\begin{Prp} \label{prplamlower}
For every constant~$c_3>0$ and any parameters~$k, s$, there is $N=N(c_3, k, s) \in \N$
such that for all~$\Omega$ in the range~\eqref{ocond} with~$|\Omega|$ sufficiently large,
the~$N^\text{th}$ eigenvalue is bounded from below by
\beq \label{lamN1}
\lambda_N \geq c_3 \:|\Omega| \:.
\eeq
\end{Prp}
\Proof In order to prove~\eqref{lamN1} we consider~$\lambda \leq c_3\, |\Omega|$ for large~$|\Omega|$. Then,
due to the summand~$\Omega^2 \sin^2 u$ in~\eqref{Vdef}, the real part of the potential
is non-negative except in a neighborhood of $u=0$ and~$u=\pi$.
By symmetry, it suffices to analyze the behavior in a neighborhood of~$u=0$.
Then the estimate
\beq \label{Vgrob}
\re V \geq -\frac{1}{4u^2} - (c_3+1) |\Omega| + \frac{|\Omega|^2}{2}\: u^2 
+ |\Omega|^2\, \O(u^3)
\eeq
shows that the potential is positive if
\beq \label{ulower}
u> u_1 := 2(c_3+1)\: |\Omega|^{-\frac{1}{2}}\:.
\eeq

We begin with the case~$k \neq s$. In this case, the estimate~\eqref{Vgrob} is improved to
\[ \re V \geq \frac{3}{4u^2} - (c_3+1) |\Omega| + \frac{|\Omega|^2}{2}\: u^2 
+ |\Omega|^2\, \O(u^3)\:. \]
In particular, we conclude that for large~$|\Omega|$,
\beq \label{reVlower}
\re V \geq - 2 \,(c_3+1) |\Omega| \qquad \text{(if~$k \neq s$)}\:.
\eeq
We choose a (possibly empty) interval~$(u_-, u_+)$ such that~$\re V$
is negative inside and non-negative outside this interval. In view of~\eqref{ulower} we may
choose~$u_-, u_+ \leq u_1$. In order to count the zeros of~$\phiD$ on the
interval~$(u_-, u_+)$ we can assume that the minimum~$u_0$ of~$\re V$
lies in the interval~$(u_-, u_+)$
(because otherwise the interval~$(u_-, u_+)$ is empty, and there is nothing to do).
We consider the solution~$y$ of the Riccati equation~\eqref{Ricreal} with initial conditions
\beq \label{yinit}
y(u_0) = i \sqrt{|\re V(u_0)|}\:.
\eeq
We now apply the T-method (see~\cite[Theorem~3.2]{tinvariant}
or~\cite[Lemma~4.1]{wdecay}), choosing~$\alpha \equiv 0$.
Then $U=\re V$, $\sigma \equiv 0$ and
\[ {\mathfrak{D}} = \frac{\re V'}{2} \]
(see~\cite[eqns.~(3.3)--(3.5)]{tinvariant}).
Using that~$\re V$ is monotone increasing on~$[u_0, u_+)$
and monotone decreasing on~$(u_-, u_0]$, we obtain
\[ T(u) = \left| \frac{\re V(u_0)}{\re V(u)} \right|^{\frac{1}{2}} \:, \]
giving rise to an invariant disk estimate with center~$m(u) = i \beta(u)$ and radius~$R$ given by
\begin{align*}
\beta(u) &= \frac{1}{2} \left( |\re V(u_0)|^\frac{1}{2} + |\re V(u)|\: |\re V(u_0)|^{-\frac{1}{2}} \right) \\
R(u) &= \frac{1}{2} \left( |\re V(u_0)|^\frac{1}{2} - |\re V(u)|\: |\re V(u_0)|^{-\frac{1}{2}} \right) .
\end{align*}
In particular, one sees that
\[ \im y \leq \sqrt{|\re V(u_0)|} \qquad \text{on~$(u_-, u_+)$}\:. \]
As a consequence,
\[ \int_{u_-}^{u_+} \im y \leq \sqrt{|\re V(u_0)|} \:|u_+-u_-|
\leq \sqrt{|\re V(u_0)|}\: u_1 \leq \big( 2(c_3+1) \big)^\frac{3}{2} \:, \]
where in the last step we applied~\eqref{ulower} and~\eqref{reVlower}.

In the case~$k=s$, on the other hand, the function~$\re V$ tends to minus infinity as~$u \searrow 0$.
We choose~$u_-=0$ and~$u_+ \leq u_1$ with~$\re V(u_+)=0$.
We first apply the invariant disk estimate near the pole
as worked out in~\cite[Section~8.1]{tinvariant}. This estimate applies up to
some $\tilde{u} \lesssim |\Omega|^{-\frac{1}{2}}$.
Choosing~$\tilde{u}$ such that~$\re V$ is monotone increasing on
the interval~$(\tilde{u}, u_1)$, on the remaining interval~$[\tilde{u}, u_1]$ we can
again use the $T$-method with~$\alpha \equiv 0$. Again, this gives rise to the estimate
\[ \int_{u_-}^{u_+} \im y \leq C(c_3) \:. \]

Working out similar estimates near~$u=\pi$, we can apply Corollary~\ref{corlower}
with~$k=2$. We conclude that we can choose~$N$ such that for all sufficiently large~$|\Omega|$,
the chosen~$\lambda$ is smaller than the $N^\text{th}$ eigenvalue. This concludes the proof.
\QED

\subsection{Weyl's Asymptotics} \label{secweyl}
In the next lemma we show that our boundary conditions~\eqref{Dirichlet} give rise to the
usual Weyl asymptotics.

\begin{Lemma} \label{lemmaweyl}
The spectrum of the Hamiltonian~\eqref{Hamiltonreal} with boundary conditions~\eqref{Dirichlet}
lies on the real axis and consist of points~$\lambda_0 < \lambda_1 < \cdots$. For large~$n$,
the eigenvalues and gaps have the asymptotics
\begin{align*}
\lambda_n &= n^2 + \O(n) \\
\lambda_{n+1} - \lambda_n &= 2 n + \O(n^0)\:.
\end{align*}
\end{Lemma}
\Proof We consider the family of solutions~$y$ of the Riccati equation~\eqref{Ricreal} with
initial conditions
\[ y(\piot) = i \sqrt{\lambda} \qquad \text{for~$\lambda \in \R^+$} \:. \]
Asymptotically for large~$\lambda$, the potential~$\re V$ becomes
nearly constant according to~\eqref{Vdef}, except at the poles at~$u=0$ and~$u=\pi$.
In the case~$k \neq s$, one can control the behavior near the poles by
using the $T$-method similar as explained after~\eqref{yinit}. We thus obtain asymptotically
\beq \label{intyes}
\int_0^\pi \im y = \sqrt{\lambda} \,\pi + \O \Big(\frac{1}{\sqrt{\lambda}} \Big) \:.
\eeq
In the case~$k = \pm s$, one can use the asymptotics of the fundamental solutions
as worked out in~\cite[Sections~7 and~8]{tinvariant} to again obtain~\eqref{intyes}.

Combining~\eqref{intyes} with~\eqref{intn} gives the result.
\QED

\section{Slightly Non-Selfadjoint Perturbations} \label{secslightly}
We now prove Theorem~\ref{thmmain} under the additional assumption that~$\Omega$
is restricted to a bounded set:
\begin{Prp} \label{prpbounded}
Let~$U \subset \C$ be a bounded set. Then for any~$s$ and~$k$ in the range~\eqref{skrange},
there is a positive integer~$N$ and a family of operators~$Q_n(\Omega)$ on~$\H_k$ defined for
all~$n \in \N \cup \{0\}$ and $\Omega \in U$
which has the properties~{\rm{(i)--(v)}} in the statement of Theorem~\ref{thmmain}.
\end{Prp} \noindent
This proposition differs from Theorem~\ref{thmmain} by the fact that here the
parameter~$N$ may depend on the set~$U$, whereas in Theorem~\ref{thmmain} the
parameter~$N$ is to be chosen uniformly for all~$\Omega$ in the unbounded strip~\eqref{ocond}.
This uniformity in~$\Omega$ is the main difficulty of the present paper; its proof will be the
concern of the remaining Sections~\ref{seclambda}--\ref{secdeform}.

\Proof[Proof of Proposition~\ref{prpbounded}]
We again consider the Hamiltonian~\eqref{Hamiltonreal} with a real potential
with boundary conditions~\eqref{Dirichlet}
Choosing contours which enclose each of the eigenvalues~$\lambda_0, \lambda_1, \ldots$
with winding number one, the contour integral~\eqref{QGamma} defines idempotent
operators~$Q_n$, $n \in 0, 1,\ldots$. Since~$H$ is formally self-adjoint, these operators are
symmetric, implying that the $Q_n$ are orthogonal projection operators onto mutually orthogonal
subspaces. For any~$\lambda \in \C \setminus \{ \lambda_0, \lambda_1, \ldots\}$,
we define the resolvent of the self-adjoint problem by
\[ s_\lambda = \sum_{n=0}^\infty \frac{1}{\lambda-\lambda_n} \:Q_n \:. \]
Here the sum converges absolutely in~$\Lin(\H)$. Moreover, the resolvent satisfies the identities
\[  \left( -\frac{d^2}{d u^2} + \re W - \lambda \1 \right) s_\lambda = \1 \qquad \text{and} \qquad
\|Q_\lambda\| = \sup_{n \in \N \cup \{0\}} \frac{1}{|\lambda-\lambda_n|} \:. \]

\begin{figure}
\begin{picture}(0,0)%
\includegraphics{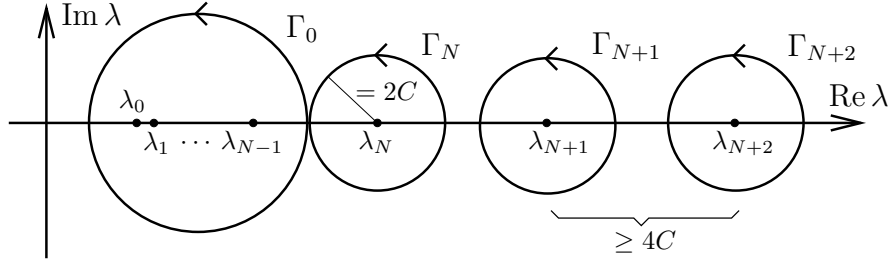}%
\end{picture}%
\setlength{\unitlength}{2072sp}%
\begingroup\makeatletter\ifx\SetFigFont\undefined%
\gdef\SetFigFont#1#2#3#4#5{%
  \reset@font\fontsize{#1}{#2pt}%
  \fontfamily{#3}\fontseries{#4}\fontshape{#5}%
  \selectfont}%
\fi\endgroup%
\begin{picture}(10326,3128)(508,-6214)
\put(6751,-4876){\makebox(0,0)[lb]{\smash{{\SetFigFont{10}{12.0}{\rmdefault}{\mddefault}{\updefault}$\lambda_{N+1}$}}}}
\put(4726,-4876){\makebox(0,0)[lb]{\smash{{\SetFigFont{10}{12.0}{\rmdefault}{\mddefault}{\updefault}$\lambda_N$}}}}
\put(1171,-3391){\makebox(0,0)[lb]{\smash{{\SetFigFont{12}{14.4}{\rmdefault}{\mddefault}{\updefault}$\im \lambda$}}}}
\put(8956,-4876){\makebox(0,0)[lb]{\smash{{\SetFigFont{10}{12.0}{\rmdefault}{\mddefault}{\updefault}$\lambda_{N+2}$}}}}
\put(3863,-3537){\makebox(0,0)[lb]{\smash{{\SetFigFont{12}{14.4}{\rmdefault}{\mddefault}{\updefault}$\Gamma_0$}}}}
\put(5472,-3745){\makebox(0,0)[lb]{\smash{{\SetFigFont{12}{14.4}{\rmdefault}{\mddefault}{\updefault}$\Gamma_N$}}}}
\put(7513,-3743){\makebox(0,0)[lb]{\smash{{\SetFigFont{12}{14.4}{\rmdefault}{\mddefault}{\updefault}$\Gamma_{N+1}$}}}}
\put(9856,-3759){\makebox(0,0)[lb]{\smash{{\SetFigFont{12}{14.4}{\rmdefault}{\mddefault}{\updefault}$\Gamma_{N+2}$}}}}
\put(10355,-4359){\makebox(0,0)[lb]{\smash{{\SetFigFont{12}{14.4}{\rmdefault}{\mddefault}{\updefault}$\re \lambda$}}}}
\put(7782,-6069){\makebox(0,0)[lb]{\smash{{\SetFigFont{10}{12.0}{\rmdefault}{\mddefault}{\updefault}$\geq 4C$}}}}
\put(1857,-4386){\makebox(0,0)[lb]{\smash{{\SetFigFont{10}{12.0}{\rmdefault}{\mddefault}{\updefault}$\lambda_0$}}}}
\put(2593,-4876){\makebox(0,0)[lb]{\smash{{\SetFigFont{10}{12.0}{\rmdefault}{\mddefault}{\updefault}$\cdots$}}}}
\put(2145,-4887){\makebox(0,0)[lb]{\smash{{\SetFigFont{10}{12.0}{\rmdefault}{\mddefault}{\updefault}$\lambda_1$}}}}
\put(3076,-4883){\makebox(0,0)[lb]{\smash{{\SetFigFont{10}{12.0}{\rmdefault}{\mddefault}{\updefault}$\lambda_{N-1}$}}}}
\put(4688,-4289){\makebox(0,0)[lb]{\smash{{\SetFigFont{10}{12.0}{\rmdefault}{\mddefault}{\updefault}$=2C$}}}}
\end{picture}%
\caption{Contour integrals for slightly non-selfadjoint perturbations.}
\label{figcircles}
\end{figure}

Our method for treating the imaginary part of the potential is to use the theory of slightly self-adjoint
perturbations (see~\cite[V.4.5]{kato}), similar as worked out in~\cite[Section~8]{angular}
or~\cite[Chapter~12]{coddington}. We first note that, since the poles in~\eqref{Wsymm}
are real-valued, the imaginary part of the potential is bounded,
\beq \label{imWes}
|\im W(u)| \leq C \qquad \text{for all~$u \in (0, \pi)$ and~$\Omega \in U$}
\eeq
(where the constant~$C$ clearly depends on~$U$). Next, using Weyl's asymptotics
of Lemma~\ref{lemmaweyl}, we can choose~$N$ so large that
\[ |\lambda_{n+1} - \lambda_n| \geq 4 C \qquad \text{for all~$n \geq N$}\:. \]
We choose contours~$\Gamma_n$ (for~$n \geq N$) as circles centered at~$\lambda_n$
with radius~$2C$. Moreover, we choose~$\Gamma_0$ as a circle which encloses the
eigenvalues~$\lambda_0, \ldots, \lambda_{N-1}$, and whose distance to the spectrum
is at least~$2C$ (see Figure~\ref{figcircles}).
Then for any~$\lambda$ on one of these contours,
\beq \label{ses}
\|s_\lambda\| \leq \frac{1}{2C} \:.
\eeq
This makes it possible to define the resolvent for the Hamiltonian~\eqref{Hamilton} with the
complex potential, which we denote for clarity by a tilde, via a Neumann series,
\beq \label{neumann}
\tilde{s}_\lambda := \sum_{k=0}^\infty \big(-s_\lambda \,\im W \big)^k  s_\lambda \:.
\eeq
We now integrate this resolvent along the contours~$\Gamma_n$,
\beq \label{cint}
\tilde{Q}_n := -\frac{1}{2 \pi i} \ointctrclockwise_{\Gamma_n} \tilde{s}_\lambda\: d\lambda
\:,\qquad n \in \{0, N, N+1, \ldots\} \:.
\eeq
As explained in~\cite[V.4.5]{kato}, these operators are idempotent and map onto the
invariant subspaces corresponding to the spectral points enclosed by the contour.
Moreover, it is shown in~\cite[V.4.5]{kato} that the spectral projections are complete.
The bound~\eqref{Qnb} follows immediately by estimating the contour integral~\eqref{cint}
and the Neumann series~\eqref{neumann} using~\eqref{ses} and~\eqref{imWes}.
\QED

\section{An A-Priori Estimate for $\im V$} \label{secimVes}
Assume that~$\lambda \in \C$ is an eigenvalue. We let~$\phiD$ be a corresponding eigenfunction. This
function satisfies the Dirichlet boundary conditions at~$u=0$ and~$u=\pi$.
Therefore, the corresponding
functions~$\phiD_L$ and~$\phiD_R$ as defined by~\eqref{phiDdef} are both multiples of~$\phiD$.
Multiplying the differential equation for~$\phiD$ by~$\overline{\phiD}$ and integrating, we obtain
\beq \label{imVexpect} 
0 = \int_0^\pi \overline{\phiD} \left( -\frac{d^2}{du^2} + V \right) \phiD
\overset{(\star)}{=}  \int_0^\pi \overline{\left( -\frac{d^2}{du^2} + \overline{V} \right)  \phiD} \phiD
= \int_0^\pi (V-\overline{V}) \:\overline{\phiD} \phiD \:,
\eeq
where in~($\star$) we integrated by parts and used the asymptotics
for the decaying solution in~\eqref{genasy} and~\eqref{ex1}, \eqref{ex2}
to conclude that the boundary terms vanish. We thus obtain the relation
\beq \label{imVint}
\int_0^\pi \im V\, |\phiD|^2 = 0 \:.
\eeq
This identity immediately gives rise to the following a-priori estimate.

\begin{Lemma} \label{lemmaimV}
Suppose that~$\lambda \in \C$ is an eigenvalue. Then
\[ |\im \lambda|,\: |\im V| \leq C\, |\Omega| \]
with a constant~$C$ which is independent of~$\lambda$ and~$\Omega$.
\end{Lemma}
\Proof Using the explicit form of the potential~\eqref{Vdef} together with~\eqref{ocond},
one sees that
\beq \label{inter}
\big| \im V - \im \lambda \big| \leq 2\,c\, |\Omega| + \text{const} \:.
\eeq
The integral equation~\eqref{imVint} implies that the function~$\im V$ must change sign
on the interval~$(0, \pi)$. As a consequence, the absolute value of~$\im \lambda$
is bounded by the right side of~\eqref{inter}. This gives the result.
\QED

\section{Overview of the Estimates for a Complex Potential} \label{secoverview}
We now enter the general estimates. Recall that our equations involve the
parameters~$k$, $s$, $\Omega$ and~$\lambda$.
We always keep~$k$ and~$s$ fixed. The parameters~$\Omega$ and~$\lambda$, however,
may vary in a certain parameter range to be specified later on,
and we must make sure that our estimates are uniform in these parameters.
In order to keep track of the dependence on~$\Omega$ and~$\lambda$,
we adopt the convention that
\[ \text{all constants are independent of~$\Omega$ and~$\lambda$} \:, \]
but they may depend on~$k$ and~$s$.
Moreover, in order to have a compact and clear notation,
we always denote constants which may be increased during our constructions
by capital letters~$\Const_1, \Const_2, \ldots$.
However, constants with small letters~$\const_1, \const_2, \ldots$
are determined at the beginning and are fixed throughout. We use the symbol
\[ \lesssim \cdots \qquad \text{for} \qquad \leq \const\, \cdots \]
with a constant~$\const$ which is independent of the capital constants~$\Const_l$
(and may thus be fixed right away, without the need to increase it later on).

When increasing the constants~$\Const_l$, we must keep track of the mutual dependences
of these constants. To this end, we adopt the convention that the constant~$\Const_l$ may depend
on all previous constants~$\Const_1, \ldots, \Const_{l-1}$, but is independent of the subsequent
constants~$\Const_{l+1}, \ldots$. In particular, we may choose the capital constants such
that~$\Const_1 \ll \Const_2 \ll \cdots$.
This dependence of the constants implies that increasing~$\Const_l$ may also
make it necessary to increase the subsequent constants~$\Const_{l+1}, \Const_{l+2}, \ldots$.
For brevity, when we write ``possibly after increasing~$\Const_l$'' 
we implicitly mean that the subsequent constants~$\Const_{l+1}, \Const_{l+2}, \ldots$
are also suitably increased.

\subsection{Different Cases and Regions}
In view of Proposition~\ref{prpbounded}, it suffices to consider
the case that~$|\Omega|$ is large. Thus in what follows we always assume that
\[ \boxed{ \quad |\Omega| \geq \Const_4\:. \quad } \]
Since the imaginary part of~$\Omega$ is bounded by~\eqref{ocond}, by increasing~$\Const_4$
we can always arrange that
\[ |\re \Omega|^2 \geq \frac{3}{4}\:|\Omega|^2\:. \]
Furthermore, Lemma~\ref{lemmaimV} gives us an a-priori bound on the
imaginary part of the eigenvalues,
\beq \label{relamright}
\boxed{ \quad \im \lambda \lesssim |\Omega|\:. \quad }
\eeq
Moreover, in view of Proposition~\ref{prplamlower}, we know in the case of a real potential
that by choosing~$N$ sufficiently large, it suffices to consider the case that~$\lambda$
is real and~$\lambda \gg |\Omega|$.
With this in mind, in our estimates we may restrict attention to the case
\beq \label{relamleft}
\boxed{ \quad \re \lambda \geq \Const_5\, |\Omega| \:. \quad }
\eeq
This inequality will be justified a-posteriori by showing that if we deform the potential
continuously starting from a real potential and ending with our complex potential~$V$, then
the inequality~\eqref{relamleft} will be preserved for all spectral points~$\lambda_n$ with~$n \geq N$
(for details see Section~\ref{sectrack}).

For large~$|\Omega|$, the real part of the potential looks qualitatively like a double-well
potential (see the left of Figure~\ref{figpot1}). More quantitatively, in the region~$[\frac{\pi}{3}, \frac{2 \pi}{3}]$
away from the poles at~$u=0$ and~$u=\pi$, according to~\eqref{Vdef} we have
\begin{align}
\re V &= \re (\Omega^2)\, \sin^2 u - \re \lambda + \O(\Omega) \label{reV} \\
\re V' &= 2 \re (\Omega^2)\, \sin u \, \cos u + \O(\Omega) \label{reVp} \\
\re V'' &= 2 \re (\Omega^2)\, \cos(2 u) + \O(\Omega)\:. \label{reVpp}
\end{align}
In particular, one sees that~$\re V$ has a unique local maximum at a point
near~$\piot$, which we denote by~$\umax$,
\[ \re V'(\umax) = 0 \qquad \text{and} \qquad \umax = \frac{\pi}{2} + \O \big( |\Omega|^{-1} \big)\:. \]
Moreover, the real part of the potential is concave near this maximum,
\beq \label{concave}
-2 |\Omega|^2 \leq \re V'' \leq -\frac{|\Omega|^2}{2} \qquad \text{on~$\Big[\frac{\pi}{3}, \frac{2 \pi}{3}\Big]$}\:.
\eeq

As the intervals~$(0, \umax]$ and~$[\umax, \pi)$ can be treated similarly, we mainly
restrict attention to the interval~$(0, \umax]$.
The value of the real part of the potential at its local maximum distinguishes different {\em{cases}}:
\beq \label{cases}
\left\{ \begin{array}{lcl} \text{WKB case} && \text{if~$\re V(\umax) < -\Const_1\, |\Omega|$} \\[0.3em]
\text{parabolic cylinder case} && \text{if~$-\Const_1\, |\Omega| \leq \re V(\umax) < \Const_1\, |\Omega|$} \\[0.3em]
\text{Airy case} && \text{if~$\re V(\umax) \geq \Const_1\, |\Omega|$}\:. \end{array} \right.
\eeq
Here~$\Const_1$ is a new constant which later on we will choose sufficiently large.

In each of the above cases, we estimate the solution by considering different {\em{regions}}, as we now
explain. First, we distinguish the {\em{pole region}} as the interval~$(0, \ul)$ with
\beq \label{uldef}
\ul := \frac{\Const_1}{\sqrt{\re \lambda}} \:.
\eeq
To the right of the pole region, there is a (possibly empty) WKB region~$(\ul, \ur)$.
The definition of~$\ur$ depends on the different cases. In the WKB case, we simply
set~$\ur=\umax$. In the parabolic cylinder case, the fact that the function~$\re V$
is concave~\eqref{concave} implies that there is a unique point~$\ur \in (\frac{\pi}{3}, \umax)$
with~$\re V(\ur) = -\Const_1 \,|\Omega|$. In the Airy case, we make use of the following result.
\begin{Lemma} \label{lemma81}
In the Airy case, there are unique points~$\ur, u_+$ in the
interval
\beq \label{Airyinterval}
\underline{u} < u < \min \big( \umax, \overline{u} \big)
\eeq
with
\beq \label{urpdef}
\re V(\ur) = -\nu\:, \qquad \re V(u_+) = \nu \:,
\eeq
where~$\underline{u}$, $\overline{u}$ and~$\nu$ are defined by
\begin{gather}
\underline{u} = \frac{\sqrt{\re \lambda}}{2\,|\Omega|} \:,\qquad
\overline{u} = 4\:\frac{\sqrt{\re \lambda}}{|\Omega|} \label{ubardef} \\
\nu = \min\Big( \frac{1}{4} \big( \Const_5^2 \,|\Omega|^2\: \re \lambda \big)^\frac{1}{3},  \;\frac{1}{2}
\big( \Const_1^2\, |\Omega|^2\, \re V(\umax) \big)^\frac{1}{3} \Big) \:.
\label{nudef}
\end{gather}
\end{Lemma}
\Proof We first show that~$\re V$ is strictly increasing on the interval~$(\underline{u}, \umax)$.
First, according to~\eqref{reVp}, by increasing~$\Const_4$ we can arrange that
the function~$\re V$ is strictly increasing on the interval~$(\frac{\pi}{8}, \frac{3 \pi}{8})$.
Moreover, the concavity of~$\re V$ implies that~$\re V$ is also monotone increasing
on the interval~$[\frac{3 \pi}{8}, \umax)$.
On the remaining interval~$(\underline{u}, \frac{\pi}{8}]$, we have the estimate
\begin{align}
\re V' &\geq \frac{|\Omega|^2}{2}\: u - \frac{\const}{u^3}
\geq \left( \frac{|\Omega|^2}{2} - \frac{\const}{\underline{u}^4} \right) u \notag \\
&\geq \left( \frac{1}{2} - \frac{16\,\const\,|\Omega|^2}{\re^2 \lambda} \right) |\Omega|^2\, u
\geq  \left( \frac{1}{2} - \frac{16\,\const}{\Const_5^2} \right) |\Omega|^2\, u \geq \frac{|\Omega|^2}{4}\: u > 0 \:,
\label{Vplower}
\end{align}
where in the last step we possibly increased~$\Const_5$. We conclude that~$\re V$
is strictly increasing on the whole interval~$(\underline{u}, \umax)$.

\begin{figure}
\psscalebox{1.0 1.0} 
{
\begin{pspicture}(0,-2.3015387)(14.275,2.3015387)
\psline[linecolor=black, linewidth=0.04, arrowsize=0.02cm 8.0,arrowlength=1.4,arrowinset=0.4]{->}(0.03,0.72349185)(3.83,0.72349185)
\psline[linecolor=black, linewidth=0.04, arrowsize=0.02cm 8.0,arrowlength=1.4,arrowinset=0.4]{->}(0.02,-2.3015082)(0.03,2.3234918)
\psline[linecolor=black, linewidth=0.02, linestyle=dashed, dash=0.17638889cm 0.10583334cm](0.01,1.3234918)(3.43,1.3234918)
\psline[linecolor=black, linewidth=0.02, linestyle=dashed, dash=0.17638889cm 0.10583334cm](0.01,0.12349182)(3.43,0.12349182)
\psline[linecolor=black, linewidth=0.02, linestyle=dashed, dash=0.17638889cm 0.10583334cm](10.1,1.6634918)(13.61,1.6634918)
\psline[linecolor=black, linewidth=0.02, linestyle=dashed, dash=0.17638889cm 0.10583334cm](10.275,-0.21650818)(13.61,-0.21650818)
\rput[bl](0.29,2.0434918){$\re V$}
\psline[linecolor=black, linewidth=0.04, arrowsize=0.02cm 8.0,arrowlength=1.4,arrowinset=0.4]{->}(4.83,0.72349185)(9.23,0.72349185)
\psline[linecolor=black, linewidth=0.04, arrowsize=0.02cm 8.0,arrowlength=1.4,arrowinset=0.4]{->}(9.63,0.72349185)(13.83,0.72349185)
\psline[linecolor=black, linewidth=0.02, linestyle=dashed, dash=0.17638889cm 0.10583334cm](5.21,0.12349182)(9.21,0.12349182)
\psline[linecolor=black, linewidth=0.02, linestyle=dashed, dash=0.17638889cm 0.10583334cm](5.23,1.3234918)(9.21,1.3234918)
\rput[bl](1.575,0.8984918){$\ur$}
\rput[bl](1.035,-2.1215081){WKB case}
\rput[bl](5.02,-2.171508){parabolic cylinder case}
\rput[bl](11.365,-2.1765082){Airy case}
\psbezier[linecolor=black, linewidth=0.04](0.04,-1.6565082)(0.89,-1.1065081)(0.79,-0.20650817)(1.79,-0.20650817)(2.79,-0.20650817)(2.89,-0.9065082)(3.39,-1.6565082)
\psline[linecolor=black, linewidth=0.02](1.825,0.6984918)(1.825,-0.17150818)
\psbezier[linecolor=black, linewidth=0.04](5.025,-1.6215081)(5.64,-0.5408185)(5.775,1.0084919)(6.775,1.0084919)(7.775,1.0084919)(7.97,-0.32374957)(8.375,-1.6215081)
\rput[bl](6.355,0.34349182){$\umax$}
\psline[linecolor=black, linewidth=0.02](6.785,0.98849183)(6.785,0.72349185)
\psline[linecolor=black, linewidth=0.02](5.775,0.7084918)(5.78,0.13849182)
\rput[bl](5.505,0.8884918){$\ur$}
\psbezier[linecolor=black, linewidth=0.04](9.785,-1.5765082)(10.77,0.23534663)(11.31,0.8518602)(11.805,1.3786497)(12.3,1.9054391)(12.95,2.1634917)(13.63,1.9234918)
\psline[linecolor=black, linewidth=0.02](13.07,2.0234919)(13.075,0.73349184)
\psline[linecolor=black, linewidth=0.02](12.125,1.6634918)(12.13,0.73349184)
\psline[linecolor=black, linewidth=0.02](10.57,0.7184918)(10.575,-0.21150818)
\rput[bl](12.675,0.34349182){$\umax$}
\rput[bl](10.275,0.8934918){$\ur$}
\rput[bl](11.865,0.34349182){$u_+$}
\rput[bl](3.855,1.1634918){$\Const_1\, |\Omega|$}
\rput[bl](3.7,-0.05650818){$-\Const_1\, |\Omega|$}
\rput[bl](9.74,1.6034918){$\nu$}
\rput[bl](9.585,-0.32150817){$-\nu$}
\end{pspicture}
}
\caption{The different cases.}
\label{figcases}
\end{figure}
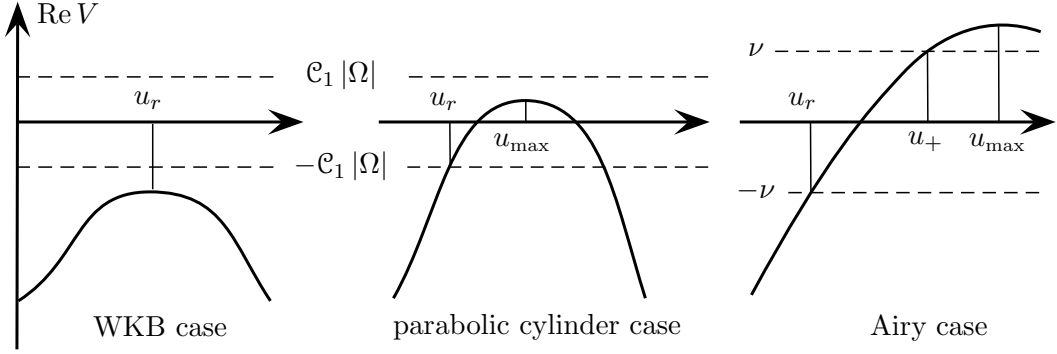

Next, at~$\underline{u}$ we have the estimate
\begin{align*}
\re V(\underline{u}) &\leq |\Omega|^2\, \underline{u}^2 + \frac{\const}{\underline{u}^2} - \re \lambda
= \frac{\re \lambda}{4} + \frac{4 \const\,|\Omega|^2}{\re \lambda} - \re \lambda \\
&\leq \frac{\re \lambda}{4} + \frac{4 \const}{\Const_5^2}\: \re \lambda - \re \lambda \leq -\frac{\re \lambda}{2} \:,
\end{align*}
where in the last step we possibly again increased~$\Const_5$.
Moreover, if~$\overline{u} < \umax$, we have the estimate
\begin{align*}
\re V(\overline{u}) &\geq \frac{1}{4}\: |\Omega|^2\, \underline{u}^2 - \frac{\const}{\underline{u}^2} - \re \lambda
\geq 4\,\re \lambda - \frac{4 \const\,|\Omega|^2}{\re \lambda} - \re \lambda \\
&\geq 4\,\re \lambda - \frac{4 \const}{\Const_5^2}\: \re \lambda - \re \lambda \geq \re \lambda \:,
\end{align*}
where in the last step we possibly again increased~$\Const_5$.
Next, it follows from the definition of~$\nu$ in~\eqref{nudef} and~\eqref{relamleft}
that~$\nu < \re \lambda/4$. Hence
\[ \re V(\underline{u}) < -\nu \qquad \text{and} \qquad \re V\big( 
\min(\overline{u}, \umax) \big) > \nu \:. \]
Now the existence of solutions~$\ur$ and~$u_+$ of~\eqref{urpdef} follows from the
intermediate value theorem. Uniqueness is an immediate consequence of the above strict monotonicity
of~$\re V$.
\QED
To summarize, the point~$\ur$ is defined by
\beq \label{urdef}
\left\{ \begin{array}{cl} \ur = \umax & \text{in the WKB case} \\[0.3em]
\re V(\ur) = -\Const_1\, |\Omega| & \text{in the parabolic cylinder case} \\[0.3em]
\re V(\ur) = -\nu & \text{in the Airy case} \:.
\end{array} \right.
\eeq
In the Airy case, the interval~$(u_+, \umax)$ with~$u_+$ as in~\eqref{urpdef}
is another WKB region to the right of the zero of~$\re V$.
We thus obtain the following regions:
\beq \label{regions}
\left\{ \begin{array}{lll} \text{pole region} &(0, \ul) \qquad& \text{in all cases} \\[0.3em]
\text{WKB region} &(\ul, \ur) & \text{in all cases} \\[0.3em]
\text{parabolic cylinder region} &(\ur, \umax) & \text{in the parabolic cylinder case} \\[0.3em]
\text{Airy region} &(\ur, u_+) & \text{in the Airy case} \\[0.3em]
\text{WKB region with~$\re V > 0$} &(u_+, \umax) & \text{in the Airy case}\:. \\[0.3em]
\end{array} \right.
\eeq
The different cases are illustrated in Figure~\ref{figcases}.

\subsection{Locating the Eigenvalues} \label{seclocate}
Our general strategy is to construct a special solution~$y_L$ of the Riccati equation~\eqref{riccati}
on the interval~$(0, \umax]$, and a special solution~$y_R$ on the interval~$[\umax, \pi)$.
These solutions are defined by the initial conditions
\beq \label{yinitLR}
y_L \big( u_0^L \big) = y_0^L \:,\qquad y_R \big( u_0^R \big) = y_0^R \:,
\eeq
where~$u_0^L$ and~$u_0^R$ are chosen near the poles at~$u=0$ respectively~$u=\pi$
(for details see Section~\ref{secpole1} and~\ref{secpole2} below).
We choose these special solutions in such a way that our estimates become as simple as possible.
This means in particular that these solutions have no singularities.
Then we introduce corresponding smooth solutions of the Sturm-Liouville equation~\eqref{5ode}
by integration (cf.~\eqref{phidef}),
\beq \label{phiLRdef}
\phi_L(u) := \exp \left( \int^u y_L \right) \:,\qquad \phi_R(u) := \exp \left( \int^u y_R \right) ,
\eeq
both normalized according to~\eqref{initial}.
These solutions will {\em{not}} satisfy
the Dirichlet boundary conditions~\eqref{Dirichlet}.
By introducing the functions~$\phiD_L$ and~$\phiD_R$ again by~\eqref{phiDdef}, we obtain
solutions which do satisfy the Dirichlet boundary conditions.
In order to locate the eigenvalues, we must analyze the eigenvalue condition~\eqref{evalcond}.
It is most convenient to evaluate the Wronskian at~$\umax$,
\beq \label{evalcond2}
w\big( \phiD_L, \phiD_R \big)\big|_{\umax} = 0 \:.
\eeq
Similar to~\eqref{zetadef} and~\eqref{phiDzeta} we set
\begin{align}
\zeta_L(u) &= \int_0^u \frac{1}{\phi_L^2} \:, \!\!\!\!\!\!\!\!\!\!\!\!\!\!\!\!\!\!\!\!\!\!\!\! & \phiD_L &= \phi_L\: \zeta_L \:, \label{zetaLdef} \\
\zeta_R(u) &= -\int_u^\pi \frac{1}{\phi_R^2} \:, \!\!\!\!\!\!\!\!\!\!\!\!\!\!\!\!\!\!\!\!\!\!\!\! & \phiD_R &= \phi_R\: \zeta_R \:. \label{zetaRdef}
\end{align}
Differentiating these relations, we obtain
\[ \frac{(\phiD_L)'}{\phiD_L} =  \frac{\phi_L'}{\phi_L} + \frac{\zeta_L'}{\zeta_L}
= y_L + \frac{1}{\phi_L^2 \zeta_L} \]
and thus
\begin{align}
w(\phiD_L, \phiD_R) &= \phiD_L \:\phiD_R \left( \frac{(\phiD_L)'}{\phiD_L} - \frac{(\phiD_R)'}{\phiD_R}\right) \notag \\
&= \phiD_L \:\phiD_R \left( y_L + \frac{1}{\phi_L^2 \zeta_L} - y_R - \frac{1}{\phi_R^2 \zeta_R}\right) \notag \\
&= \phi_L\,\zeta_L \: \phi_R \,\zeta_R 
\left( y_L + \frac{1}{\phi_L^2 \zeta_L} - y_R - \frac{1}{\phi_R^2 \zeta_R}\right) . \label{wronskirel}
\end{align}
Therefore, the eigenvalue condition~\eqref{evalcond2} can be written alternatively as
\beq \label{evalcond3}
\big( y_L - y_R \big) + \frac{1}{\phi_L^2 \zeta_L} - \frac{1}{\phi_R^2 \zeta_R}  = 0\:.
\eeq
Indeed, in this form the eigenvalue condition is most suited for our analysis.
Our main task is to analyze the behavior of the functions~$y_L$ and~$y_R$
as well as the derived functions~$\phi_L, \phi_R$
and~$\zeta_L, \zeta_R$ (obtained by~\eqref{phiLRdef} and~\eqref{zetaLdef}, \eqref{zetaRdef}).

\section{Estimates in Different Regions} \label{secesregion}
\subsection{Estimates in the WKB Region~$(\ul, \ur)$} \label{secWKBes}
The name ``WKB region'' suggests that in these regions the WKB solutions
should be a good approximation. This really is the case, in the following sense:

\begin{Prp} \label{prpWKB}
For any~$\delta>0$ and for sufficiently large~$\Const_1$, the WKB conditions
\beq \label{CVnew}
\frac{|V'|}{|V|^\frac{3}{2}}, \;\frac{|V''|}{|V|^2}, \;\frac{|V'''|}{|V|^\frac{5}{2}} < \delta
\eeq
hold in the WKB regions~$(\ul, \ur)$ and~$(u_+, \umax)$ (see~\eqref{regions}),
uniformly in~$\Omega$ and~$\lambda$.
\end{Prp} \noindent
For the significance of the inequalities~\eqref{CVnew} we also refer to~\cite[eq.~(4.1)]{tinvariant}
and the estimates in~\cite[Section~4]{tinvariant}.

The proof of this proposition is split up into several lemmas. The proof will be completed
at the end of this section.

\begin{Lemma} \label{lemma102}
Possibly by increasing~$\Const_1$, we can arrange that in the region~$u>\ul$,
the potential and its derivatives are bounded by
\begin{align*}
|W| &\leq |\Omega|^2 \,u^2 + \frac{\re \lambda}{2} \:, &
|V'| &\lesssim |\Omega|^2 \,u + \frac{(\re \lambda)^\frac{3}{2}}{\Const_1^3} \\
|V''| &\lesssim |\Omega|^2 + \frac{(\re \lambda)^2}{\Const_1^4} \:, &
|V'''| &\lesssim |\Omega|^2
+ \frac{(\re \lambda)^{\frac{5}{2}}}{\Const_1^5} \:.
\end{align*}
\end{Lemma}
\Proof
Using the explicit form of the potentials in~\eqref{Wsymm}, \eqref{Wasy} and~\eqref{Vdef}, we obtain
\begin{align*}
|W| &\leq |\Omega|^2 \,u^2 + \frac{\const}{\ul^2} \leq |\Omega|^2 \,u^2 + \frac{\const}{\Const_1^2}\: \re \lambda 
\overset{(\star)}{\leq} |\Omega|^2 \,u^2 + \frac{\re \lambda}{2} \\
|V'| &\leq 2 |\Omega|^2 \,u + \frac{\const}{\ul^3} \lesssim |\Omega|^2 \,u
+ \frac{(\re \lambda)^\frac{3}{2}}{\Const_1^3} \\
|V''| &\lesssim |\Omega|^2 + \frac{1}{\ul^4} \lesssim |\Omega|^2 + \frac{(\re \lambda)^2}{\Const_1^4} \:,
\quad
|V'''| \lesssim |\Omega|^2 + \frac{\const}{\ul^5} \lesssim |\Omega|^2
+ \frac{(\re \lambda)^{\frac{5}{2}}}{\Const_1^5} \:,
\end{align*}
where in~$(\star)$ we possibly increased~$\Const_1$.
\QED

\begin{Lemma} The statement of Proposition~\ref{prpWKB} holds in the region
\beq \label{leftregion}
\ul < u < \underline{u} \:,
\eeq
where~$\underline{u}$ is again defined by~\eqref{ubardef}.
\end{Lemma}
\Proof Using~\eqref{Vdef}, we obtain
\beq \label{absVlower}
|V(u)| \geq |\re \lambda| - |W| \geq \frac{\re \lambda}{2} - |\Omega|^2 u^2 \geq \frac{\re \lambda}{4}
\gtrsim \re \lambda \:.
\eeq
Hence
\begin{align*}
|V'| &\lesssim |\Omega|^2 \,u + \frac{(\re \lambda)^\frac{3}{2}}{\Const_1^3}
\lesssim |\Omega| \,\sqrt{\re \lambda} + \frac{(\re \lambda)^\frac{3}{2}}{\Const_1^3} \\
\frac{|V'|}{|V|^\frac{3}{2}} &
\lesssim \frac{|\Omega|}{\re \lambda} + \frac{1}{\Const_1^3}
\lesssim \frac{1}{\Const_5} + \frac{1}{\Const_1^3} \\
\frac{|V''|}{|V|^2} &\lesssim \frac{|\Omega|^2}{(\re \lambda)^2} + \frac{1}{\Const_1^4}
\lesssim \frac{1}{\Const_5^2} + \frac{1}{\Const_1^4} \\
\frac{|V'''|}{|V|^\frac{5}{2}} &\lesssim \frac{|\Omega|^2}{(\re \lambda)^\frac{5}{2}} + \frac{1}{\Const_1^5}
\lesssim \frac{1}{\Const_5^\frac{5}{2}\, \Const_4^\frac{1}{2}} + \frac{1}{\Const_1^5} \:,
\end{align*}
giving the result.
\QED

It remains to consider the complement of the region~\eqref{leftregion}. This complement
is empty unless
\[ \frac{\sqrt{\re \lambda}}{2\, |\Omega|} \leq \frac{\pi}{2} \:. \]
Therefore, in what follows we can assume that
\beq \label{lamlower}
\re \lambda \lesssim |\Omega|^2 \:.
\eeq
Moreover,
\beq \label{ulower2}
u \geq \frac{\sqrt{\re \lambda}}{2\, |\Omega|} \geq
\frac{\sqrt{\re \lambda|}}{2\, |\Omega|} \geq \frac{\sqrt{\Const_5}}{2\, \sqrt{|\Omega|}} 
\gtrsim \frac{1}{\sqrt{|\Omega|}}\:.
\eeq

\begin{Lemma} Under the assumptions of Proposition~\ref{prpWKB},
\[ \frac{|V''|}{|V|^2}, \;\frac{|V'''|}{|V|^\frac{5}{2}} < \delta \:. \]
\end{Lemma}
\Proof 
From~\eqref{urpdef}, \eqref{nudef}, \eqref{relamright} and~\eqref{relamleft} we know that
\[ |V| \geq \nu \gtrsim \min \big( \Const_5 \,|\Omega|,
\Const_1 \,|\Omega| \big) = \Const_1\: |\Omega| \:.\]
Combining this inequality with~\eqref{lamlower} and~\eqref{ulower2},
the derivatives of the potential can be estimated by
\begin{align*}
|V''| &\lesssim |\Omega|^2 + \frac{1}{u^4} \lesssim |\Omega|^2 + |\Omega|^2 \lesssim |\Omega|^2 \\
|V'''| &\lesssim |\Omega|^2 + \frac{\const}{u^5} \lesssim |\Omega|^2
+ |\Omega|^\frac{5}{2} \lesssim |\Omega|^\frac{5}{2} \:.
\end{align*}
As a consequence,
\[ \frac{|V''|}{|V|^2} \lesssim \frac{|\Omega|^2}{\Const_1^2 \,|\Omega|^2} = \frac{1}{\Const_1^2} \:,\qquad
\frac{|V'''|}{|V|^\frac{5}{2}} \lesssim \frac{|\Omega|^\frac{5}{2}}{\Const_1^\frac{5}{2} \,|\Omega|^\frac{5}{2}}
= \frac{1}{\Const_1^\frac{5}{2}} \:, \]
completing the proof.
\QED

It remains to estimate the term involving the first derivatives in~\eqref{CVnew}.
\begin{Lemma} The statement of Proposition~\ref{prpWKB} holds in the region
\[ u \geq \overline{u} \:, \]
where~$\overline{u}$ is again defined by~\eqref{ubardef}.
\end{Lemma}
\Proof We can bound the potential from below by
\[ \re V \geq \frac{|\Omega|^2}{4}\, u^2 - \re \lambda \gtrsim \frac{|\Omega|^2}{8}\, u^2 \:. \]
Hence
\begin{align*}
\frac{|V'|}{|V|^\frac{3}{2}} &\lesssim 
\frac{|\Omega|^2 \,u}{|\Omega|^3\, u^3} + \frac{(\re \lambda)^\frac{3}{2}}{\Const_1^3\, |\Omega|^3\, u^3} 
\lesssim \frac{|\Omega|}{\re \lambda} + \frac{1}{\Const_1^3} \leq
\frac{1}{\Const_5} + \frac{1}{\Const_1^3} \:,
\end{align*}
giving the result.
\QED

It remains to estimate the term involving the first derivatives in~\eqref{CVnew} in the region
\beq \label{intervalleft}
\underline{u} < u < \overline{u} \:.
\eeq
We begin with a lemma in the Airy case.
\begin{Lemma} The statement of Proposition~\ref{prpWKB} holds in the Airy case if
\beq \label{Airy1}
\big( \Const_5^2 \,|\Omega|^2\: \re \lambda \big)^\frac{1}{3}
\leq 2 \:\big( \Const_1^2\, |\Omega|^2\, \re V(\umax) \big)^\frac{1}{3} \:.
\eeq
\end{Lemma}
\Proof In view of~\eqref{nudef}, the assumptions imply that
\beq \label{nuform}
\nu = \frac{1}{4} \big( \Const_5^2 \,|\Omega|^2\: \re \lambda \big)^\frac{1}{3} \:.
\eeq
Moreover, according to~\eqref{urpdef}, we know that~$|V| \geq \nu$. Hence
\begin{align*}
|V'| &\lesssim  |\Omega|^2 \,u + \frac{\const}{u^3}
\lesssim |\Omega|^2 \,u \left( 1 + \frac{1}{|\Omega|^2\, u^4} \right) \\
&\lesssim |\Omega|^2 \,u \left( 1 + \frac{|\Omega|^2}{\re^2 \lambda} \right) 
\overset{\eqref{relamleft}}{\lesssim} |\Omega|^2 \,u
\overset{\eqref{ubardef}}{\lesssim} |\Omega|\, \sqrt{\re \lambda} \\
\frac{|V'|}{|V|^\frac{3}{2}} &\lesssim \frac{|\Omega|\, \sqrt{\re \lambda}}{\nu^\frac{3}{2}}
\overset{\eqref{nuform}}{\lesssim} \frac{1}{\Const_5} \:.
\end{align*}
This concludes the proof.
\QED
Thus in the Airy case, in what follows we may assume that~\eqref{Airy1} is violated.
Since in the WKB region under consideration, we stay away from the zeros of~$\re V$
in the sense that~$|\re V| \geq \nu$ with~$\nu$ as in~\eqref{nudef}, it follows that
\beq \label{Airyspecial}
|\re V| \geq \frac{1}{2} \big( \Const_1^2\, |\Omega|^2\, \re V(\umax) \big)^\frac{1}{3}
\qquad \text{in the Airy case}\:.
\eeq

We now return to the analysis of the region~\eqref{intervalleft}, without specifying
whether we are in the WKB, the Airy or the parabolic cylinder case.
As a consequence of~\eqref{intervalleft},
\[ |V'| \lesssim |\Omega|^2 \, u + \frac{1}{u^3} \lesssim |\Omega| \, \sqrt{\re \lambda} 
\lesssim |\Omega|^2 \:, \]
where in the last step we again applied~\eqref{lamlower}.
From this inequality, we obtain the desired estimate provided that one of the following two
inequalities holds:
\beq \label{inequalt}
|V(u)| \geq \Const_2\, |\Omega|^\frac{4}{3}
\qquad \text{or} \qquad \re V(\umax) \geq \frac{|\Omega|^2}{\Const_1} \:.
\eeq
Namely, the first inequality implies that
\beq \label{firstes}
\frac{|V'|}{|V|^\frac{3}{2}} \lesssim \frac{|\Omega|^2}{\Const_2^\frac{3}{2}\, |\Omega|^2}
= \frac{1}{\Const_2^\frac{3}{2}} \:.
\eeq
On the other hand, if the second inequality in~\eqref{inequalt} holds,
we are in the Airy case (possibly after increasing~$\Const_4$), so that~\eqref{Airyspecial} yields
the estimate
\[ |V| \geq \big( \Const_1^2\, |\Omega|^2\, \re V(\umax) \big)^\frac{1}{3}
\geq \Const_1^\frac{1}{3}\, |\Omega|^\frac{4}{3} \:. \]
This implies that the first inequality in~\eqref{inequalt} again holds (for~$\Const_2=\Const_1^\frac{1}{3}$),
making it possible to again use the estimate~\eqref{firstes}.

It remains to consider the case that both inequalities in~\eqref{inequalt} are violated, i.e.
\[ |V(u)| < \Const_2\, |\Omega|^\frac{4}{3}
\qquad \text{and} \qquad \re V(\umax) < \frac{|\Omega|^2}{\Const_1} \:. \]
In this case, $|\re V(u) - \re V(\umax)| \lesssim |\Omega|^2/\Const_1$,
implying that~$u \approx \umax$
(more precisely, by increasing~$\Const_1$ we can make~$|u-\umax|$ arbitrarily small,
uniformly in~$\Omega$ and~$\lambda$).
Since~$\re V$ is concave near~$\umax$~\eqref{concave}, we may integrate
this inequality to obtain
\begin{gather*}
\frac{|\Omega|^2}{2} \, |u-\umax| \leq |V'(u)| \leq |\Omega|^2 \, |u-\umax| \\
\frac{|\Omega|^2}{4}\, (u-\umax)^2 \leq \re V(\umax) - \re V(u) \leq \frac{|\Omega|^2}{2}\, (u-\umax)^2 \:.
\end{gather*}
Hence
\begin{align*}
|V'(u)| &\lesssim |\Omega| \:\sqrt{\re V(\umax) - \re V(u)}
\leq |\Omega| \Big( \sqrt{|\re V(\umax)|} + \sqrt{|V(u)|} \Big)
\end{align*}
In the case~$|\re V(\umax)| \leq |V(u)|$, it follows that
\[ \frac{|V'(u)|}{|V(u)|^\frac{3}{2}} \leq \frac{2\,|\Omega|}{|V(u)|} \lesssim
\frac{|\Omega|}{\Const_1\, |\Omega|} = \frac{1}{\Const_1} \:, \]
giving the result.
In the remaining case~$|\re V(\umax)| > |V(u)|$, we know from~\eqref{urdef}
and~\eqref{cases} that we are in the Airy case.
Hence, again using~\eqref{Airyspecial}, we obtain
\[ \frac{|V'(u)|}{|V(u)|^\frac{3}{2}} \leq \frac{2\,|\Omega|\, \sqrt{\re V(\umax)}}{|V(u)|^{\frac{3}{2}}} \lesssim
\frac{2^\frac{3}{2}\, |\Omega| \,\sqrt{|\re V(\umax)|}}{\sqrt{ \Const_1^2\, |\Omega|^2\, |\re V(\umax)|}}
= \frac{2^\frac{3}{2}}{\Const_1} \:. \]
This concludes the proof of Proposition~\ref{prpWKB}.

\subsection{Estimates in the Pole Region in the Case~$k=s$} \label{secpole1}
In this section, we analyze the pole region~$(0, \ul)$ in the case~$k=s$.
We consider the solution~$\phi_L$ as defined by~\eqref{yinitLR} and~\eqref{phiLRdef}.
For ease in notation, we omit all subscripts~$L$. The parameter~$u_0$ in~\eqref{yinitLR} is chosen as
\beq \label{u00def}
u_0 := \frac{1}{2} \:|\Omega|^{-\frac{1}{2}} \:.
\eeq
Possibly by increasing~$\Const_5$, we can arrange
that~$u_0 > \ul$ (cf.~\eqref{uldef} and~\eqref{relamleft}), so that~$u_0$ lies in the WKB region.
We choose the initial values
at~$u_0$ in~\eqref{yinitLR} equal to the value~$\phi_\WKB'/\phi_\WKB$ for the WKB approximation,
\beq \label{y0WKB}
y_0 = \sqrt{V(u_0)} - \frac{V'(y_0)}{4 V(y_0)} \qquad \text{and} \qquad
\im \sqrt{V(u_0)} > 0 \:.
\eeq

We expand the potential near~$u=0$,
\beq \label{V0ex}
V(u) = -\frac{1}{4 u^2} - \mu + \Omega^2 \,u^2 + \O \big( |\Omega| u^2 \big)
+ \O \big( |\Omega|^2\, u^4 \big) \:.
\eeq

\begin{Lemma} \label{lemmaremu}
For any~$\delta>0$, we can arrange by increasing~$\Const_5$ that
\[ \int_0^{\ul} \frac{1}{|\phi|^2} \leq \delta\:, \]
uniformly in~$\Omega$ and~$\lambda$.
\end{Lemma}
\Proof Using the asymptotics as worked out in~\cite[Section~7.1]{tinvariant},
on the interval~$(0, u_0)$ the solution~$\phi$ has the form
\beq \label{K0}
\phi(u) \approx -c \sqrt{u} \:\Big( K_0(\sqrt{\mu} u) + \big( \arg \sqrt{\mu} - \log(2) + \gamma + i \big) \,I_0(\sqrt{\mu} u) \Big) \:,
\eeq
where~$c$ is the constant (see~\cite[eqn.~(7.3)]{tinvariant})
\[ c = \frac{\sqrt{2 \pi}}{\arg \sqrt{\mu} - \log(2) + \gamma + i} \]
(and~$\mu$ is related to~$\lambda$ and~$\Omega$ by~\eqref{mudef}).
As specified in~\cite[Section~8.1]{tinvariant}, the error in~\eqref{K0} becomes
arbitrarily small for large~$|\Omega|$.
Note that~$c$ is bounded uniformly in~$\mu$. For small~$u$, the function~$\phi$
has the asymptotics (see~\cite[Section~7.1]{tinvariant})
\beq \label{phi0asy}
\phi(u) = -c \left( \sqrt{u}\: \log |\sqrt{\mu} u| + i \sqrt{u} + \O(\sqrt{\mu} u) \right) \:.
\eeq
In particular, using that
\[ \int_0^u \frac{1}{u\, (1+ \log^2(\sqrt{\mu} u)} = \arctan \big( \log(\sqrt{\mu} u) \big) + \frac{\pi}{2} \:, \]
one sees that~$1/|\phi|^2$ is integrable. Hence
\[ \lim_{v \searrow 0} \int_0^{v} \frac{1}{|\phi|^2} = 0 \:, \qquad \text{uniformly in~$\mu$}\:.\]
In view of~\eqref{uldef}, by increasing~$\Const_5$ we can make~$\sqrt{\mu} \,\ul$ as small as we like.
This gives the result.
\QED

\subsection{Estimates in the Pole Region in the Case~$k \neq s$} \label{secpole2}
We now analyze the pole region~$(0, \ul)$ in the case~$k \neq s$.
We again consider the solution~$\phi_L$ as defined by~\eqref{yinitLR} and~\eqref{phiLRdef}
and omit all subscripts~$L$. The parameter~$u_0$ in~\eqref{yinitLR} is chosen as the
minimum of~$\re V$,
\beq \label{u0ks}
\re V'(u_0) = 0 \:.
\eeq
Introducing the abbreviation
\[ \Lambda = (k-s)^2 - \frac{1}{4} \:, \]
the potential near~$u=0$ has the expansion
\beq \label{Vn0ex}
V(u) = \frac{\Lambda}{u^2} - \mu + \Omega^2 \,u^2 + \O \big( |\Omega| u^2 \big)
+ \O \big( |\Omega|^2\, u^4 \big) \:.
\eeq
Computing the zero of the derivative, we obtain
\beq \label{u0Ldef}
u_0 = \Lambda^{\frac{1}{4}}\, |\Omega|^{-\frac{1}{2}} + \O \big( |\Omega|^{-\frac{3}{2}} \big) \:,
\eeq
so that for large~$\Omega$ we have the estimates
\beq \label{u0es}
\frac{1}{2}\: \Lambda^{\frac{1}{4}}\, |\Omega|^{-\frac{1}{2}}
\leq u_0 \leq 2 \Lambda^{\frac{1}{4}}\, |\Omega|^{-\frac{1}{2}} \:.
\eeq
Possibly by increasing the constant~$\Const_5$ in~\eqref{relamleft}, we can again arrange
that~$u_0 > \ul$, so that~$u_0$ lies in the WKB region. We again choose the initial values
at~$u_0$ in agreement with the WKB approximation~\eqref{y0WKB}.

\begin{Lemma} \label{lemmaremu2} 
For any~$\delta>0$, we can arrange by increasing~$\Const_5$ that
\[ \int_0^{\ul} \frac{1}{|\phi|^2} \leq \delta \:, \]
uniformly in~$\Omega$ and~$\lambda$.
\end{Lemma}
\Proof Using the asymptotics as worked out in~\cite[Section~7.2]{tinvariant},
the solution~$\phi$ has the form
\beq \label{KL}
\phi(u) \approx c  \sqrt{u}\, K_{|k-s|} \big( -\sqrt{\mu} u \big) \:,
\eeq
where~$c$ is the constant
\[ c = \sqrt{-\frac{2}{\pi}} \:. \]
As specified in~\cite[Section~8.2]{tinvariant}, the error in~\eqref{KL} becomes
arbitrarily small for large~$|\Omega|$. 
For small~$u$, the function~$\phi$ has the asymptotics (see~\cite[Section~7.2]{tinvariant})
\beq \label{phias2}
\begin{split}
\phi(u) &= c\:\frac{(n-1)!}{\sqrt{2}\, \mu^{\frac{1}{4}}} \left( -\frac{\sqrt{\mu} u}{2} \right)^{\frac{1}{2}-{|k-s|}}
(1 + \O(u)) \\
& \quad-c\:\frac{\sqrt{2}}{{|k-s|}!\, \mu^{\frac{1}{4}}} \left( \frac{\sqrt{\mu} u}{2} \right)^{\frac{1}{2}+{|k-s|}}
(1 + \O(u)) \:.
\end{split} 
\eeq
In particular, one sees that~$|\phi|$ has a pole at~$u=0$ and is thus bounded from below near~$u=0$. Hence
\[ \lim_{v \searrow 0} \int_0^{v} \frac{1}{|\phi|^2} = 0 \:, \qquad \text{uniformly in~$\mu$}\:.\]
In view of~\eqref{uldef},
by increasing~$\Const_5$ we can make~$\sqrt{\mu} \,\ul$ as small as we like. This gives the result.
\QED

\subsection{Estimates in the Parabolic Cylinder Region} \label{secesPCR}
In the next proposition we estimate the Riccati solution in the parabolic cylinder region.
\begin{Prp} \label{prpTparabolic}
Assume that in the parabolic cylinder region~$[\ur, \umax]$, one of the following two conditions hold:
\begin{itemize}
\item[(a)] The potential has a positive imaginary part, $\im V|_{[\ur, u_+]} \geq 0$.
\item[(b)] The imaginary part of the potential has a zero on~$[\ur^L, \ur^R]$.
\end{itemize}
Moreover, assume that the Riccati solution begins in the upper half plane, $\im y(\ur) \geq 0$.
Then there is a constant~$\Const_2$ (depending on~$\Const_1$) such that
for large~$|\Omega|$, the solution on the interval~$[\ur, \umax]$ can be estimated in terms of~$y(\ur)$ by
\begin{gather*}
|y(u)| \leq \Const_2\,|y(\ur)| \\
\im y(u) \geq \frac{\im y(\ur)}{\Const_2} \\
\frac{|\phi(\ur)|}{\Const_2} \leq |\phi(u)| \leq \Const_2\, |\phi(\ur)| \:.
\end{gather*}
\end{Prp}
\Proof We set~$\nu=\Const_1 |\Omega|$.
Using that the function~$\re V$ is concave near~$\umax$, we obtain
\beq \label{umaxures}
(\umax - \ur)^2 \lesssim \frac{\nu}{|\Omega|^2}
\eeq
and thus
\[ \nu\, (\umax - \ur)^2 \lesssim \Const_1^2 \:. \]
Our strategy is to estimate~$y$ using the $T$-method
as introduced in~\cite[Section~3.2]{tinvariant} choosing
\beq \label{abchoice}
\alpha = \sqrt{2 \nu} \qquad \text{and} \qquad \tilde{\beta}=0 \:.
\eeq
Hence
\beq \label{tVUform}
\tilde{V} = \alpha^2 = 2 \nu \qquad \text{and} \qquad
U = \re V - \alpha^2 \leq -\nu \:.
\eeq

In case~(a), our method is to apply~\cite[Theorem~3.3]{tinvariant} for~$g \equiv 0$. 
The terms~$E_1, \ldots, E_4$ are estimated as follows,
\begin{align*}
E_2 &= E_4 = 0 \\
|E_1| &\lesssim \frac{1}{\sqrt \nu}\: |\re V - \re \tilde{V}| + \frac{\re V'}{\nu} 
\lesssim \sqrt{\nu} + \frac{\re V'}{\nu} \\
|E_3| &\lesssim \frac{|\im V|}{\sqrt{\nu}} \overset{(\star)}{\lesssim} \frac{|\Omega|}{\sqrt{\nu}} \:,
\end{align*}
where in~$(\star)$ we used~\eqref{relamright} as well as the fact that
\[ \big| \im (\Omega^2) \big| \lesssim |\Omega| \big| \im \Omega \big|
\overset{\eqref{ocond}}{\lesssim} |\Omega| \:. \]
As a consequence, we can apply Lemma~\ref{lemma99} to obtain
\begin{align*}
\int_{\ur}^{\umax} &\Big( |E_1| + |E_3| \Big) \lesssim \sqrt{\nu}\, (\umax-\ur) + \frac{1}{\nu} \int_{\ur}^{\umax} \re V'
+ \frac{|\Omega|}{\sqrt{\nu}}\:  (\umax-\ur) \\
&= \sqrt{\nu}\, (\umax-\ur) + \frac{1}{\nu} \Big( \re V(\umax) - \re V(\ur) \Big)
+ \frac{|\Omega|}{\nu}\: \sqrt{\nu}\, (\umax-\ur) \\
&\lesssim \Const_1 + 2 + \Const_1\: \frac{|\Omega|}{\nu} \leq \Const_1 + 3\:.
\end{align*}
This concludes the proof in case~(a).

In case~(b), the imaginary part of~$V$ could be negative.
Therefore, in order to apply~\cite[Theorem~3.3]{tinvariant} we need to choose the function~$g$ positive
in accordance with the inequality
\beq \label{gneces}
g \geq T-1 \:.
\eeq
We choose~$\alpha$ and~$\tilde{\beta}$ as in~\eqref{abchoice} and~$g = |\Omega|^{\frac{1}{2}}$.
Then the error terms~$E_1$, $E_2$ and~$E_3$ estimated just as above. 
Estimating~$\im V$ with the help of the mean value theorem by
\[ |\im V| \leq (\ur^+-\ur^-) \:\sup_{[\ur^L , \ur^R]} |V'| \lesssim |\Omega|\: (\ur^R-\ur^L) \]
(where in the last step we used the explicit form of the potential~\eqref{Vdef}), 
the error term~$E_4$, is estimated by
\begin{align*}
\int_{\ur}^{\umax} |E_4| &= g \int_{\ur}^{\umax}  \frac{|\im V|}{\sqrt{|U|}}
\lesssim g \int_{\ur}^{\umax}  \frac{|\Omega|\, (\ur^R-\ur^L)}{\sqrt{\nu}} \\
&\lesssim \frac{g\, |\Omega|}{\sqrt{\nu}}\: (\ur^R-\ur^L)^2
\overset{\eqref{umaxures}}{\lesssim} \frac{g\, \sqrt{\nu}}{|\Omega|}
= \frac{g\, \sqrt{\Const_1}}{\sqrt{|\Omega|}}
= \frac{\sqrt{\Const_1}}{|\Omega|^{\frac{1}{4}}} \:.
\end{align*}
This can be made arbitrarily small by increasing~$|\Omega|$, implying that the inequality~\eqref{gneces} holds.
This concludes the proof.
\QED

\subsection{Estimates in the Airy Region}
We proceed with estimates in the Airy region. We first recall that it remains to
consider the interval~\eqref{Airyinterval}. For this interval to be non-empty, we
can again assume that~\eqref{lamlower} holds,
\beq \label{lamlowerrep}
\re \lambda \lesssim |\Omega|^2 \qquad \text{and thus} \qquad
|\re V| \lesssim |\Omega|^2\:.
\eeq
We begin with a preparatory lemma.

\begin{Lemma} \label{lemma99}
In the Airy region, the function~$\re V$ is strictly monotone. Moreover,
\beq \label{Airyes}
\nu\, (u_+-\ur)^2 \eqsim \Const_1^2 \:.
\eeq
\end{Lemma}
\Proof The strict monotonicity was already shown in the proof of Lemma~\ref{lemma81}.
In preparation for the estimate~\eqref{Airyes}, we recall that the region~$(\ur, u_+)$
is contained in the interval~$(\underline{u}$, $\overline{u})$
(see Lemma~\ref{lemma81}), and thus
\beq \label{upriori}
\ur, u_+ \eqsim \frac{\sqrt{\re \lambda}}{|\Omega|} \:.
\eeq

We consider the regions~$(\underline{u}, \frac{3 \pi}{8})$
and~$[\frac{3 \pi}{8}, \umax)$ separately. In the region~$(\underline{u}, \frac{3 \pi}{8})$, we
know from~\eqref{Vplower} that~$\re V' \gtrsim |\Omega|^2 u$. Moreover,
the method in~\eqref{Vplower} also gives the reverse inequality,
\begin{align*}
\re V' &\leq |\Omega|^2\: u + \frac{\const}{u^3}
\lesssim \left( 1 + \frac{1}{|\Omega|^2 \underline{u}^4} \right) |\Omega|^2 u \notag \\
&\eqsim \left( 1 + \frac{|\Omega|^2}{\re^2 \lambda} \right) |\Omega|^2 u \lesssim |\Omega|^2 u \:,
\end{align*}
where in the last step we used~\eqref{upriori} and~\eqref{relamright}. We conclude that
\[ \re V'|_{(\ur, u_+)} \eqsim |\Omega|^2 u \eqsim |\Omega|\, \sqrt{\re \lambda} \:. \]
As a consequence, the mean value theorems
\[ (u_+ - \ur) \inf_{(\ur, u_+)} \re V' \leq 2 \nu \leq (u_+ - \ur) \sup_{(\ur, u_+)} \re V' \]
give rise to the estimate
\[ u_+ - \ur \eqsim \frac{\nu}{|\Omega|\, \sqrt{\re \lambda}} \]
and thus
\beq \label{nurel}
\nu\, (u_+-\ur)^2 \eqsim \frac{\nu^3}{|\Omega|^2\, \re \lambda} \:.
\eeq
In order to estimate this further, we need to determine the scaling of~$\nu$.
Using the estimate for the second derivative in~\eqref{reVpp} with
the fact that~$\re V$ has a maximum at~$\umax$ and no zero on the interval~$(\frac{3 \pi}{8}, \umax)$,
we conclude that
\[ \re V(\umax) \gtrsim |\Omega|^2 \:. \]
Combining this inequality with the first inequality in~\eqref{lamlowerrep}, we find that
the first term in~\eqref{nudef} can be bounded in terms of the second term. More precisely,
we obtain the inequality
\[ \nu^3 \eqsim \min\big( \Const_5^2, \Const_1^2 \big)\, |\Omega|^2 \, \re \lambda
= \Const_1^2\, |\Omega|^2 \:. \]
Using this inequality in~\eqref{nurel} gives
\[ \nu\, (u_+-\ur)^2 \eqsim \Const_1^2 \:. \]
This concludes the proof on the interval~$(\underline{u}, \frac{3 \pi}{8})$.

It remains the consider the region~$(\frac{3 \pi}{8}, \umax)$.
We make use of the concavity of~$\re V$, \eqref{reVpp}.
Denoting the zero of~$\re V$ by~$u_1$, we obtain
\beq \label{reVsim}
\re V(\umax) \eqsim |\Omega|^2\, (\umax-u_1)^2 \:.
\eeq
On the other hand, for the potential to have a zero near~$\piot$,
the spectral parameter must scale like~$\lambda \eqsim |\Omega|^2$.
Therefore, the second term in~\eqref{nudef} can be estimated in terms of the first, so that
\beq \label{nu1}
\nu^3 \eqsim \min \big( \Const_1^2, \Const_5^2 \big)\: \, |\Omega|^2\, \re V(\umax) 
= \Const_1^2 \, |\Omega|^2\, \re V(\umax)\:.
\eeq
Since~$\nu < \re V(\umax)/2$ (cf.~\eqref{nudef} and~\eqref{cases}), the estimate~\eqref{reVsim}
can be extended to
\beq \label{nu2}
\re V(\umax) \eqsim |\Omega|^2\, (\umax-u)^2 \qquad \text{for all~$u \in [\ur, u_+]$}\:.
\eeq
Moreover,
\beq \label{nu3}
\re V'(u) \eqsim |\Omega|^2\, (\umax-u) \qquad \text{for all~$u \in [\ur, u_+]$}\:.
\eeq
We now combine the estimates~\eqref{nu1}, \eqref{nu2} and~\eqref{nu3} to obtain
\begin{align*}
u_+ - \ur &\eqsim \frac{\nu}{\re V'}  \\
\nu\, (u_+-\ur)^2 &\eqsim \frac{\nu^3}{\re^2 V'}
\eqsim \frac{\Const_1^2\, |\Omega|^2\, \re V(\umax)}{\re^2 V'} \\
&\eqsim \frac{\Const_1^2\, |\Omega|^4\, (\umax-u)^2}{ |\Omega|^4\, (\umax-u)^2}
= \Const_1^2 \:.
\end{align*}
This concludes the proof.
\QED

We now estimate the Riccati solution in the Airy region.
\begin{Prp} \label{prpTairy}
Assume that in in the Airy region~$[\ur, u_+]$, one of the following two conditions hold:
\begin{itemize}
\item[(a)] The potential has a positive imaginary part, $\im V|_{[\ur, u_+]} \geq 0$.
\item[(b)] The imaginary part of the potential is small in the sense that
\[ |\im V| \leq |\Omega|^{1-\delta} \qquad \text{for a suitable constant~$\delta>0$}\:. \]
\end{itemize}
Moreover, assume that the Riccati solution begins in the upper half plane, $\im y(\ur) \geq 0$.
Then there is a constant~$\Const_2$ (depending on~$\Const_1$) such that
for large~$|\Omega|$,
the solution on the interval~$[\ur, u_+]$ can be estimated in terms of~$y(\ur)$ by
\begin{gather}
|y(u)| \leq \Const_2\,|y(\ur)| \label{yes1} \\
\im y(u) \geq \frac{\im y(\ur)}{\Const_2} \label{yes2} \\
\frac{|\phi(\ur)|}{\Const_2} \leq |\phi(u)| \leq \Const_2\, |\phi(\ur)| \:. \label{pes}
\end{gather}
\end{Prp}
\Proof 
As in the proof of Proposition~\ref{prpTparabolic} we use the $T$-method
choosing~$\alpha$ and~$\tilde{\beta}$ as in~\eqref{abchoice}.
Then~$\tilde{V}$ and~$U$ are again estimated by~\eqref{tVUform}.

In case~(a), we apply~\cite[Theorem~3.3]{tinvariant} for~$g \equiv 0$.
The error terms~$E_1, \ldots, E_4$ are estimated as follows,
\begin{align*}
E_2 &= E_4 = 0 \\
|E_1| &\lesssim \frac{1}{\sqrt \nu}\: |\re V - \re \tilde{V}| + \frac{\re V'}{\nu} 
\lesssim \sqrt{\nu} + \frac{\re V'}{\nu} \\
|E_3| &\lesssim \frac{|\im V|}{\sqrt{\nu}} \lesssim \frac{|\Omega|}{\sqrt{\nu}} \:.
\end{align*}
As a consequence, we can apply Lemma~\ref{lemma99} to obtain
\begin{align*}
\int_{\ur}^{u_+} &\Big( |E_1| + |E_3| \Big) \lesssim \sqrt{\nu}\, (u_+-\ur) + \frac{1}{\nu} \int_{\ur}^{u_+} \re V'
+ \frac{|\Omega|}{\sqrt{\nu}}\:  (u_+-\ur) \\
&= \sqrt{\nu}\, (u_+-\ur) + \frac{1}{\nu} \Big( \re V(u_+) - \re V(\ur) \Big)
+ \frac{|\Omega|}{\nu}\: \sqrt{\nu}\, (u_+-\ur) \\
&\lesssim \Const_1 + 2 + \Const_1\: \frac{|\Omega|}{\nu} \:.
\end{align*}
Finally, by combining~\eqref{nudef} with~\eqref{relamright}, \eqref{relamleft} and~\eqref{cases},
we conclude that~$\nu \gtrsim |\Omega|$. This concludes the proof in case~(a).

In the remaining case~(b), we choose~$g$ as
\[ g = |\Omega|^{\frac{\delta}{2}} \:. \]
Then the error term~$E_4$ is estimated by
\[ \int_{\ur}^{\umax} |E_4| = g \int_{\ur}^{u_+}  \frac{|\im V|}{\sqrt{|U|}}
\lesssim g \,|\Omega|^{1-\delta} \;\frac{(u_+-\ur)}{\sqrt{\nu}}
\overset{\eqref{Airyes}}{\lesssim} 
g \,|\Omega|^{1-\delta} \;\frac{\sqrt{\Const_1}}{\nu} \lesssim
\frac{|\Omega|^{-\frac{\delta}{2}}}{\sqrt{\Const_1}} \:. \]
This concludes the proof.
\QED

In the next lemma we compare the imaginary part of the potential
on the two Airy regions~$[\ur^L, u_+^L]$ and~$[\ur^R, u_+^R]$.
\begin{Lemma} \label{lemmaairy123}
In the Airy case, one of the following three statements holds:
\begin{itemize}
\item[(i)] \hspace*{2cm} $\displaystyle \im V|_{[\ur^L, u_+^L]} \geq 0 \qquad \text{and} \qquad
\im V|_{[\ur^R, u_+^R]} \geq 0$ \\[-0.5em]
\item[(ii)] \hspace*{2cm} $\displaystyle \im V|_{[\ur^L, u_+^L]} \leq 0 \qquad \text{and} \qquad
\im V|_{[\ur^R, u_+^R]} \leq 0$ \\[-0.5em]
\item[(iii)]  \hspace*{1cm} $\displaystyle \big|\im V|_{[\ur^L, u_+^L]} \big|
\lesssim \sqrt{|\Omega|} \qquad \text{and} \qquad
\big| \im V|_{[\ur^R, u_+^R]} \big| \lesssim \sqrt{|\Omega|}$.
\end{itemize}
\end{Lemma}
\Proof We first consider the case that the Airy regions are near the poles
in the sense that~$u_+^L, (\pi-u_+^R) \lesssim |\Omega|^{-\frac{1}{4}}$.
Then the factor~$\sin^2 u$ in~\eqref{Vdef} is bounded by~$|\Omega|^{-\frac{1}{2}}$,
implying that~$\im (V+\lambda)$ is bounded by~$\sqrt{|\Omega|}$.
Therefore, depending on the value of~$\im \lambda$, we are in one of the above cases~(i)--(iii).

It remains to consider the case that the Airy regions are away from the poles.
Then the factors~$1/(\sin^2 u)$ in~\eqref{Vdef} is bounded by~$\sqrt{|\Omega}$.
As a consequence, the imaginary part of~$V$ can be related to its real part by
\[ \im V = \frac{2\, \im \Omega}{\re \Omega}\: \re V + \text{(const)} + \O\big( \sqrt{|\Omega|} \big) \:. \]
The function~$\re V$ has a zero~$v_{L\!/\!R}$ in each of the intervals~$[\ur^L, u_+^L]$ and~$[\ur^R, u_+^R]$.
It follows that at these zeros, the imaginary part of the potential has the form
\[ \im V(v_{L\!/\!R}) = \text{(const)} + \O\big( \sqrt{|\Omega|} \big) \:. \]

It remains to show that on each of the intervals~$[\ur^L, u_+^L]$ and~$[\ur^L, u_+^L]$,
the total variation of the function~$\im V$ is~$\lesssim \sqrt{\Omega|}$.
By symmetry, it suffices to consider the interval~$[\ur^L, u_+^L]$. For ease in notation,
we omit then index~$L$. We then obtain
\begin{align*}
\text{TV}|_{[\ur, u_+]} \im V &\leq (u_+^L-\ur^L) \, \sup_{[\ur^L, u_+^L]} \big| \im V' \big| \\
&\lesssim
\sqrt{\nu} \, (u_+^L-\ur^L)\: \frac{|\Omega|}{\sqrt{\nu}}
\overset{\eqref{Airyes}}{\lesssim} \frac{|\Omega|}{\sqrt{\nu}} \lesssim \frac{\sqrt{|\Omega|}}{\Const_1}\:.
\end{align*}
This concludes the proof.
\QED

\subsection{Estimates on the Interval~$[u_+, \umax]$} \label{secupumax}
It remains to estimate the solution in the Airy case
on the interval~$[u_+, \umax]$ (see~\eqref{regions} and~\eqref{urpdef}).
\begin{Lemma} \label{lemma89}
For any~$\varepsilon>0$, by increasing~$\Const_1$ one can arrange that
\[ \int_{u_+}^\umax \frac{1}{|\phi|^2} \leq \varepsilon \:. \]
Moreover,
\beq \label{philowerWKB}
|\phi(u)| \gtrsim \frac{1}{(\re V(u))^\frac{1}{4}} \: e^{\int_{u_+}^u \re \sqrt{V}} \:.
\eeq
\end{Lemma}
\Proof In Proposition~\ref{prpWKB} it was shown that the WKB conditions~\eqref{CVnew}
are satisfied on the interval~$[u_+, \umax]$.
Thus the solution is well-approximated by the WKB solution
\beq \label{WKBgen}
\phi \approx \frac{1}{(\re V)^\frac{1}{4}} \left( C_1\: e^{\int_{u_+}^u \sqrt{V}}
+ C_2\: e^{-\int_{u_+}^u \sqrt{V}} \right) ,
\eeq
with error terms which are under control in view of the estimates in~\cite{tinvariant}.
Note that one of the fundamental solutions in~\eqref{WKBgen} is exponentially increasing,
whereas the other is exponentially decaying.

Combining the estimate of Proposition~\ref{prpWKB} at~$u=\ur$
with the estimates~\eqref{yes1} and~\eqref{yes2} on the interval~$[\ur, u_+]$
and taking into account the normalization~\eqref{initial},
one sees that the coefficient of the exponentially increasing fundamental solution in~\eqref{WKBgen}
is bounded away from zero, and that~$|\phi(u_+)| \eqsim |\re V(u_+)|^{-\frac{1}{4}}$.
This gives~\eqref{philowerWKB}.
Next, we increase~$\Const_5$ to a new constant~$\tilde{\Const}_3$.
Denoting the corresponding boundary of the Airy region by~$\tilde{u}_+$, we obtain
\[ |\phi| \gtrsim \frac{1}{(\re V)^\frac{1}{4}} \: e^{\int_{u_+}^{\tilde{u}_+} \re \sqrt{V}} 
\: e^{\int_{\tilde{u}_+}^u \re \sqrt{V}}  \:. \]
As a consequence,
\begin{align}
\int_{\tilde{u}_+}^\umax \frac{1}{|\phi|^2} &\lesssim
-e^{-2 \int_{u_+}^{\tilde{u}_+} \re \sqrt{V}} 
\int_{\tilde{u}_+}^\umax \frac{d}{du} \left( e^{-2 \int_{\tilde{u}_+}^u \re \sqrt{V}} \right) \notag \\
&= e^{-2 \int_{u_+}^{\tilde{u}_+} \re \sqrt{V}}  \left( 1 - e^{-2 \int_{\tilde{u}_+}^{u_{\max}} \re \sqrt{V}} \right) 
\leq e^{-2 \int_{u_+}^{\tilde{u}_+} \re \sqrt{V}} \:. \label{expsmall}
\end{align}
The integral in the last exponent can be estimated from above by~$\sqrt{\nu}\, (\tilde{u}_+ - u_+)$.
Applying Lemma~\ref{lemma99}, this term can be made arbitrarily large by
increasing~$\Const_1$ (and consequently~$\Const_5$).
As a consequence, the last exponent in~\eqref{expsmall} can be made arbitrarily small.
This gives the result.
\QED

\section{Integral Estimates of $\im V$} \label{secimVes2}
In this section we shall derive the following estimates.
\begin{Prp} \label{prpimVsqrtV}
For all eigenvalues~$\lambda$, the following inequality holds,
\[ \int_{\ul}^{\ur} \frac{|\im V|}{\sqrt{|V|}} \lesssim 1 \:. \]
\end{Prp}

\begin{Prp} \label{prpdelta}
For any~$\delta>0$, by increasing~$\Const_1$
one can arrange that that for all eigenvalues~$\lambda$
in the WKB case and the parabolic cylinder case the following inequality holds:
\beq \label{intdelta}
\bigg| \int_{\ul}^{\ur} \frac{\im V}{\sqrt{|V|}} \bigg| < \delta \:.
\eeq
\end{Prp}

\subsection{Elementary Estimates of the Potential}
We begin with integral estimates of our potential.
\begin{Lemma} \label{lemmamonotone}
The function~$\re V$ is monotone increasing on the interval~$[u_0, \ur]$. Moreover,
\beq \label{intReV}
\int_{u_0}^{\ur} \frac{1}{|\re V|^\frac{3}{2}} \lesssim \frac{1}{\Const_1\, |\Omega|^2} \:.
\eeq
\end{Lemma}
\Proof In order to prove the monotonicity of~$\re V$, we first recall that in the proof of Lemma~\ref{lemma81}
we already showed that~$\re V$ is monotone increasing on the interval~$(\uu, \umax)$.
On the remaining interval~$[u_0, \uu]$, we need to consider the cases~$k=s$ and~$k \neq s$
separately. In the case~$k=s$, the monotonicity of~$\re V$ is obvious from~\eqref{V0ex}.
In the case~$k \neq s$, we see from~\eqref{Vn0ex} that $\re V$ is convex.
Combining this with the fact that~$u_0$ is chosen as a minimum of~$\re V$ (see~\eqref{u0ks}),
we conclude again that~$\re V$ is monotone increasing on the interval~$[u_0, \uu]$.

For the integral estimate~\eqref{intReV}, we first consider the interval~$(\ul, u_2)$ with
\beq \label{u2def7}
u_2= \min \big( \ur, \sqrt{\Const_1} \,|\Omega|^{-\frac{1}{2}} \big) \:.
\eeq
Then the desired estimate is obtained by using that the
integration range scales like~$|\Omega|^{-\frac{1}{2}}$.
On the remaining interval~$(u_2, \ur)$ we consider the
regions~$(u_2, \frac{\pi}{6})$ and~$(\frac{\pi}{6}, \ur)$ separately.
In the first region, we approximate~$\re V$ by the quadratic potential
\beq \label{poly0}
|\re V| \geq a - c\, (u-u_2)^2 \:,
\eeq
with parameters~$a,c>0$ to be specified below.
Applying Lemma~\ref{lemmaimV}, it follows by explicit computation that
\begin{align*}
\int_{u_2}^{\min(\ur, \frac{\pi}{6})} \frac{1}{|V|^\frac{3}{2}} &\leq
\int_{u_2}^{\min(\ur, \frac{\pi}{6})} \frac{1}{\big(a - c\, (u-u_2)^2 \big)^\frac{3}{2}} \\
&= \frac{u-u_2}{a\: \sqrt{a - c\, (u-u_2)^2 }} \bigg|_{\min(\ur, \frac{\pi}{6})}
\leq \frac{u-u_2}{a\: \sqrt{\Const_1 \,|\Omega|}} \bigg|_{\min(\ur, \frac{\pi}{6})} \:,
\end{align*}
where in the last step we used that we are in the WKB region (see~\eqref{regions}, \eqref{urdef}
and~\eqref{nudef}). Since the quadratic polynomial~$a-c(u-u_2)^2$ is positive, we know that
\[ u-u_2 \leq \sqrt{\frac{a}{c}} \:, \]
implying that
\[ \int_{u_2}^{\min(\ur, \frac{\pi}{6})} \frac{1}{|V|^\frac{3}{2}}
\leq \frac{1}{\sqrt{\Const_1 a c \,|\Omega|}} \:. \]
Using that~$a \geq \Const_1 \,|\Omega|$ and~$c \eqsim |\Omega|^2$, we obtain the desired estimate.

On the remaining interval~$[\frac{\pi}{6}, \ur]$, we estimate~$\re V$ 
on the interval~$[u_0, \ur]$ by the quadratic polynomial
\beq \label{poly}
|\re V| \geq a + b \, (\ur-u) + c\, (\ur-u)^2 \:,
\eeq
where the values of the positive coefficients~$a$, $b$ and~$c$ will be estimated below.
It follows by explicit computation that
\begin{align*}
\int_{u_0}^{\ur} \frac{1}{|\re V|^\frac{3}{2}} &\leq
\int_{u_0}^{\ur} \Big( a + b \, (\ur-u) + c\, (\ur-u)^2 \Big)^{-\frac{3}{2}} \\
&\leq \int_{-\infty}^{\ur} \Big( a + b \, (\ur-u) + c\, (\ur-u)^2 \Big)^{-\frac{3}{2}} 
= \frac{2}{\sqrt{a} \, b + 2 a \sqrt{c}} \:.
\end{align*}
It remains to analyze the coefficients~$a$, $b$ and~$c$.
In view of~\eqref{nudef}, \eqref{urdef} and~\eqref{nudef}, we can choose~$a\geq \Const_5 |\Omega|$.
Moreover, 
the expansions~\eqref{reV}--\eqref{reVpp} show that
we can choose either~$b \eqsim |\Omega|^2$ or~$c \eqsim |\Omega|^2$.
This concludes the proof.
\QED

\begin{Lemma} \label{lemmaeles}
For any~$\delta>0$ there is a constant~$C=C(\delta)$ such that the following inequality holds,
\beq \label{awayes}
\int_{\ul}^{\min\big(\ur, \piot-\delta \big)} \frac{1}{\sqrt{|V|}} \leq \frac{C}{|\Omega|} \:.
\eeq
Moreover, there is a constant~$C$ such that
\beq \label{alles}
\int_{\ul}^{\ur} \frac{1}{\sqrt{|V|}} \leq C\: \frac{\log |\Omega|}{|\Omega|} \:.
\eeq
\end{Lemma}
\Proof As in the previous lemma, we first consider the interval~$(\ul, u_2)$ with~$u_2$
according to~\eqref{u2def7}.
Then, according to~\eqref{cases} and~\eqref{regions}, \eqref{urdef}, \eqref{nudef} and~\eqref{absVlower},
we know that
\[ |\re V| \big|_{(\ul, \ur)} \gtrsim \Const_1\, |\Omega| \]
and thus
\[ \int_{u_0}^{u_2} \frac{1}{\sqrt{|V|}} \lesssim \frac{u_2-u_0}{\sqrt{\Const_1 \,|\Omega|}}\:. \]
The desired estimate is obtained by using that the
integration range scales like~$|\Omega|^{-\frac{1}{2}}$.

It remains to consider the interval~$(u_2, \ur)$. We consider the regions~$(u_2, \frac{\pi}{6})$
and $(\frac{\pi}{6}, \ur)$ separately. In the first region, the first and second derivatives of~$\re V$
are non-negative. We again estimate~$\re V$ by the quadratic polynomial~\eqref{poly0},
where the values of the positive coefficients~$a$ and~$c$ will be estimated below.
It follows by explicit computation that
\begin{align*}
\int_{u_2}^{\min(\ur, \frac{\pi}{6})} \frac{1}{\sqrt{|V|}} &\leq
\int_{u_2}^{\min(\ur, \frac{\pi}{6})} \frac{1}{\sqrt{a - c\, (u-u_2)^2}} \\
&= \frac{1}{\sqrt{c}} \: \arctan \bigg( \frac{\sqrt{c}\: (u-u_2)}{\sqrt{a - c\, (u-u_2)^2}} \bigg)
\bigg|_{\min(\ur, \frac{\pi}{6})} \:.
\end{align*}
Since the $\arctan$ is bounded, it suffices to note that~$c \eqsim |\Omega|^2$
to obtain the desired $1/|\Omega|$ behavior.

On the remaining interval~$(\frac{\pi}{6}, \ur)$, we again estimate~$\re V$ by 
the quadratic polynomial~\eqref{poly},
where the values of the positive coefficients~$a$, $b$ and~$c$ will be estimated below.
An explicit computation yields
\begin{align}
&\int_{u_2}^{\ur} \frac{1}{\sqrt{|V|}} \leq \int_{u_2}^{\ur} \frac{1}{\sqrt{a + b \, (\ur-u) + c\, (\ur-u)^2}} \notag \\
&\;\;= \frac{1}{\sqrt{c}}\: \log \Big( b + 2c\, (\ur-u) + 2 \sqrt{c}\: \sqrt{a + b \, (\ur-u) + c\, (\ur-u)^2} \Big) \Big|_{\ur}^{u_2}\:.
\label{logterm}
\end{align}
It remains to analyze the argument of the logarithm. From the explicit form of the potential~\eqref{Vdef}
we know that we may choose~$b, c \eqsim |\Omega|^2$.
If~$a \gg |\Omega|^2$ is large, the logarithm in~\eqref{logterm}
is given approximately by~$\log(2 \sqrt{c}\: \sqrt{a})$. As a consequence, the difference of the logarithms
at the upper and lower boundary points is uniformly bounded. Using that~$c \eqsim |\Omega|^2$,
we obtain the desired estimate.

It remains to consider the case that~$a \lesssim |\Omega|^2$.
Then the arguments of the logarithm scale like~$|\Omega|^2$, both at the upper and lower
boundary point. As a consequence, the logarithm is again uniformly bounded,
giving the desired estimate. This concludes the proof of the inequality~\eqref{awayes}.

In order to prove~\eqref{alles}, we first note that
the estimate~\eqref{awayes} fails to hold in general if~$\delta=0$.
The problem is that in this case, the parameter~$b$ in~\eqref{logterm} could be small,
leading to a factor~$\log |\Omega|$. This gives~\eqref{alles}.
\QED
Applying Lemma~\ref{lemmaimV} to~\eqref{alles}, we obtain the estimate
\[ \int_{\ul}^{\ur} \frac{|\im V|}{\sqrt{|V|}} \lesssim \log |\Omega| \:. \]
Unfortunately, the factor~$\log |\Omega|$ is not good enough for our purposes.
The next lemma shows that, if~$\im V$ vanishes at the right boundary point,
then we get an estimate without such a logarithmic factor.

\begin{Lemma} \label{lemmaelfin}
Assume that~$\tilde{u} \in (\ul, \ur)$ is a point where~$\im V(\tilde{u})=0$. Then
\[ \int_{\ul}^{\tilde{u}} \frac{|\im V|}{\sqrt{|V|}} \lesssim 1 \:. \]
\end{Lemma}
\Proof Combining Lemma~\ref{lemmaimV} with Lemma~\ref{lemmaeles},
it remains to consider the case~$\ur > \piot-\delta$. Moreover, it
remains to estimate the integral over the interval~$(\frac{\pi}{6}, \ur)$.
We again estimate~$\re V$ by the quadratic polynomial~\eqref{poly}
with positive coefficients~$a$, $b$ and~$c$. Moreover, we estimate~$\im V$ by
\[ |\im V(u)| \lesssim |\Omega|\: (\tilde{u}-u) \:. \]
We thus obtain
\begin{align*}
&\int_{u_2}^{\tilde{u}} \frac{|\im V|}{\sqrt{|V|}} \lesssim \int_{u_2}^{\tilde{u}}
\frac{|\Omega| \,(\tilde{u}-u)}{\sqrt{a + b \, (\ur-u) + c\, (\ur-u)^2}}
\leq \int_{u_2}^{\ur} \frac{|\Omega| \,(\ur-u)}{\sqrt{a + b \, (\ur-u) + c\, (\ur-u)^2}}  \\
&\;\;\leq \frac{\Omega}{2c} \int_{u_2}^{\ur} \frac{2c\,(\ur-u) + b}{\sqrt{a + b \, (\ur-u) + c\, (\ur-u)^2}}
= \frac{\Omega}{c} \sqrt{a + b \, (\ur-u) + c\, (\ur-u)^2} \Big|_{\ur}^{u_2} \:.
\end{align*}
The result follows because~$c \eqsim |\Omega|^2$ and the argument of the square
root is bounded by~$\lesssim |\Omega|^2$.
\QED

\subsection{Estimates on the Interval~$(0, \ul)$}
\begin{Lemma} \label{lemmazul}
For any~$\tilde{\delta}>0$, we can arrange by increasing~$\Const_5$ that
\[ \int_0^{\ul} |\im V|\:  \big| \phiD_L \big|^2 \leq \tilde{\delta}\:, \]
uniformly in~$\Omega$ and~$\lambda$.
\end{Lemma}
\Proof We begin with the case~$k=s$. Using the asymptotics as in the proof of Lemma~\ref{lemmaremu},
we find that on the interval~$(0, u_\ell)$,
\begin{align*}
|\phi_L(u)| &\leq 2\, |c|\: \sqrt{u}\: \log(\sqrt{\mu} u) \\
|\zeta_L(u)| &\leq \frac{1}{|c|^2} \left( \arctan \big( \log(\sqrt{\mu} u) \big) + \frac{\pi}{2} \right)
\lesssim \frac{1}{\log(\sqrt{\mu} u)} \:,
\end{align*}
uniformly in~$\mu$. As a consequence,
\[ |\phiD_L|^2 \lesssim u \qquad \text{uniformly in~$\mu$} \:. \]
Applying Lemma~\ref{lemmaimV}, we obtain
\[ \int_0^{\ul}  |\im V|\:  \big| \phiD_L \big|^2 \lesssim
|\Omega| \, \ul^2 \overset{\eqref{uldef}}{\leq} \frac{\Const_1^2\, |\Omega|}{\re \lambda} 
\overset{\eqref{relamleft}}{\leq} \frac{\Const_1^2}{\Const_5} \:. \]
This gives the result.

In the case~$k \neq s$, we work similarly with the asymptotics~\eqref{phias2},
\begin{align*}
|\phi| &\eqsim |\mu|^{-\frac{L}{2}}\: u^{\frac{1}{2}-L} \\
\int_0^u \frac{1}{|\phi|^2} &\lesssim |\mu|^L\: u^{2L} \\
|\phiD_L| &\lesssim |\mu|^{\frac{L}{2}}\: u^{L+\frac{1}{2}} \\
\int_0^{\ul} |\im V|\: |\phiD_L|^2 &\lesssim |\Omega| \: |\mu|^L\: \ul^{2L+2} 
\overset{\eqref{uldef}}{\lesssim} \frac{\Const_1^{2L+2}\,|\Omega|}{|\re \lambda|}
\overset{\eqref{relamleft}}{\leq} \frac{\Const_1^{2L+2}}{\Const_5}  \:.
\end{align*}
This concludes the proof.
\QED

\subsection{WKB Representation of~$\phiD_L$} \label{secWKBrep}
We again let~$\phiD_L$ be the solutions~\eqref{phiDdef}
with~$\phi_L$ defined by~\eqref{phiLRdef} and~\eqref{yinitLR}).
In this section, we shall approximate~$\phiD_L$ on the interval~$(\ul, \ur)$
by a suitable WKB wave function.
Our starting point is the WKB approximation of~$\phi_L$
\beq \label{phiWKB}
\phi_L \approx \phi_\WKB = \frac{c_\WKB}{\sqrt[4]{V}} \: e^{\int_{u_0}^u \sqrt{V}} \:.
\eeq
In order to comply with the initial conditions~\eqref{initial}, we must choose
\[ c_\WKB^2 = \frac{\sqrt{V(u_0)}}{\im \sqrt{V(u_0)}}\:. \]
In the next proposition, we compute what this approximation means
for the osculating circles as introduced in Section~\ref{secosc}.
\begin{Prp} \label{prppRapprox}
On the interval~$[\ul, \ur]$, the radius and center of the osculating circle are given by
\beq \label{pRapprox}
\p \approx \zeta(u_0) + \frac{1}{2 c_\WKB^2} \qquad \text{and} \qquad
R \approx \frac{1}{2}\: e^{-2 \int_{u_0}^u \re \sqrt{V}} \:,
\eeq
where the error can be made arbitrarily small by increasing~$\Const_1$.
\end{Prp}
\Proof Using the WKB approximation~\eqref{phiWKB}, we obtain
\[ y = \frac{\phi'}{\phi} \approx \sqrt{V} - \frac{V'}{4 V} \]
and thus
\beq \label{zetaWKB}
\begin{split}
\zeta(u) -\zeta(u_0) &= \int_{u_0}^u \frac{1}{\phi^2} \approx \frac{1}{c_\WKB^2}
\int_{u_0}^u \sqrt{V(\tau)} \:e^{-2
\int_{u_0}^\tau \sqrt{V}} \:d\tau \\
&=  -\frac{1}{2 c_\WKB^2} \int_{u_0}^u \frac{d}{d \tau} e^{-2 \int_{u_0}^\tau \sqrt{V}} \:d\tau = -\frac{1}{2 c_\WKB^2}
\left( e^{-2 \int_{u_0}^u \sqrt{V}} - 1 \right) .
\end{split}
\eeq
It is remarkable that the integral can be carried out explicitly, giving a simple
expression for~$\zeta(u)$.
\QED

We next compute~$\phiD_L$. Using~\eqref{phiWKB} and~\eqref{zetaWKB}, we obtain
\begin{align*}
\phiD_L(u) &= \phi_L(u)\: \zeta_L(u) = \phi_L \left( \int_0^{u_0} \frac{1}{\phi^2}
+ \big( \zeta(u) - \zeta(u_0) \big) \right) \\
&\approx \frac{c_\WKB}{\sqrt[4]{V}}\:e^{\int_{u_0}^u \sqrt{V}}
\left( \int_0^{u_0} \frac{1}{\phi^2}
-\frac{1}{2 c_\WKB^2} \left(e^{-2 \int_{u_0}^u \sqrt{V}} - 1 \right) \right) .
\end{align*}
Hence
\beq
\phiD_L(u) \approx \frac{1}{2 c_\WKB} \left( \frac{\alpha}{\sqrt[4]{V}}\: e^{\int_{u_0}^u \sqrt{V}}
- \frac{1}{\sqrt[4]{V}}\: e^{-\int_{u_0}^u \sqrt{V}} \right) \:,
\label{phiDWKB}
\eeq
where~$\alpha$ is the constant
\[ \alpha = 1 + 2 \,c_\WKB^2 \int_0^{u_0} \frac{1}{\phi^2} \:. \]

\begin{Lemma} \label{lemmaalphaabs}
By choosing~$\Const_5$ sufficiently large, we can make the expression
\[ \Big| |\alpha| - 1 \Big| \]
arbitrarily small.
\end{Lemma}
\Proof Choosing~$\Const_5$ sufficiently large, we can arrange that the potential at~$u_0$
is approximately real and negative (cf.~\eqref{u00def}, \eqref{V0ex} and~\eqref{u0Ldef}, \eqref{Vn0ex}).
Hence
\[ c_\WKB^2 \approx i \:, \]
implying that
\beq \label{absalpha}
|\alpha| \approx \Big| 1 + 2 i \int_0^{u_0} \frac{1}{\phi^2} \Big| =
\big| 1 + 2 i \zeta(u_0) \big| \:,
\eeq
with an arbitrarily small error. Next, using the initial conditions~\eqref{initial} in~\eqref{KPrel},
we obtain
\[ \p(u_0) = \zeta(u_0) -\frac{i}{2} \qquad \text{and} \qquad R(u_0) = \frac{1}{2} \:. \]
Solving for~$\zeta(u_0)$ and substituting into~\eqref{absalpha}, one finds that~$|\alpha| \approx 2\, |\p(u_0)|$,
and thus
\[ \Big| |\alpha| - 1 \Big| \leq 2\, \big| |\p(u_0)|-R(u_0) \big| \:. \]
Since~$\zeta(0)=0$, we know that~$|\p(0)|=R(0)$. Hence
\[ \big| |\p(u_0)|-R(u_0) \big| \leq \int_0^{u_0} \big( |\p'|+ |R'| \big) \:. \]
In view of~\eqref{KPrel} and~\eqref{Rprel}, we know that~$|\p'|=|R'|$.
Hence our task is to show that the total variation of~$R$ on the interval~$(0, u_0)$
is arbitrarily small. Since~$R(u_0)=\frac{1}{2}$, it suffices to show that the
total variation of~$\log R$ is small. Thus, according to~\eqref{Rprel}, it remains to estimate the integral
\[ \int_0^{u_0} \frac{|R'|}{R} = \int_0^{u_0} \frac{|\im V|}{\im y}\:. \]
In view of the estimates in Sections~\ref{secpole1} and~\ref{secpole2}, we know that
\[ \im y \gtrsim \re \mu\:. \]
Hence
\[ \int_0^{u_0} \frac{|\im V|}{\im y} \lesssim \frac{u_0}{\re \mu} \lesssim \frac{|\Omega|}{\re \mu}\:. \]
By choosing~$\Const_5$ sufficiently large, we can make this expression arbitrarily small.
\QED

Using this lemma in formula~\eqref{phiDWKB}, we obtain the estimate
\beq \label{phiDes2}
|\phiD_L(u)|^2 \lesssim \frac{1}{\sqrt{|V|}}\: \cosh \left( 2 \int_{u_0}^u \re \sqrt{V} \right) \:.
\eeq

\subsection{Integral Estimates of WKB Solutions}
\begin{Lemma} \label{lemmaleft}
Assume that~$\tilde{u} \in (\ul, \ur)$ is a point where~$\im V(\tilde{u})=0$. Then
\[ \int_{\ul}^{\tilde{u}} |\im V| \:|\phiD_L|^2 \lesssim 1 \:. \]
\end{Lemma}
\Proof We can work with the WKB approximation~\eqref{phiDes2}.
According to Lemma~\ref{lemmaelfin}, we know that
\[ \bigg| \int_{u_0}^{\tilde{u}} \re \sqrt{V} \bigg| \lesssim \int_{\ur}^{\tilde{u}} \frac{|\im V|}{\sqrt{|V|}} \lesssim 1 \:, \]
giving uniform control of the absolute value of the hyperbolic cosine in~\eqref{phiDes2}.
As a consequence,
\[ \int_{\ul}^{\tilde{u}} |\im V| \:|\phiD_L|^2 \lesssim \int_{\ul}^{\tilde{u}} \frac{|\im V|}{\sqrt{|V|}} \lesssim 1\:, \]
where in the last step we again applied Lemma~\ref{lemmaelfin}.
This concludes the proof.
\QED

Next, we take the absolute square of the WKB approximation~\eqref{phiDWKB},
\begin{align}
\big|\phiD_L(u) \big|^2
&\approx \frac{|\alpha|^2}{4 \,|c_\WKB|^2} \: \frac{1}{\sqrt{|V|}}
\: e^{2 \int_{u_0}^u \re \sqrt{V}}
+ \frac{1}{4 \,|c_\WKB|^2} \: \frac{1}{\sqrt{|V|}}\: e^{-2 \int_{u_0}^u \re \sqrt{V}} \label{absphiD1} \\
&\qquad + \frac{1}{2 \,|c_\WKB|^2} \: \frac{1}{\sqrt{|V|}}\:
\re \left( \alpha \:e^{2i \int_{u_0}^u \im \sqrt{V}} \right) \:. \label{absphiD2}
\end{align}
The integrand of the last term is oscillatory. As a consequence, the resulting integral
is small, as quantified in the next lemma.

\begin{Lemma} \label{lemmaoscillate}
For any~$\delta>0$, by increasing~$\Const_1$
one can arrange that that for all eigenvalues~$\lambda$ the following inequality holds:
\[ \int_{\ul}^{\ur} \frac{|\im V|}{\sqrt{|V|}}\; e^{2i \int_{u_0}^u \im \sqrt{V}} < \delta \:. \]
\end{Lemma}
\Proof Since on the interval~$(0, \piot)$, the function~$\im V$ changes signs only once,
it suffices to show that for any~$\tilde{u}_1, \tilde{u}_2 \in [\ul, \ur]$,
\[ \left| \int_{\tilde{u}_1}^{\tilde{u}_2} \frac{\im V}{\sqrt{|V|}}\; e^{2i \int_{u_0}^u \im \sqrt{V}} \right| < \delta \:. \]
Integrating by parts,
\begin{align*}
&\int_{\tilde{u}_1}^{\tilde{u}_2} \frac{\im V}{\sqrt{|V|}}\; e^{2i \int_{u_0}^u \im \sqrt{V}}
= \int_{\tilde{u}_1}^{\tilde{u}_2}  \frac{\im V}{\sqrt{|V|}}\: \frac{1}{2 i\,\im \sqrt{V}}\; 
\frac{d}{du}\: e^{2i \int_{u_0}^u \im \sqrt{V}} \\
&\;= \frac{\im V}{\sqrt{|V|}}\: \frac{1}{2 i\,\im \sqrt{V}}\; 
e^{2i \int_{u_0}^u \im \sqrt{V}} \bigg|_{\tilde{u}_1}^{\tilde{u}_2}
- \int_{\tilde{u}_1}^{\tilde{u}_2}  \frac{d}{du} \left( \frac{\im V}{\sqrt{|V|}}\: \frac{1}{2 i\,\im \sqrt{V}} \right)
e^{2i \int_{u_0}^u \im \sqrt{V}} \:,
\end{align*}
we obtain the estimate
\begin{align*}
& \left| \int_{\tilde{u}_1}^{\tilde{u}_2} \frac{\im V}{\sqrt{|V|}}\; e^{2i \int_{u_0}^u \im \sqrt{V}} \right| \\
&\leq \frac{|\im V|}{2\, |V|}\: \bigg|_{\tilde{u}_1} 
+ \frac{|\im V|}{2\, |V|}\: \bigg|_{\tilde{u}_2}
+ \int_{\tilde{u}_1}^{\tilde{u}_2}  \left| \frac{d}{du} \bigg( \frac{\im V}{2\, \sqrt{|V|} \,\im \sqrt{V}} \bigg) \right| \\
&\lesssim \frac{|\im V|}{|\re V|}\: \bigg|_{\tilde{u}_1} 
+ \frac{|\im V|}{|\re V|}\: \bigg|_{\tilde{u}_2}
+ \int_{\tilde{u}_1}^{\tilde{u}_2} \frac{|\im V'|}{|\re V|}
+ \int_{\tilde{u}_1}^{\tilde{u}_2} \frac{|\im V'|^2}{|\re V|^2}
+ \int_{\tilde{u}_1}^{\tilde{u}_2} \frac{|\re V'| \,|\im V|}{|\re V|^2} \:.
\end{align*}
All the terms except for the last summand can immediately be estimated in the desired way
using the explicit form of our potential.
For the last term we use the monotonicity of~$\re V$ (see Lemma~\ref{lemmamonotone})
to obtain
\[ \int_{\tilde{u}_1}^{\tilde{u}_2} \frac{|\re V'| \,|\im V|}{|\re V|^2} 
\lesssim |\Omega| \int_{\tilde{u}_1}^{\tilde{u}_2} \frac{|\re V'|}{|\re V|^2} 
\leq  |\Omega|\, \sup_{[\tilde{u}_1, \tilde{u}_2]} |V|^{-1} \:. \]
This gives the result.
\QED

Keeping track of the constants, we now write~$\phiD$ as
\beq \label{phiDrep}
\phiD(u) = \left\{ \begin{array}{cl} c_L \,\phiD_L & \text{if~$u \leq \piot$} \\[0.3em]
c_R \,\phiD_R & \text{if~$u > \piot$}\:, \end{array} \right.
\eeq
where~$c_L$ and~$c_R$ are non-zero complex numbers 
(and~$\phiD_L$ and~$\phiD_R$ are again the solutions~\eqref{phiDdef}
with~$\phi_L$ and~$\phi_R$ defined by~\eqref{phiLRdef} and~\eqref{yinitLR}).

\begin{Lemma} \label{lemma119} If~$|c_L/c_R| \leq 1$, then
\beq \label{intphiD}
\int_\piot^\pi |\im V|\, |\phiD|^2 \lesssim |c_R|^2 \:.
\eeq
\end{Lemma}
\Proof We again denote the zeros of~$\im V$ by~$\tilde{u}_L$ and~$\tilde{u}_R$.
Lemma~\ref{lemmazul} and Lemma~\ref{lemmaleft} imply that
\[ \left( \int_0^{\tilde{u}_L} + \int_{\tilde{u}_R}^\piot \right)  |\im V|\, |\phiD|^2
\lesssim |c_R|^2 \:. \]
Moreover, using the representation~\eqref{phiDrep} in~\eqref{imVint}, we obtain
\[ \int_\piot^{\tilde{u}_R} |\im V|\, |\phiD|^2 \lesssim |c_R|^2 \:. \]
This gives the result.
\QED
\begin{Lemma} \label{lemma1110} If~$|c_L/c_R| \leq 1$, then
\[ \int_{u_0}^{\ur} \frac{|\im V|}{\sqrt{|V|}} \lesssim 1 \:. \]
\end{Lemma}
\Proof The strategy is to combine Lemma~\ref{lemma119} with the fact that
the potential is approximately symmetric with respect to reflections at~$u=\piot$.
In order to make this approximate symmetry precise, we consider the homotopy
\beq \label{Vhomo}
V_\tau(u) := \tau \,V(u) + (1-\tau)\, V(\pi-u) \qquad \text{for} \qquad \tau \in [0,1] \:.
\eeq
Then the mean value theorem implies that
\[ \big| \re V_1(u) - \re V_2(u) \big| \leq \sup_{\tau \in [0,1]} \left|
\frac{d}{d\tau} \re V_\tau(u) \right| \:, \]
and similarly for the imaginary part. Using that the function~$\sin^2 u$
in~\eqref{Vdef} is reflection symmetric, one finds that
\beq \label{Vdiff}
\big| \re V_1(u) - \re V_2(u) \big| \lesssim |\Omega| \qquad \text{and} \qquad
\big| \im V_1(u) - \im V_2(u) \big| \lesssim 1 \:.
\eeq
This implies that the WKB approximation holds on the ``reflected
WKB-region''~$[\pi-\ur, \pi-u_0]$.
Using the WKB approximation~\eqref{absphiD1} and~\eqref{absphiD2} in~\eqref{intphiD},
the oscillatory contribution~\eqref{absphiD2} was estimated in Lemma~\ref{lemmaoscillate}.
Noting that one of the factors~$\exp(\pm 2 \int_{u_0}^u \re \sqrt{V})$ in~\eqref{absphiD1}
is greater than one, Lemma~\ref{lemma119} implies that
\[ \int_{\pi-\ur}^{\pi-u_0} \frac{|\im V|}{\sqrt{|V|}} \lesssim 1 \:. \]

Again applying the reflection argument and the mean value theorem, it remains to show that
\[ \int_{u_0}^{\ur} \left| \frac{d}{d\tau} \left( \frac{\im V_\tau}{\sqrt{V_\tau}} \right) \right| \lesssim 1 \:. \]
Again using the explicit form of the potential~\eqref{Vdef},
\begin{align*}
\left|\frac{d}{d\tau} \left( \frac{\im V_\tau}{\sqrt{V_\tau}} \right) \right|
= \left| \frac{\im \partial_\tau V_\tau}{\sqrt{V_\tau}} \right| + \left| \frac{(\partial_\tau V_\tau)\: \im V_\tau}
{V_\tau^\frac{3}{2}} \right|
\lesssim \frac{1}{\sqrt{|\Omega|}} + \frac{|\Omega|^2}{|V|^\frac{3}{2}} \:.
\end{align*}
Integrating this inequality from~$u_0$ to~$\ur$, we can apply
Lemma~\ref{lemmamonotone} to obtain the result.
\QED

\Proof[Proof of Proposition~\ref{prpimVsqrtV}]
It suffices to consider the case~$|c_L/c_R| \leq 1$ and to show that
\[ \int_{u_l^L}^{\ur^L} \frac{|\im V|}{\sqrt{|V|}} + \int_{\ur^R}^{u_l^R} \frac{|\im V|}{\sqrt{|V|}} \lesssim 1\:. \]
Then case~$|c_L/c_R| > 1$ can be treated similarly by exchanging the
left and right subintervals with the reflection~$u \leftrightarrow \pi - u$.

On the interval~$[\ur^R, u_l^R]$, we argue as in the proof of Lemma~\ref{lemma119}
to obtain
\[ \int_{\ur^R}^{u_l^R} \frac{|\im V|}{\sqrt{|V|}} \lesssim 1\:. \]
On the interval~$[u_l^L, \ur^L]$ we consider different subintervals:
The region from~$\ul$ to~$u_0$ is estimated in Lemma~\ref{lemmaeles}.
The region from~$u_0$ to~$\ur$, on the other hand, is estimated
in Lemma~\ref{lemma1110}. This concludes the proof.
\QED

\subsection{Estimates in the WKB and Parabolic Cylinder Cases} \label{secreflect}
We now give the proof of Proposition~\ref{prpdelta}.
Our strategy is to refine the method of Section~\ref{secimVes}
and to combine it with the ``reflection argument'' which was
already used in the proof of Lemma~\ref{lemma1110}.
Another ingredient is Proposition~\ref{prpimVsqrtV} (whose proof 
was completed in the previous section).

We will apply Proposition~\ref{prpimVsqrtV} in the following way.
In the WKB region, we know from~\eqref{relamright}, \eqref{relamleft} and~\eqref{urdef} that
\[ |\im V| \lesssim |\Omega| \qquad \text{and} \qquad \re V \lesssim -\Const_1\, |\Omega| \]
and thus
\[ \frac{|\im V|}{|\re V|} \lesssim \frac{1}{\Const_1} \qquad \text{in the WKB region}\:. \]
As a consequence, we may expand the square root of the potential as
\[ \sqrt{V} = \sqrt{\re V + i \im V} = i \sqrt{|\re V| - i \im V} = i \sqrt{|\re V|} + \frac{\im V}{\sqrt{|\re V|}} + \cdots \:, \]
showing that
\beq \label{resqrtapprox}
\re \sqrt{V} = \frac{\im V}{\sqrt{|V|}} \Big(1 + \O\Big(\frac{1}{\Const_1} \Big) \Big)\:.
\eeq
In particular,
\[ \big| \re \sqrt{V} \big| \lesssim \frac{|\im V|}{\sqrt{|V|}} \:, \]
and applying Proposition~\ref{prpimVsqrtV}, we conclude that
\beq \label{intsqrtV}
\int_{\ul}^{\ur} \big| \re \sqrt{V} \big| \lesssim 1 \:.
\eeq
This shows that that the exponentials and hyperbolic cosine in the WKB approximation
(see~\eqref{phiWKB}, \eqref{phiDWKB}, \eqref{phiDes2} and~\eqref{absphiD1}, \eqref{absphiD2})
are uniformly bounded.

We again assume that~$\lambda \in \C$ is an eigenvalue and~$\phiD$ the corresponding
eigenfunction. Moreover, assume that we are in the 
WKB case or the parabolic cylinder case (but not in the Airy case, which is
excluded in Proposition~\ref{prpdelta}).
Using the representation~\eqref{phiDrep} in~\eqref{imVint}, we obtain the identity
\beq \label{cLRrel}
|c_L|^2 \int_0^\piot \im V\: |\phiD_L|^2 + |c_R|^2 \int_\piot^\pi \im V\: |\phiD_R|^2 = 0 \:.
\eeq
Denoting the integrands by~$f$,
\[ f(u) := \left\{ \begin{array}{cl} \im V\: |\phiD_L|^2 & \text{if~$u \leq \piot$} \\[0.3em]
\im V\: |\phiD_R|^2 & \text{if~$u > \piot$\:,} \end{array} \right. \]
we decompose~$f$ into its even and odd parts,
\beq \label{fpmdef}
f = f_+ + f_- \qquad \text{with} \qquad f_\pm(u) := \frac{1}{2} \Big( f(u) \pm f(\pi-u) \Big) \:.
\eeq
Then we can rewrite~\eqref{cLRrel} as
\begin{align*}
0 &= |c_L|^2 \int_0^\piot (f_+ + f_-)  + |c_R|^2 \int_\piot^\pi (f_+ + f_-) \\
&= |c_L|^2 \int_0^\piot (f_+ + f_-)  + |c_R|^2 \int_0^\piot (f_+ - f_-) \\
&= \Big( |c_L|^2 + |c_R|^2 \Big) \int_0^\piot f_+ \:+\: 
\Big( |c_L|^2 - |c_R|^2 \Big) \int_0^\piot f_- \:.
\end{align*}
Since the constants~$c_L$ and~$c_R$ are non-zero, we obtain the inequality
\[ \bigg| \int_0^\piot f_+ \bigg| \leq \bigg| \int_0^\piot f_- \bigg| \:. \]
Combining this estimate with~\eqref{fpmdef}, we obtain the inequality
\beq \label{startineq}
\bigg| \int_0^\piot \im V\: |\phiD_L|^2 \bigg| = \bigg| \int_0^\piot (f_+ + f_-) \bigg|
\leq 2 \:\bigg| \int_0^\piot f_- \bigg|\:.
\eeq

We now estimate the right side of this inequality obtain the following result.
\begin{Lemma} \label{lemmadelta}
For any~$\delta>0$, by increasing~$\Const_1$
one can arrange that that for all eigenvalues~$\lambda$
in the WKB case and the parabolic cylinder case the following inequality holds:
\[ \bigg| \int_{\ul}^{\ur} \im V\: |\phiD_L|^2 \bigg| < \delta \:. \]
\end{Lemma}
\Proof
We introduce the ``parity transformation''~$P$ which reflects at the point~$\piot$,
\[ P \::\: (0,\pi) \rightarrow (0,\pi)\:,\qquad Pu = \pi - u \:. \]
Then~\eqref{startineq} can be written as
\[ \bigg| \int_0^\piot \im V\: |\phiD_L|^2 \bigg|
\leq \bigg| \int_0^\piot \Big( \im V\: |\phiD_L|^2 - \im (V \circ P)  \: \big|\phiD_R \circ P \big|^2 \Big) \bigg| \:. \]
On the interval~$(0,\ul)$, this integral can be estimated by Lemma~\ref{lemmazul},
where we choose~$\tilde{\delta}=\delta/16$. Moreover, in view of the 
estimates of Lemma~\ref{prpTparabolic}, in the parabolic cylinder case
we may estimate the functions~$\phiD$ and~$\zeta$ at the point~$\umax$ in terms
of their values at~$\ur$. This shows that the integral over~$[\ur,\piot]$ can be made
smaller than~$\delta/8$. We conclude that
\beq \label{interint}
\bigg| \int_0^\piot \im V\: |\phiD_L|^2 \bigg|
\leq \bigg| \int_{\ul}^{\ur} \Big( \im V\: |\phiD_L|^2 - \im (V \circ P)  \: \big|\phiD_R \circ P \big|^2 \Big) \bigg| +
\frac{\delta}{4} \:.
\eeq

We next specify the wave functions~$\phiD_L$ and~$\phiD_R$.
Using~\eqref{absphiD1} and~\eqref{absphiD2} together with Lemma~\ref{lemmaalphaabs}
and Lemma~\ref{lemmaoscillate}, we know that in the integral on the right side of~\eqref{interint},
the factor~$|\phiD_L|^2$ may be replaced by the function
\[ \frac{1}{2 \sqrt{|V|}}\: \cosh \left( 2 \int_{u_0}^u \re \sqrt{V} \right) \:, \]
making an error which can be made arbitrarily small
by increasing~$\Const_1$
(here we use Proposition~\ref{prpimVsqrtV} to conclude
that a small pointwise error gives rise to a small error of the integral). 
Moreover, using~\eqref{resqrtapprox}, we may replace~$\re \sqrt{V}$
by~$\im V/\sqrt{|V|}$, again making an arbitrarily small error. Therefore, setting
\beq \label{rhodef}
\rhoD_L(u) := \frac{1}{2 \,\sqrt{|V|}}\: \cosh \bigg( 2 \int_{u_0}^u \frac{\im V}{\sqrt{|V|}} \bigg) \:,
\eeq
we can arrange that
\beq \label{coshest}
\bigg| \int_{\ul}^{\ur} \im V\: |\phiD_L|^2 - \int_{\ul}^{\ur} \im V\: \rhoD_L \bigg| \leq \frac{\delta}{8}\:.
\eeq
Using the same argument on the interval~$[\piot, \pi]$, we conclude that
\begin{align}
\bigg|& \int_0^\piot \im V\: |\phiD_L|^2 \bigg|
\leq \bigg| \int_{\ul}^{\ur} \Big( \im V\: \rhoD_L - \im (V \circ P)  \: \big(\rhoD_R \circ P \big) \Big) \bigg| +
\frac{\delta}{2} \notag \\
&\leq \bigg| \int_{\ul}^{\ur} \im \big(V-V\circ P \big)\: \rhoD_L \bigg| \label{term1} \\
&\qquad + \bigg| \int_{\ul}^{\ur} \im (V \circ P)  \: \Big(\rhoD_L - \big(\rhoD_R \circ P \big) \Big) \bigg| +
\frac{\delta}{2} \:. \label{term2}
\end{align}

We next estimate the integrals in~\eqref{term1} and~\eqref{term2} after each other.
In order to estimate~\eqref{term1}, we first note that, from the explicit form of
the potential~\eqref{Vdef}, it is obvious that
\[ \big| \im \big(V-V\circ P \big) \big| \lesssim 1 \:. \]
As a consequence,
\[ \int_{\ul}^{\ur} \big|\im \big(V-V\circ P \big)\big|\: |\rhoD_L|
\lesssim \int_{\ul}^{\ur} |\rhoD_L| \lesssim
\int_{\ul}^{\ur} \frac{1}{\sqrt{|V|}} \:, \]
where in the last step we again used Proposition~\ref{prpimVsqrtV} to conclude
that the hyperbolic cosine in~\eqref{rhodef}
is uniformly bounded. Using the estimate~\eqref{alles} in Lemma~\ref{lemmaeles},
we conclude that
\[ \int_{\ul}^{\ur} \big|\im \big(V-V\circ P \big)\big|\: |\rhoD_L|
\lesssim \frac{\log|\Omega|}{|\Omega|} \:, \]
which tends to zero for large~$|\Omega|$ and can thus be made smaller than~$\delta/4$.

In order to estimate~\eqref{term2}, we again use the homotopy~\eqref{Vhomo}. Setting
\begin{align*}
u_0(\tau) &= \tau u_0^L + (1-\tau) u_0^R \\
\rhoD_\tau &= \frac{1}{2 \sqrt{|V_\tau|}}\: \cosh \left( 2 \int_{u_0(\tau)}^u
\frac{\im V_\tau}{\sqrt{|V_\tau|}} \right) \:,
\end{align*}
we again use the mean value theorem to obtain
\begin{align*}
\big| & \rhoD_L - \big(\rhoD_R \circ P \big) \big| = \big| \rhoD_1 - \rhoD_0 \big|
\leq \sup_{\tau \in [0,1]} \Big| \frac{d}{d\tau} \rhoD_\tau \Big| \\
&\lesssim \sup_\tau \frac{1}{\sqrt{|V_\tau|}} \left(
\frac{|\partial_\tau V_\tau|}{|V_\tau|}
+ \frac{|\im V_\tau|}{\sqrt{|V_\tau|}} \: \big|\partial_\tau u_0(\tau) \big|
+ \int_{u_0(\tau)}^u \bigg( \frac{|\partial_\tau \im V_\tau|}{\sqrt{|V_\tau|}}
+ \frac{|\partial_\tau \im V_\tau\, \partial_\tau V_\tau|}{|V_\tau|^\frac{3}{2}}
\bigg) \right)
\end{align*}
(where we again used~\eqref{intsqrtV} to conclude that the hyperbolic cosine is uniformly bounded).
Using that (see also~\eqref{Vdiff})
\[ \big|\partial_\tau \re V\big| \lesssim |\Omega| \:,\qquad
\big|\partial_\tau \im V\big| \lesssim 1 
\qquad \text{and} \qquad \big|\partial_\tau u_0(\tau) \big| \lesssim \frac{1}{\sqrt{|\Omega|}} \:, \]
a straightforward computation shows that for any~$\tilde{\delta}>0$, we can arrange that
\[ \big| \rhoD_L - \big(\rhoD_R \circ P \big) \big| \leq \frac{\tilde{\delta}}{\sqrt{|V|}}\:. \]
Hence
\[ \bigg| \int_{\ul}^{\ur} \im (V \circ P)  \: \Big(\rhoD_L - \big(\rhoD_R \circ P \big) \Big) \bigg|
\leq \tilde{\delta} \int_{\ul}^{\ur} \frac{|\im (V \circ P)|}{\sqrt{|V|}} \lesssim \tilde{\delta} \:, \]
where in the last step we again applied Proposition~\ref{prpimVsqrtV}.
By choosing~$\tilde{\delta}$ sufficiently small, we can arrange that
\[ \bigg| \int_{\ul}^{\ur} \im (V \circ P)  \: \Big(\rhoD_L - \big(\rhoD_R \circ P \big) \Big) \bigg| \leq \frac{\delta}{4}\:. \]
This concludes the proof.
\QED

\Proof[Proof of Proposition~\ref{prpdelta}] The remaining task is to estimate the integral in~\eqref{intdelta}
from above by the integral in the statement of Lemma~\ref{lemmadelta}.
Applying~\eqref{coshest}, we obtain
\[ \bigg| \int_{\ul}^{\ur} \im V\: |\phiD_L|^2 \bigg| \geq 
\bigg| \int_{\ul}^{\ur} \im V\: \rhoD_L \bigg| - \delta \:. \]
Moreover, using~\eqref{rhodef}, we obtain
\begin{align*}
\int_{\ul}^{\ur} \im V\: \rhoD_L &= \int_{\ul}^{\ur} 
\frac{\im V}{2 \sqrt{|V|}}\: \cosh \bigg( 2 \, \int_{u_0}^u \frac{\im V}{\sqrt{|V|}} \bigg)
= \frac{1}{4} \int_{\ul}^{\ur} \frac{d}{du} \sinh \bigg( 2 \, \int_{u_0}^u \frac{\im V}{\sqrt{|V|}} \bigg) \\
&= \frac{1}{4} \sinh \bigg( 2 \, \int_{u_0}^{\ur} \frac{\im V}{\sqrt{|V|}} \bigg)
- \frac{1}{4} \sinh \bigg( 2 \, \int_{u_0}^{\ul} \frac{\im V}{\sqrt{|V|}} \bigg) \:.
\end{align*}
In the last summand we can use the estimate
\[ \int_{u_0}^{\ul} \frac{|\im V|}{\sqrt{|V|}} \lesssim \frac{|\Omega|}{\sqrt{\Const_1 |\Omega|}}
\: (u_0-\ul) \lesssim \frac{1}{\sqrt{\Const_1}} \]
to conclude that this summand can be made arbitrarily small.
In the first summand, on the other hand,
we apply the inequality~$|\sinh x| \geq |x|$. This gives the result.
\QED

\section{The $\lambda$-Dependence of the Osculating Circles} \label{seclambda}
In view of the result of Proposition~\ref{prpbounded}, it remains to consider the
situation for large~$|\Omega|$. In this regime, Weyl's asymptotics as worked out in
Lemma~\ref{lemmaweyl} is of no use, because we have no control of how the
error terms~$\O(n)$ and~$\O(n^0)$ depend on~$\Omega$.
In particular, we cannot expect that the gaps between the
eigenvalues for a real potential are so large that the imaginary part of the potential can be treated
as a slightly non-selfadjoint perturbation. As a consequence, we must analyze the spectrum
of the Hamiltonian with the complex potential~\eqref{Hamilton} in detail.
As a technical tool, we will again work with the osculating circle estimates
as developed in Section~\ref{secosc}. As a refinement, we need to analyze in detail
how the osculating circles depend on the spectral parameter~$\lambda$.
In view of~\eqref{Vdef} and~\eqref{mudef}, we know that
\beq \label{Vlam}
V_\lambda := \partial_\lambda V \equiv -1\:.
\eeq
Moreover, we choose~$u_0$ in~\eqref{ybound} independent of~$\lambda$
(where~$y_0$ clearly depends on~$\lambda$).

In the following computations, we treat~$\re \lambda$ and~$\im \lambda$ as two independent
real variables. Then for any function~$f(\lambda)$ which is complex differentiable, we have
\[ \frac{\partial}{\partial \re \lambda}\:f = f_\lambda \:,\qquad
\frac{\partial}{\partial \im \lambda}\:f = \lim_{h \searrow 0}
\frac{f(\lambda+i h) - f(\lambda)}{h} =  i f_\lambda \:. \]
In the next lemma we compute the $\lambda$-derivatives of~$\phi$.
\begin{Lemma} \label{lemmaDlamphi}
Choosing the initial conditions~\eqref{initial},
\begin{align*}
\frac{\partial}{\partial \re \lambda} \: \log \phi(u)
&= -\frac{\im y_\lambda(u_0)}{2 \im y(u_0)} + \int_{u_0}^u y_\lambda \\
\frac{\partial}{\partial \im \lambda} \: \log \phi(u)
&= -\frac{\re y_\lambda(u_0)}{2 \im y(u_0)} + i \int_{u_0}^u y_\lambda\:.
\end{align*}
\end{Lemma}
\Proof Differentiating~\eqref{initial} and~\eqref{phidef} with respect to~$\lambda$ gives
\begin{align*}
\frac{\partial}{\partial \re \lambda} \: \log \phi(u_0) &= -\frac{1}{2} \: \frac{\im y_\lambda(u_0)}{\im y(u_0)} \\
\frac{\partial}{\partial \im \lambda} \: \log \phi(u_0) &= -\frac{1}{2} \: \frac{\re y_\lambda(u_0)}{\im y(u_0)} \\
\frac{\partial}{\partial u} \frac{\partial}{\partial \re \lambda} \: \log \phi(u)
&= \frac{\partial}{\partial \re \lambda} y(u) = y_\lambda(u) \\
\frac{\partial}{\partial u} \frac{\partial}{\partial \im \lambda} \: \log \phi(u)
&= \frac{\partial}{\partial \im \lambda} y(u) = i y_\lambda(u) \:.
\end{align*}
Integrating the last two differential equations from~$u_0$ to~$u$ gives the result.
\QED

We next compute the second mixed derivatives of~$K$ and~$p$.
\begin{Lemma} \label{lemma42} Choosing the initial conditions~\eqref{initial},
\begin{align}
\frac{\partial}{\partial \re \lambda} \frac{\partial}{\partial u}\,\log K(u, \lambda)
&= -\frac{\im V \:\im y_\lambda}{\im^2 y} \label{DuDRelogK} \\
\frac{\partial}{\partial \re \lambda} \frac{\partial}{\partial u}\,p(u, \lambda)
&= -i \: \frac{\im V}{\phi^2 \im^2 y} \left( \frac{\im y_\lambda}{\im y} - \frac{\im y_\lambda(u_0)}{2 \im y(u_0)}
+ \int_{u_0}^u y_\lambda \right) \label{DuDRep} \\
\frac{\partial}{\partial \im \lambda} \frac{\partial}{\partial u}\,\log K(u, \lambda)
&= -\frac{1}{\im y} - \frac{\im V \,\re y_\lambda}{\im^2 y} \label{DuDImlogK} \\
\frac{\partial}{\partial \im \lambda} \frac{\partial}{\partial u}\,p(u, \lambda) 
&=  -\frac{i}{2 \,\phi^2 \im^2 y} \nonumber \\
&\quad -i\:\frac{\im V}{\phi^2 \im^2 y} \left( \frac{\re y_\lambda}{\im y} - \frac{\re y_\lambda(u_0)}{2 \im y(u_0)} + i \int_{u_0}^u y_\lambda \right) . \label{DuDImp} 
\end{align}
\end{Lemma}
\Proof
Comparing~\eqref{rey} and~\eqref{imy}, one sees that
\[ \frac{\partial}{\partial u} \log K(u, \lambda) = \frac{\im V}{\im y} \:. \]
Differentiating with respect to~$\lambda$ gives
\begin{align*}
\frac{\partial}{\partial \re \lambda} & \frac{\partial}{\partial u}\,\log K(u, \lambda)
= \frac{\partial}{\partial \re \lambda} \frac{\im V}{\im y} \\
\frac{\partial}{\partial \im \lambda} & \frac{\partial}{\partial u}\,\log K(u, \lambda)
= \frac{\partial}{\partial \im \lambda} \frac{\im V}{\im y}\:,
\end{align*}
and using~\eqref{Vlam} gives~\eqref{DuDRelogK} and~\eqref{DuDImlogK}.
In order to derive~\eqref{DuDRep} and~\eqref{DuDImp}, we first apply~\eqref{KPrel} to obtain
\begin{align*}
K\, \frac{\partial}{\partial \re \lambda} & \frac{\partial}{\partial u}\,p(u, \lambda)
= K\,\frac{\partial}{\partial \re \lambda} \left( \frac{i \, \im V}{2 \,\phi^2 \,\im^2 y} \right) \\
&= i e^{-2 i \vartheta} \: \frac{\im V}{\im y} \left( -2\, \frac{\partial}{\partial \re \lambda} \log \phi
-2\: \frac{\im y_\lambda}{\im y} \right) \\
K\, \frac{\partial}{\partial \im \lambda} & \frac{\partial}{\partial u}\,p(u, \lambda)
= K\,\frac{\partial}{\partial \im \lambda} \left( \frac{i \, \im V}{2 \,\phi^2 \,\im^2 y} \right) \\
&= i e^{-2 i \vartheta} \: \frac{1}{\im y} \left( -1
-2 \im V\, \frac{\partial}{\partial \im \lambda} \log \phi -2 \im V\:\frac{\re y_\lambda}{\im y} \right) .
\end{align*}
Applying Lemma~\ref{lemmaDlamphi} gives
\begin{align*}
K\, \frac{\partial}{\partial \re \lambda} \frac{\partial}{\partial u}\,p(u, \lambda)
&= -2 i e^{-2 i \vartheta}\; \frac{\im V}{\im y} 
\left( \frac{\im y_\lambda}{\im y} - \frac{\im y_\lambda(u_0)}{2 \im y(u_0)}
+ \int_{u_0}^u y_\lambda \right) \\
K\, \frac{\partial}{\partial \im \lambda} \frac{\partial}{\partial u}\,p(u, \lambda)
& = -\frac{i e^{-2 i \vartheta}}{\im y} 
-2 i e^{-2 i \vartheta}\; \frac{\im V}{\im y} \left( \frac{\re y_\lambda}{\im y} - \frac{\re y_\lambda(u_0)}{2 \im y(u_0)}
+ i \int_{u_0}^u y_\lambda \right) .
\end{align*}
Using~\eqref{KPrel} and~\eqref{phiarg} gives the result.
\QED

\subsection{General Estimates in the WKB Region} \label{secWKB}
In the previous section, we derived general formulas for the $\lambda$-derivatives of the
center and radius of the osculating circles.
We want to use these formulas in order to control the behavior of the osculating
circles in the WKB region~$(\ul, \ur)$.
We now work out the corresponding estimates in general,
stating the assumptions needed for the estimates to work
(see eqs~\eqref{WKB2}, \eqref{WKB1}, \eqref{WKB3} and~\eqref{monotone} below).
We also derive explicit formulas for the error terms
(see eqs~\eqref{E1def}--\eqref{E3def} below).
The remaining task will be to justify the above assumptions and to control the error terms.
This will be done subsequently in the different cases in the following
Sections~\ref{secright}--\ref{secpoles}.

Consider an interval~$I=[u_a, u_b] \subset (0, \piot]$. We assume that on~$I$ the following inequalities hold for
suitable constants~$C, c>1$,
\begin{gather}
\int_{u_a}^{u_b} \frac{|\im V|}{\im y} \leq c \label{WKB2} \\
\frac{\sqrt{|\re V|}}{C} \leq \im y \leq C\: \sqrt{|\re V|} \:. \label{WKB1}
\end{gather}

\begin{Lemma} Assume that~$u_0 \in I$ and that~\eqref{WKB2} and~\eqref{WKB1} hold.
Then the function~$K(u)$ (cf.~\eqref{KPrel}) is bounded on the interval~$I$ by
\[ \frac{1}{\tilde{c}} \leq K(u) \leq {\tilde{c}} \:, \]
where~$\tilde{c} = e^c$.
\end{Lemma}
\Proof Comparing~\eqref{KPrel} with~\eqref{initial}, we know that~$K(u_0)=2$.
Moreover, from~\eqref{KPrel} and~\eqref{KPprel} we obtain
\[ \frac{d}{du} \log K = \frac{K'}{K} = \frac{\im V}{\im y} \:. \]
Integrating and using~\eqref{WKB2} gives the result.
\QED
We now estimate the terms appearing in Lemma~\ref{lemma42}.
We again consider an interval~$I=[u_a, u_b]$.
We assume that on~$I$ the inequalities~\eqref{WKB2} and~\eqref{WKB1}
hold and that the following inequalities hold for suitable constants~$C, c>1$,
\begin{align}
\int_{u_a}^{u} \frac{|\re y|}{\sqrt{|\re V|}} &\leq \frac{c}{\sqrt{|\re V(u)|}} \label{WKB3} \\
|\re V(u)| &\leq C^2 \inf_{[u_a, u]} |\re V| \:. \label{monotone}
\end{align}

\begin{Lemma} \label{lemma44}
Under the assumptions~\eqref{WKB2}--\eqref{monotone}, there is a constant~$c_1=c_1(c,C)$ such that on~$I$,
\[ | y_\lambda(u)| \leq \: c_1 \bigg( \frac{1}{\sqrt{|\re V(u)|}} + |\partial_\lambda y_0| \bigg) \:. \]
More precisely,
\beq \label{ylames}
\begin{split}
\left| y_\lambda(u) - \frac{i}{2 \im y(u)} \right| &\leq c_1\: \left( \frac{1}{\sqrt{|\re V_0|}}
+ |\partial_\lambda y_0| \right) \frac{\sqrt{|\re V|}}{\sqrt{|\re V_0|}} \\
& \quad + c_1 \:\sqrt{|\re V(u)|} \int_{u_a}^u \frac{1}{|\re V|} \:\bigg(
\frac{|\im V|}{\im y} + |\re y| \bigg) \:. 
\end{split}
\eeq
\end{Lemma}
\Proof
We integrate by parts to obtain
\begin{align*}
\int_{u_a}^u \phi^2 &= \int_{u_a}^u \frac{|\phi|^2}{2 i \,\im y(u)}\: \frac{d}{du} \bigg( \frac{\phi^2}{|\phi|^2} \bigg) \\
&= \frac{\phi^2}{2 i \,\im y(u)} \bigg|_{u_a}^u
- \int_{u_a}^u \frac{d}{du} \bigg( \frac{|\phi|^2}{2 i \,\im y(u)} \bigg) \frac{\phi^2}{|\phi|^2} \\
&= \frac{\phi^2}{2 i \,\im y} \bigg|_{u_a}^u + 2 i \int_{u_a}^u \frac{\phi^2}{\im y}
\left(\re y - \frac{\im V}{4 \im y} \right) .
\end{align*}
Using this relation in~\eqref{ylam} gives
\begin{align*}
y_\lambda(u) &= \frac{\phi^2(u_a)}{\phi^2(u)} \left( \partial_\lambda y_0 + \frac{1}{2 i \,\im y(u_a)} \right) \\
&\quad - \frac{1}{2 i \,\im y(u)} - \frac{2i}{\phi^2(u)} \int_{u_a}^u \frac{\phi^2}{\im y}
\left(\re y - \frac{\im V}{4 \im y} \right) \:.
\end{align*}
We now estimate the resulting terms:
\begin{align*}
\left| \frac{\phi^2(u_a)}{\phi^2(u)} \: \partial_\lambda y_0 \right| &\leq {\tilde{c}}\:
\frac{\im y(u)}{\im y(u_a)} \: |\partial_\lambda y_0| \leq C^2 {\tilde{c}}\:
\frac{|\partial_\lambda y_0|}{\sqrt{|\re V_0|}}\: \sqrt{|\re V(u)|}
\leq C^3 {\tilde{c}}\: |\partial_\lambda y_0| \\
\left| \frac{\phi^2(u_a)}{\phi^2(u)}\: \frac{1}{2 i \,\im y(u_a)} \right| &\leq
{\tilde{c}}\: \frac{\im y(u)}{\im^2 y(u_a)} \leq \frac{C^3 {\tilde{c}}}{|\re V_0|}\: \sqrt{|\re V(u)|}
\leq \frac{C^5 {\tilde{c}}}{\sqrt{|\re V(u)|}} \\
\left| \frac{1}{2 y(u) } \right| &\leq \frac{C}{2}\: \frac{1}{\sqrt{|\re V(u)|}} \:.
\end{align*}
Moreover,
\begin{align*}
\bigg| & \frac{2}{\phi^2(u)} \int_{u_a}^u \frac{\phi^2}{\im y}
\left(\re y - \frac{\im V}{4 \im y} \right) \bigg| \leq 2 {\tilde{c}}^2
\im y \int_{u_a}^u \frac{1}{\im^2 y} \left| \re y - \frac{\im V}{4 \im y} \right| \\
&\leq 2 C^3 {\tilde{c}}^2 \:\sqrt{|\re V(u)|} \int_{u_a}^u \frac{1}{|\re V|}
\left| \re y - \frac{\im V}{4 \im y} \right| \\
&\leq \frac{5}{2}\: c \,C^3 {\tilde{c}}^2 \:\frac{\sqrt{|\re V(u)|}}{\inf_I |\re V|}
\leq \frac{5}{2}\: \frac{c \,C^5 {\tilde{c}}^2}{\sqrt{|\re V(u)|}}\:.
\end{align*}
Combining all the terms gives the results.
\QED

\begin{Prp} \label{prp85} Under the assumptions~\eqref{WKB2}--\eqref{monotone},
the following inequalities hold,
\begin{align*}
\bigg| \frac{\partial}{\partial \im \lambda} \log K(u, \lambda) \big|_{u_a}^{u_b}
+ \int_{u_a}^{u_b} \frac{1}{\im y} \bigg|
&\leq \mathscr{E} \\
\left| \frac{\partial}{\partial \re \lambda} \log K(u, \lambda) \Big|_{u_a}^{u_b} \right|
&\leq \mathscr{E} \\
\left| \frac{\partial \p(u, \lambda)}{\partial \re \lambda}  \Big|_{u_a}^{u_b} \right|
+ \left| \frac{\partial \p(u, \lambda)}{\partial \im \lambda} \Big|_{u_a}^{u_b} \right|
&\leq \mathscr{E} \: ,
\end{align*}
where the error term~${\mathscr{E}}$ is given by
\begin{align}
\mathscr{E} =\;& c_2 \int_{u_a}^{u_b}
\left( \frac{|\im V|}{|\re V|}+ \frac{|\im V(u_b)|}{|\re V(u_b)|} \right)
\bigg( \frac{1}{\sqrt{|\re V(u)|}} + |\partial_\lambda y_0| \bigg) \label{E1def} \\
&+\frac{c_2}{|\re V(u_b)|} + c_2 \int_{u_a}^{u_b} \frac{|\re y|}{|\re V|} \label{E2def} \\
&+c_2 \int_{u_a}^{u_b} \bigg\{ \frac{\im^2 V}{|\re V|^\frac{3}{2}}
+ \frac{|\im V'| + |\re y|\, |\im V|}{|\re V|}
\bigg\} \notag \\
&\qquad\qquad\qquad\qquad\qquad
\times \bigg[ \int_{u_a}^u \bigg( \frac{1}{\sqrt{|\re V|}} + |\partial_\lambda y_0| \bigg) 
\bigg] \,du \label{E3def}
\end{align}
with a constant~$c_2=c_2(c,C)$.
\end{Prp}
\Proof We integrate the formulas of Lemma~\ref{lemma42} from~$u_a$ to~$u_b$.
Possibly after integrating by parts, we apply~\eqref{WKB1}--\eqref{monotone} and
Lemma~\ref{lemma44}. More precisely, we estimate the individual terms as follows:
\begin{align*}
\int_{u_a}^{u_b} & \left| \frac{\im V \:\im y_\lambda}{\im^2 y} \right|
\leq C^2 \,c_1 \int_{u_a}^{u_b} \frac{|\im V|}{|\re V|} \:\bigg( \frac{1}{\sqrt{|\re V(u)|}} + |\partial_\lambda y_0| \bigg) \\
\int_{u_a}^{u_b} & \left| \frac{\im V}{\phi^2 \im^2 y} \left( \frac{\im y_\lambda}{\im y}
- \frac{\im y_\lambda(u_a)}{2 \im y(u_a)} \right) \right| \\
\leq\; &  {\tilde{c}} \int_{u_a}^{u_b} \left| \frac{\im V}{\im y} \left( \frac{\im y_\lambda}{\im y}
- \frac{\im y_\lambda(u_a)}{2 \im y(u_a)} \right) \right|
\leq C^2\,{\tilde{c}}\,c_1 \int_{u_a}^{u_b} \frac{|\im V|}{|\re V|} \bigg( \frac{1}{\sqrt{|\re V|}}
+ |\partial_\lambda y_0| \bigg) \\
\int_{u_a}^{u_b} & \left| \frac{\im V}{\phi^2 \im^2 y} \left( \frac{\re y_\lambda}{\im y}
- \frac{\re y_\lambda(u_a)}{2 \im y(u_a)} \right) \right| 
\leq C^2\,{\tilde{c}}\,c_1 \int_{u_a}^{u_b} \frac{|\im V|}{|\re V|} \bigg( \frac{1}{\sqrt{|\re V|}}
+ |\partial_\lambda y_0| \bigg) \\
\int_{u_a}^{u_b} &\frac{1}{\phi^2 \im^2 y}  = 
\int_{u_a}^{u_b} \frac{1}{|\phi|^2 \,\im^2 y}\: \frac{|\phi|^2}{\phi^2}
= \int_{u_a}^{u_b} \frac{i}{2 \,|\phi|^2 \,\im^3 y}\:
 \frac{d}{du} \bigg( \frac{|\phi|^2}{\phi^2} \bigg) \\
&=\frac{i}{2 \,\phi^2 \,\im^3 y} \bigg|_{u_a}^{u_b}
-\frac{i}{2} \int_{u_a}^{u_b} \frac{d}{du} \bigg( \frac{1}{|\phi|^2 \,\im^3 y} \bigg) \:\frac{|\phi|^2}{\phi^2} \\
&= \frac{i}{2 \,\phi^2 \,\im^3 y} \bigg|_{u_a}^{u_b}
+\frac{3i}{2} \int_{u_a}^{u_b} \frac{1}{\phi^2 \im^3 y} \left( \frac{\im V}{\im y} - \frac{4}{3}\: \re y \right) \\
\Longrightarrow \bigg| &\int_{u_a}^{u_b} \frac{1}{\phi^2 \im^2 y} \bigg| \leq
\frac{C^2 {\tilde{c}}\,(1+C^2)}{2 \,|\re V(u_b)|} 
+ 2 C^2 {\tilde{c}} \int_{u_a}^{u_b} \frac{1}{|\re V|} \left( \frac{|\im V|}{\im y} + |\re y| \right) \\
\int_{u_a}^{u_b} & \frac{\im V}{\phi^2 \im^2 y} \int_{u_a}^u y_\lambda
= \int_{u_a}^{u_b} \frac{|\phi|^2}{\phi^2} \:\frac{\im V}{|\phi|^2 \im^2 y} \int_{u_a}^u y_\lambda \\
=\;& \int_{u_a}^{u_b} \frac{d}{du} \bigg( \frac{|\phi|^2}{\phi^2} \bigg)
\frac{i \im V}{2 |\phi|^2 \im^3 y} \int_{u_a}^u y_\lambda \\
=\;& \frac{i \im V(u_b)}{2 \phi^2(u_b) \im^3 y(u_b)} \int_{u_a}^{u_b} y_\lambda
- \int_{u_a}^{u_b} \frac{|\phi|^2}{\phi^2} \frac{d}{du}
\bigg(  \frac{i \im V}{2 |\phi|^2 \im^3 y} \int_{u_a}^u y_\lambda  \bigg) \\
=\;&\;-\; \frac{i}{2} \int_{u_a}^{u_b} \frac{\im V \,\im y_\lambda}{\phi^2 \im^3 y} \\
&+ \frac{i}{2} \int_{u_a}^{u_b} \frac{3 \im^2 V - \im y \,\im V' \,
- 4 \im V \,\im y \,\re y}{\phi^2 \im^4 y} \int_{u_a}^u y_\lambda \\
\Longrightarrow \bigg| &\int_{u_a}^{u_b} \frac{\im V}{\phi^2 \im^2 y} \int_{u_a}^u y_\lambda \bigg|
\leq \frac{c_1 C^2 {\tilde{c}}}{2} \:\frac{|\im V(u_b)|}{|\re V(u_b)|} \int_{u_a}^{u_b}
\bigg( \frac{1}{\sqrt{|\re V(u)|}} + |\partial_\lambda y_0| \bigg) \\
&+\frac{c_1 C^2 {\tilde{c}}}{2} \int_{u_a}^{u_b} \frac{|\im V|}{|\re V|}
\:\bigg( \frac{1}{\sqrt{|\re V(u)|}} + |\partial_\lambda y_0| \bigg) \\
&+ \frac{c_1 C^2 {\tilde{c}}}{2} \int_{u_a}^{u_b} du \left( \frac{3 C \im^2 V}{|\re V|^\frac{3}{2}}
+ \frac{|\im V'|}{|\re V|} + \frac{4 |\im V|\, |\re y|}{|\re V|} \right) \\
&\qquad\qquad\qquad \times \int_{u_a}^u 
\:\bigg( \frac{1}{\sqrt{|\re V(u)|}} + |\partial_\lambda y_0| \bigg) .
\end{align*}
Combining all the terms gives the result.
\QED

\subsection{Estimates on the Interval~$[u_0, \ur]$} \label{secright}
We shall now apply the estimates of Section~\ref{secWKB} on the interval~$[u_0, \ur]$
(see~\eqref{urdef}, \eqref{u00def} and~\eqref{u0ks}).
Our task is to show that the inequalities~\eqref{WKB2}, \eqref{WKB1}, \eqref{WKB3}
and~\eqref{monotone} hold. Moreover, we need to estimate the error terms~${\mathcal{E}}$
in~\eqref{E1def}--\eqref{E3def}. We begin by collecting a few properties of our potential.

\begin{Lemma} \label{lemmachange}
At the points~$u_0$ and~$\underline{u}$
(see~\eqref{u00def}, \eqref{u0Ldef} and~\eqref{ubardef}), for sufficiently large~$\Const_5$
the potential satisfies the inequalities
\begin{gather}
\big| \re V(u_0) - \re \lambda \big| \lesssim |\Omega| \label{Vu0es} \\
|V'(u_0)| \lesssim |\Omega|^\frac{3}{2} \:,\qquad
|V''(u_0)| \lesssim |\Omega|^2 \label{Vu0pes} \\
\re V(u_0) \leq \frac{9}{8}\, \re V(\underline{u}) < 0 \:. \label{Vchange}
\end{gather}
\end{Lemma}
\Proof The inequalities~\eqref{Vu0es} and~\eqref{Vu0pes}
follow immediately from~\eqref{u00def}, \eqref{V0ex}
and~\eqref{u0Ldef}, \eqref{Vn0ex}. Combining~\eqref{Vu0es} with~\eqref{relamleft},
we find that for sufficiently large~$\Const_5$,
\[ \re V(u_0) \leq -\frac{15}{16} \: \re \lambda \:. \]
Moreover, similar as in the proof of Lemma~\ref{lemma81}, again for large~$\Const_5$
one obtains
\begin{align*}
\re V(\underline{u}) &\geq \frac{16}{17}\: |\Omega|^2\, \underline{u}^2
- \frac{\const}{\underline{u}^2} - \re \lambda
= \frac{16}{17}\: \frac{\re \lambda}{4}
- \frac{4 \const\, |\Omega|^2}{\re \lambda} - \re \lambda \\
&= -\frac{13}{17}\: \frac{\re \lambda}{4}
- \frac{4 \const\, |\Omega|^2}{\re \lambda} \geq -\frac{14}{17}\: \re \lambda\:.
\end{align*}
Combining these inequalities gives~\eqref{Vchange}.
\QED

\begin{Lemma} \label{lemmaEestim} Assume that the condition~\eqref{WKB2} holds.
Then the conditions~\eqref{WKB1}, \eqref{WKB3} and~\eqref{monotone} are satisfied on the interval~$[u_0, \ur]$. Moreover,
choosing~$c_1$ and~$c_2$ sufficiently large, we can arrange that
the error terms~${\mathscr{E}}$ \eqref{E1def}--\eqref{E3def} are bounded by
\beq \label{Ein}
\mathscr{E} \leq \frac{1}{20} \int_{u_0}^{\ur} \frac{1}{\sqrt{|\re V|}}\:.
\eeq
\end{Lemma}
\Proof
The inequality~\eqref{monotone} follows immediately from Lemma~\ref{lemmamonotone}.
Since the WKB conditions are satisfied by Proposition~\ref{prpWKB}, we know that
\beq \label{yapprox}
y \approx \sqrt{V} - \frac{V'}{4V} \:.
\eeq
The results of the analysis in~\cite{tinvariant, special} gives us rigorous bounds
for the error of this approximation. This implies~\eqref{WKB1} for sufficiently large~$C$.

In preparation for proving the other inequalities, we need a few estimates for the potential.
First, from the explicit form of the potential~\eqref{Vdef}, it is obvious that
\beq \label{c0in}
|V''|, |V'''| \lesssim |\Omega|^2 \qquad \text{and} \qquad |\im V'| \lesssim\, |\Omega|\:.
\eeq
Estimating the potential without derivatives is a bit more subtle because the
constant~$\lambda$ comes into play.
We first note that by construction of~$\ur$ (see~\eqref{urdef}, \eqref{nudef}, \eqref{cases}
and~\eqref{relamleft}), we know that
\beq \label{Vlower}
\re V \leq -\Const_1\: |\Omega| \:.
\eeq
Moreover, using~\eqref{relamright} in~\eqref{inter}, the imaginary part of the potential is uniformly bounded,
\beq \label{imVbound}
|\im V| \lesssim |\Omega| \:.
\eeq
Therefore, by increasing~$\Const_1$ we can arrange that the real part of the potential dominates
its imaginary part in the sense that
\beq \label{redom}
\frac{|\re V|}{|\im V|} \gtrsim \Const_1\:.
\eeq

We now come to the proof of~\eqref{WKB3}. We first need to estimate the real part of~$y$.
Using~\eqref{redom}, we may express~\eqref{yapprox} in terms of the real and imaginary parts of~$V$
to obtain
\[ y \approx i \sqrt{- \re V- i \im V} - \frac{V'}{4V} \\
\approx i \sqrt{-\re V} +\frac{\im V}{\sqrt{-\re V}} - \frac{V'}{4V} \:. \]
Hence
\beq \label{reyid}
|\re y| \lesssim \frac{|\im V|}{\sqrt{-\re V}} + \frac{|V'|}{|V|}
\eeq
(where the errors are again under control in view of~\eqref{redom} and the fact that
the WKB conditions are satisfied).
The last estimate implies that
\beq \label{intform}
\int_{u_0}^u \frac{|\re y|}{\sqrt{|\re V|}} \lesssim
\int_{u_0}^u \frac{|\im V|}{|\re V|} + \int_{u_0}^u \frac{|V'|}{|V|^\frac{3}{2}} \:.
\eeq
In the first integral on the right, we apply~\eqref{monotone} to obtain
\[ \int_{u_0}^u \frac{|\im V|}{|\re V|} \leq \frac{C}{\sqrt{|\re V(u)|}} \int_{u_0}^u \frac{|\im V|}{\sqrt{|\re V|}} 
\lesssim \frac{1}{\sqrt{|\re V(u)|}} \:, \]
where in the last step we again used the WKB approximation~\eqref{yapprox}
together with the inequality~\eqref{WKB2}.
The second integral in~\eqref{intform} can be estimated by
\begin{align*}
\int_{u_0}^u \frac{|V'|}{|V|^\frac{3}{2}} &\lesssim
\int_{u_0}^u \frac{|V'|}{|\re V|^\frac{3}{2}} \leq
\int_{u_0}^u \frac{\re V'}{|\re V|^\frac{3}{2}} + \int_{u_0}^u \frac{|\im V'|}{|\re V|^\frac{3}{2}} \\
&\!\!\!\!\overset{\eqref{c0in}}{\lesssim} \int_{u_0}^u \Big( \frac{1}{\sqrt{|\re V|}} \Big)'
+ \frac{|\Omega|}{|\re V(u)|^\frac{3}{2}} \lesssim
\frac{1}{\sqrt{|\re V(u)|}} + \frac{|\Omega|}{|\Omega|\: \sqrt{|\re V(u)|}} \:,
\end{align*}
where in the last line we also used again the monotonicity statement
of Lemma~\ref{lemmamonotone} together with~\eqref{Vlower}. 
Combining the obtained inequalities gives~\eqref{WKB3}.

In order to prove~\eqref{Ein}, we first estimate the term~$\partial_\lambda y_0$.
Differentiating~\eqref{y0WKB} and using the monotonicity of~$\re V$ as well
as the results of Lemma~\ref{lemmachange}, we obtain
\[ |\partial_\lambda y_0| \lesssim \frac{1}{\sqrt{|\re V(u)|}} \:. \]
For this reason, we may disregard the term~$\partial_\lambda y_0$ in~\eqref{E1def}
and~\eqref{E3def}. In order to estimate the integral~\eqref{E1def}, we simply bound the
first bracket in the integrand by
\beq
\frac{|\im V|}{|\re V|} \overset{\eqref{redom}}{\lesssim} \frac{1}{\Const_1} \:.  \label{imVes2}
\eeq
The first summand in~\eqref{E2def} can be the estimated with the help
the monotonicity of~$\re V$ by
\begin{align*}
\frac{1}{|\re V(\ur)|} &\leq \frac{9}{|\re V(\ur)|} - \frac{8}{|\re V(\underline{u})|}
\overset{\eqref{Vchange}}{\leq} \frac{9}{|\re V(\ur)|} - \frac{9}{|\re V(u_0)|} \\
&= 9 \int_{u_0}^{\ur} \Big( \frac{1}{\re V} \Big)' 
= 9 \int_{u_0}^{\ur} \frac{\re V'}{\re V^2} 
\leq \int_{u_0}^{\ur} \frac{10}{\sqrt{|\re V|}}\: \frac{|V'|}{|V|^\frac{3}{2}} \:, 
\end{align*}
where in the last step we used again that the real part of~$V$ dominates the imaginary part
and that~$\re V$ is monotone. Now the factor~$|V'|/|V|^\frac{3}{2}$ is small
in view of the WKB property established in Proposition~\ref{prpWKB}.

To estimate the integrand in~\eqref{E2def}, we fist use~\eqref{reyid},
\[ \frac{|\re y|}{|\re V|} \lesssim \frac{|\im V|}{|\re V|^\frac{3}{2}} + \frac{|V'|}{|\re V|\: |V|}
\lesssim \frac{2}{\sqrt{|\re V|}} \left[ \frac{|\im V|}{|\re V|} + \frac{|V'|}{|V|^\frac{3}{2}} \right] . \]
Now the square bracket can be made arbitrarily small by applying again~\eqref{imVes2}
and Proposition~\ref{prpWKB}.

It remains to estimate the nested integral~\eqref{E3def}. We first exchange the orders
of integration,
\begin{align}
\int_{u_0}^{\ur} & du \;\bigg\{ \frac{\im^2 V}{|\re V|^\frac{3}{2}}
+ \frac{|\im V'| + |\re y|\, |\im V|}{|\re V|}
\bigg\} \int_{u_0}^u \frac{d\tau}{\sqrt{|\re V(\tau)|}} \notag \\
&=\int_{u_0}^{\ur} \frac{d\tau}{\sqrt{|\re V(\tau)|}}  \int_{\tau}^{\ur}
\bigg\{ \frac{\im^2 V}{|\re V|^\frac{3}{2}} + \frac{|\im V'| + |\re y|\, |\im V|}{|\re V|} \bigg\} \: du \:.
\label{Reyint}
\end{align}
Now we estimate the inner integral term by term.
The first term can be estimated with the help of~\eqref{imVbound}
and Lemma~\ref{lemmamonotone},
\[ \int_\tau^{\ur} \frac{\im^2 V}{|\re V|^\frac{3}{2}} \lesssim
\int_\tau^{\ur} \frac{|\Omega|}{|\re V|^\frac{3}{2}} \lesssim \frac{1}{\Const_1}\:. \]
Next, using~\eqref{c0in} and~\eqref{Vlower}, we find that
\[ \int_\tau^{\ur} \frac{|\im V'|}{|\re V|} \lesssim \frac{1}{\Const_1}\: (\ur-\tau) \lesssim \frac{1}{\Const_1}\:. \]
In order to estimate the remaining term involving~$\re y$ in~\eqref{Reyint},
we first apply~\eqref{reyid}. Then the first summand on the right of~\eqref{reyid}
gives rise to a contribution which is precisely of the form of the first summand 
in the curly brackets in~\eqref{E3def}. Hence it remains to consider the term
\[ \int_\tau^{\ur} \frac{|\im V|}{|\re V|}\:\frac{|V'|}{|V|}
\leq \int_\tau^{\ur} |\im V|\; \frac{|\im V'| + \re V'}{|V|^2} \:, \]
where in the last step we applied Lemma~\ref{lemmamonotone}.
The first summand on the right can be estimated with the help of~\eqref{imVbound}
and~\eqref{c0in} by
\[ \int_\tau^{\ur} |\im V|\; \frac{|\im V'|}{|V|^2} \lesssim
\int_\tau^{\ur} \frac{|\Omega|^2}{|V|^2}
\overset{\eqref{Vlower}}{\lesssim} \frac{|\Omega|^\frac{3}{2}}{\sqrt{\Const_1}}
\int_\tau^{\ur} \frac{1}{|V|^\frac{3}{2}} \lesssim
\frac{1}{\sqrt{|\Omega|\, \Const_1^3}} \:, \]
where in the last step we again applied Lemma~\ref{lemmamonotone}.
The remaining second summand on the right is estimated as follows,
\[ \int_\tau^{\ur} |\im V|\; \frac{\re V'}{|V|^2}
\lesssim |\Omega| \int_\tau^{\ur} \frac{\re V'}{(\re V)^2}
\leq \frac{|\Omega|}{|\re V(\ur)|} \lesssim \frac{1}{\Const_1} \:, \]
where we again applied Lemma~\ref{lemmamonotone}. This concludes the proof.
\QED

\subsection{Estimates on the Interval~$[\ul, u_0]$} \label{secpoles}
In preparation, we note that,  using~\eqref{relamright} in~\eqref{inter}, we again obtain the following
uniform bound for~$\im V$,
\beq \label{imVbound2}
|\im V| \lesssim |\Omega| \:.
\eeq
We treat the cases~$k=s$ and~$k \neq s$ separately.

\begin{Lemma} {\bf{(Case $k=s$)}} \label{lemmaEestim3}
Assume that~$k=s$.
On the interval~$[\ul, u_0]$, the conditions~\eqref{WKB2}--\eqref{monotone} are satisfied.
Moreover, by choosing~$\Const_1$ sufficiently large, we can arrange that
the error terms~${\mathscr{E}}$ \eqref{E1def}--\eqref{E3def} are bounded by
\[ \mathscr{E} \leq \frac{1}{20} \int_{\ul}^{u_0} \frac{1}{\sqrt{|V}}\:. \]
\end{Lemma}
\Proof Near the pole, the potential has the following asymptotic expansion (cf.~\eqref{u00def} and~\eqref{V0ex}),
\begin{align}
\re V &\leq -\frac{3}{4}\:|\Omega| - \re \mu + \O(1)
\leq -\frac{|\Omega|}{2} - \re \mu \leq -\re \mu \lesssim \re \lambda \label{Vrel} \\
V'(u) &= \frac{1}{2 u^3} + 2 \Omega^2 u + \O \big( |\Omega| u \big)
+ \O \big( |\Omega|^2 u^3 \big) \:. \label{Vprel}
\end{align}
Since~\eqref{Vprel} has a positive real part, we know that~$\re V$ is monotone
increasing. This implies~\eqref{monotone} if we also keep in mind
that the imaginary part of~$V$ is dominated by the real part in view of~\eqref{Vrel} and~\eqref{c0in}.
Moreover, using the results of the analysis in~\cite{tinvariant, special}, we know that~\eqref{yapprox}
holds with rigorous error bounds. This implies~\eqref{WKB1} for sufficiently large~$C$.
The inequality~\eqref{WKB2} follows from the estimate
\begin{align}
\int_{\ul}^{u_0} \frac{|\im V|}{\im y} &\lesssim \int_{\ul}^{u_0} \frac{|\im V|}{\sqrt{|\re V|}} \notag \\
&\!\!\!\!\!\overset{\eqref{Vrel}}{\lesssim} \frac{1}{\sqrt{\re \mu}} \int_{\ul}^{u_0} |\im V| 
\overset{\eqref{imVbound2}}{\lesssim} \frac{u_0\, |\Omega|}{|\re \lambda|}
\lesssim \frac{1}{\sqrt{\Const_5}} \:, \label{esin}
\end{align}
where in the last step we used~\eqref{relamleft} and~\eqref{u00def}.
In order to prove~\eqref{WKB3}, we again apply~\eqref{intform} and estimate
the two resulting integrals by
\begin{align*}
\int_{\ul}^u \frac{|\im V|}{|\re V|} &\leq \frac{1}{\sqrt{|\re V(u)|}} \int_{\ul}^u \frac{|\im V|}{\sqrt{|\re V|}}
\overset{\eqref{esin}}{\leq} \frac{1}{\sqrt{\Const_5\, |\re V(u)|}} \\
\int_{\ul}^u \frac{|V'|}{|V|^\frac{3}{2}} &\leq
\int_{\ul}^u \frac{\re V'}{(-\re V)^\frac{3}{2}} + \int_{\ul}^u \frac{|\im V'|}{|\re V|^\frac{3}{2}} \\
&\leq 2 \int_{\ul}^u \frac{d}{du} \left( \frac{1}{\sqrt{-\re V}} \right)
+ \frac{1}{\sqrt{|\re V(u)|}} \int_{\ul}^u \frac{|\im V'|}{|\re V|} \\
&\lesssim \frac{1}{\sqrt{-\re V(u)}} + \frac{1}{\sqrt{|\re V(u)|}} \: \frac{u_0\, |\Omega|}{\re \lambda}
\lesssim \frac{1}{\sqrt{|V(u)|}} + \frac{1}{\sqrt{|V(u)|}} \: \frac{1}{\Const_5 \sqrt{|\Omega|}}\:.
\end{align*}

In order to estimate~${\mathscr{E}}$, we proceed exactly as in the proof of Lemma~\ref{lemmaEestim}, using
the following estimates:
\begin{align*}
\int_{\ul}^{u_0} \frac{|\im V|^2}{|\re V|^\frac{3}{2}} &\overset{\eqref{Vrel}}{\lesssim}
\frac{u_0\, |\Omega|^2}{(\re \lambda)^\frac{3}{2}}  \lesssim \frac{1}{(\Const_5)^\frac{3}{2}} \\
\int_{\ul}^{u_0} \frac{|\im V'|}{|V|} &\overset{\eqref{Vrel}}{\lesssim} \frac{u_0\, |\Omega|}{\re \lambda}
\lesssim \frac{1}{\Const_5 \sqrt{|\Omega|}} \\
|V'| \;&\overset{\eqref{Vprel}}{\lesssim} \frac{1}{\ul^3} + |\Omega|^2 \,u_0
\overset{\eqref{uldef}}{\lesssim} \frac{(\re \lambda)^\frac{3}{2}}{\Const_1^3} + |\Omega|^\frac{3}{2}
\overset{\eqref{relamleft}}{\lesssim} \frac{(\re \lambda)^\frac{3}{2}}{\Const_1^3} \\
\int_{\ul}^{u_0} \frac{|\im V|\, |V'|}{|V|^2} &\overset{\eqref{Vrel}}{\lesssim}
\frac{u_0\,|\Omega|}{(\re \lambda)^2}\: \frac{(\re \lambda)^\frac{3}{2}}{\Const_1^3}
\overset{\eqref{relamleft}}{\lesssim} \frac{1}{\Const_1^3 \sqrt{\Const_5}}\:.
\end{align*}
This concludes the proof.
\QED

In the case~$k \neq s$, the function~$\re V$ is monotone {\em{decreasing}} on the
interval~$[\ul, u_0]$. Therefore, when applying the estimates in Section~\ref{secWKB}
we need to proceed backwards in~$u$, starting from~$u_0$.
Therefore, the conditions~\eqref{WKB3} and~\eqref{monotone} need to be replaced by
\begin{gather}
\int_{u}^{u_0} \frac{|\re y|}{\sqrt{|V|}} \leq \frac{c}{\sqrt{|V(u)|}} \label{WKB3new} \\
|\re V(u)| \leq C^2 \inf_{[u, u_0]} |\re V| \:. \label{monotonenew}
\end{gather}

\begin{Lemma} {\bf{(Case $k \neq s$)}} \label{lemmaEestim4}
Assume that~$k \neq s$. Then on the interval~$[\ul, u_0]$, the conditions~\eqref{WKB2}, \eqref{WKB1}
and~\eqref{WKB3new}, \eqref{monotonenew} are satisfied.
Moreover, by choosing~$\Const_1$ sufficiently large, we can arrange that
the error terms~${\mathscr{E}}$ \eqref{E1def}--\eqref{E3def} are bounded by
\[ \mathscr{E} \leq \frac{1}{20} \int_{\ul}^{u_0} \frac{1}{\sqrt{|V}}\:. \]
\end{Lemma}
\Proof Near the pole, the potential has the following asymptotic expansion (cf.~\eqref{u0es} and~\eqref{Vn0ex}),
\begin{align}
\re V &\leq - \frac{\re \mu}{2} + 8\, \sqrt{\Lambda}\, |\Omega|
\leq -\frac{\re \mu}{4} \lesssim -\re \lambda \label{Vlowernew} \\
V'(u) &= -\frac{2 \Lambda}{u^3} + 2 \Omega^2 u + \O \big( |\Omega| u \big)
+ \O \big( |\Omega|^2 u^3 \big) \label{Vprelnew} \:.
\end{align}

Since~$\re V$ is convex on~$(0, u_0]$ and has its minimum at~$u_0$,
we know that the function~$\re V$ is monotone decreasing on the interval~$(0, u_0]$.
This implies~\eqref{monotonenew}.
Moreover, using the results of the analysis in~\cite{tinvariant, special}, 
we know that~\eqref{yapprox} holds with rigorous error bounds.
This implies~\eqref{WKB1} for sufficiently large~$C$.
The inequality~\eqref{WKB2} follows from the estimate
\beq
\int_{\ul}^{u_0} \frac{|\im V|}{\im y} \lesssim \int_{\ul}^{u_0} \frac{|\im V|}{\sqrt{|\re V|}}
\leq \frac{u_0\, |\Omega|}{\sqrt{\re \lambda}}
\lesssim \frac{1}{\sqrt{\Const_5}} \:, \label{esinnew}
\eeq
where we used~\eqref{Vlowernew}, \eqref{imVbound2} and~\eqref{u0es}.
In order to prove~\eqref{WKB3new}, we again apply~\eqref{intform} and estimate
the two resulting integrals by
\begin{align*}
\int_{u}^{u_0} \frac{|\im V|}{|\re V|} &\leq \frac{1}{\sqrt{|\re V(u)|}} \int_{u}^{u_0} \frac{|\im V|}{\sqrt{|\re V|}}
\overset{\eqref{esinnew}}{\lesssim} \frac{1}{\sqrt{\Const_5\, |V(u)|}} \\
\int_{u}^{u_0} \frac{|V'|}{|V|^\frac{3}{2}} &\leq
\int_{u}^{u_0} \frac{(-\re V')}{(-\re V)^\frac{3}{2}} + \int_{u}^{u_0} \frac{|\im V'|}{|\re V|^\frac{3}{2}} \\
&\leq 2 \int_{u}^{u_0} \frac{d}{du} \left( \frac{1}{\sqrt{-\re V}} \right)
+ \frac{1}{\sqrt{|\re V(u)|}} \int_{u}^{u_0} \frac{|\im V'|}{|\re V|} \\
&\lesssim \frac{1}{\sqrt{-\re V(u)}} + \frac{1}{\sqrt{|\re V(u)|}} \: \frac{u_0 \,|\Omega|}{|\re V(u)|}
\lesssim \frac{1}{\sqrt{|V(u)|}} \:,
\end{align*}
where in the last step we applied~\eqref{u0es}, \eqref{Vlowernew} and~\eqref{relamleft}.

In order to estimate~${\mathscr{E}}$, we proceed exactly as in the proof of Lemma~\ref{lemmaEestim}, using
the following estimates:
\begin{align*}
\int_{\ul}^{u_0} \frac{|\im V|^2}{|V|^\frac{3}{2}} &\overset{\eqref{Vlowernew}}{\lesssim}
\frac{u_0\, |\Omega|^2}{(\re \lambda)^\frac{3}{2}}  \lesssim \frac{1}{(\Const_5)^\frac{3}{2}} \\ \\
\int_{\ul}^{u_0} \frac{|\im V'|}{|V|} &\overset{\eqref{Vlowernew}}{\lesssim}
\frac{u_0\, |\Omega|}{\re \lambda}
\lesssim \frac{1}{\Const_5 \sqrt{|\Omega|}} \\
|V'| \;&\overset{\eqref{Vprelnew}}{\lesssim} \frac{1}{\ul^3} + |\Omega|^2 \,u_0
\overset{\eqref{uldef}}{\lesssim} \frac{(\re \lambda)^\frac{3}{2}}{\Const_1^3} + |\Omega|^\frac{3}{2}
\overset{\eqref{relamleft}}{\lesssim} \frac{(\re \lambda)^\frac{3}{2}}{\Const_1^3} \\
\int_{\ul}^{u_0} \frac{|\im V|\, |V'|}{|V|^2} &\overset{\eqref{Vlowernew}}{\lesssim}
\frac{u_0\,|\Omega|}{(\re \lambda)^2}\: \frac{(\re \lambda)^\frac{3}{2}}{\Const_1^3}
\overset{\eqref{relamleft}}{\lesssim} \frac{1}{\Const_1^3 \sqrt{\Const_5}}\:.
\end{align*}
This concludes the proof.
\QED

\section{Annular Regions where~$|\zeta_L|$ is Bounded Below} \label{secannular}
We now construct regions in the complex $\lambda$-plane in which~$|\zeta_L|$
is bounded below. For given~$\lambda_0$ and~$\Delta \lambda>0$ we introduce the {\em{annular region}}
\beq \label{ALdef}
\begin{split}
A_L(\lambda_0) := \Big\{ \lambda \in \C \:\Big|\: & \big| \re (\lambda-\lambda_0) \big| \in (\Delta \lambda, 5 \Delta \lambda) \\
\text{and } & \big| \im (\lambda-\lambda_0) \big| \in (\Delta \lambda, 5 \Delta \lambda) \Big\}
\end{split}
\eeq
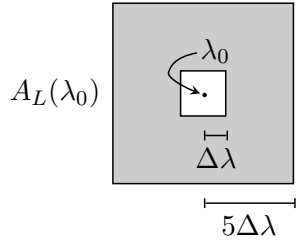
\begin{figure}%
\scalebox{1} 
{
\begin{pspicture}(0,-1.595)(5.4,1.555)
\definecolor{color542b}{rgb}{0.8,0.8,0.8}
\psframe[linewidth=0.02,dimen=outer,fillstyle=solid](3.62,0.765)(3.3,0.445)
\psframe[linewidth=0.02,dimen=outer,fillstyle=solid,fillcolor=color542b](4.57,1.555)(2.15,-0.865)
\psframe[linewidth=0.02,dimen=outer,fillstyle=solid](3.67,0.655)(3.05,0.035)
\psline[linewidth=0.02cm,tbarsize=0.07055555cm 3.0]{|*-|*}(3.38,-0.215)(3.68,-0.215)
\usefont{T1}{ptm}{m}{n}
\rput(3.53,-0.51){$\Delta \lambda$}
\psline[linewidth=0.02cm,tbarsize=0.07055555cm 3.0]{|*-|*}(3.38,-1.115)(4.58,-1.115)
\usefont{T1}{ptm}{m}{n}
\rput(3.97,-1.41){$5 \Delta \lambda$}
\psdots[dotsize=0.06](3.38,0.325)
\usefont{T1}{ptm}{m}{n}
\rput(3.52,0.91){$\lambda_0$}
\psbezier[linewidth=0.02,arrowsize=0.073cm 3.0,arrowlength=1.0,arrowinset=0.4]{->}(3.28,0.885)(2.96,0.825)(2.92,0.685)(2.9,0.625)(2.88,0.565)(3.06,0.445)(3.32,0.345)
\usefont{T1}{ptm}{m}{n}
\rput(1.39,0.35){$A_L(\lambda_0)$}
\end{pspicture} 
}
\caption{Annular regions.}
\label{figannulus}
\end{figure}%
(see Figure~\ref{figannulus}).
We again choose~$\ul$ and~$\ur$ as the boundaries of the WKB regions
(see~\eqref{uldef}, \eqref{urdef}).
With these choices, we have the following result.

\begin{Prp} \label{prp94} For sufficiently large~$|\Omega|$, the following statement holds:
Choose any~$\varepsilon<1/(1000)$. Suppose that for a given~$\lambda_0$,
\beq \label{zLumax}
|\zeta_L(\umax)| \leq \varepsilon\:.
\eeq
Then, choosing
\beq \label{Dellam}
\Delta \lambda = 75\,\varepsilon \bigg( \int_{\ul}^{\ur} \frac{1}{\sqrt{|V|}} \bigg|_{\lambda_0} \bigg)^{-1} \:,
\eeq
it follows that
\[ \varepsilon <  |\zeta_L(\umax)| < 1000 \,\varepsilon \qquad \text{for all~$\lambda \in A_L(\lambda_0)$}\:. \]
\end{Prp}
The remainder of this section is devoted to the proof of this proposition.

\subsection{Estimates in the WKB Region}
We again consider the family of solutions of the Riccati equation~\eqref{riccati}
with initial conditions~\eqref{yinitLR} and~\eqref{y0WKB}
(where~$u_0$ is given by~\eqref{u00def} or~\eqref{u0ks}).
We again let~$\phi$ be the corresponding solution of the Sturm-Liouville equation normalized
according to~\eqref{initial}.

\begin{Prp} \label{prposclam}
Assume that
\beq \label{sqVes}
\int_{\ul}^{\ur} \sqrt{|V|} \geq \Const_3 \:.
\eeq
Moreover, assume that on~$[\ul, \ur]$ the WKB conditions in Proposition~\ref{prpWKB}
as well as the inequalities~\eqref{WKB2}--\eqref{monotone} are satisfied.
Then by choosing the constants~$\Const_1, \ldots \Const_3$
sufficiently large, we can arrange that the following statement holds. Choosing~$\Delta \lambda$ according to~\eqref{Dellam},
for every~$\lambda \in A_L(\lambda_0)$,
\begin{gather*}
35\, \varepsilon \leq \Big| \vartheta(\lambda) \big|_{\ul}^{\ur} - \vartheta(\lambda_0) \big|_{\ul}^{\ur} \Big|
\leq 190\, \varepsilon \\
70\, \varepsilon \leq \Big| \log K(\lambda) \big|_{\ul}^{\ur} - \log K(\lambda_0) \big|_{\ul}^{\ur} \Big|
\leq 380\, \varepsilon \:,
\end{gather*}
with an arbitrarily small error.
\end{Prp}
\Proof We analyze what the condition~\eqref{sqVes} means. Combining the estimate
\[ \Const_3 \leq \int_{\ul}^{\ur} \sqrt{|V|} = \int_{\ul}^{\ur} \frac{|V|}{\sqrt{|V|}} \leq \sup_{[\ul, \ur]} |V|
\int_{\ul}^{\ur} \frac{1}{\sqrt{|V|}} \]
with~\eqref{Dellam}, we conclude that
\beq \label{Dellambda}
\Delta \lambda \leq \frac{64\, \varepsilon}{\Const_3} \sup_{(\ul, \ur)} |V| \:.
\eeq
This shows that varying~$\lambda$ on the scale~$\Delta \lambda$
keeps the form of the potential~$V$ unchanged, up to an arbitrarily small error.

By choosing the constants~$c_1$ and~$c_2$ sufficiently large, we can arrange that
the assumptions of Lemma~\ref{lemma44} and Proposition~\ref{prp85} hold. Possibly by further
increasing~$c_1$ and~$c_2$,
we can arrange that on the interval~$[u_0, \ur]$, the WKB approximation
\[ y \approx \sqrt{V} -\frac{V'}{4 V} \]
holds with an arbitrarily small error. Moreover, we can make the error term~${\mathscr{E}}$
in Proposition~\ref{prp85} as well as the right side in~\eqref{ylames} as small as we like.
Hence on the interval~$(u_0, \ur)$ and at~$\lambda=\lambda_0$,
\begin{align*}
\frac{\partial}{\partial \re \lambda} \vartheta \big|_{u_0}^{\ur} &=
\int_{u_0}^{\ur} \im y_\lambda \approx \int_{u_0}^{\ur} \frac{1}{2 \im y}
\approx \int_{u_0}^{\ur} \frac{1}{2 \im \sqrt{V}} \\
\frac{\partial}{\partial \im \lambda} \vartheta \big|_{u_0}^{\ur} &\approx 0 \\
\frac{\partial}{\partial \re \lambda} \log K(u, \lambda) \big|_{u_0}^{\ur} &\approx 0 \\
\frac{\partial}{\partial \im \lambda} \log K(u, \lambda) \big|_{u_0}^{\ur}
&\approx -\int_{u_0}^{\ur} \frac{1}{\im y}
\approx -\int_{u_0}^{\ur} \frac{1}{\im \sqrt{V}} \\
\frac{\partial \p(u, \lambda)}{\partial \re \lambda}  \Big|_{u_0}^{\ur}, 
\frac{\partial \p(u, \lambda)}{\partial \im \lambda} \Big|_{u_0}^{\ur} &\approx 0 \:,
\end{align*}
where~$\approx$ means ``up to an arbitrarily small error.''

Since according to~\eqref{Dellambda}, the form of the potential is nearly constant on the
scale~$\Delta \lambda$, it follows that for any~$\lambda \in A_L(\lambda_0)$,
\begin{align*}
\vartheta(\lambda) \big|_{u_0}^{\ur} - \vartheta(\lambda_0) \big|_{u_0}^{\ur}
&\approx \re(\lambda-\lambda_0)\: \int_{u_0}^u \frac{1}{2 \im \sqrt{V}} \bigg|_{\lambda_0}
\approx \frac{\re(\lambda-\lambda_0)}{32\, \Delta \lambda}  \\
\log K(\lambda) \big|_{u_0}^{\ur} - \log K(\lambda_0) \big|_{u_0}^{\ur}
&\approx -\im(\lambda-\lambda_0)\: \int_{u_0}^u \frac{1}{\im \sqrt{V}} \bigg|_{\lambda_0}
\approx -\frac{\im(\lambda-\lambda_0)}{16\, \Delta \lambda} \:.
\end{align*}
Using the form of the annular region, we obtain the result.
\QED

\subsection{Estimates of~$\partial_\lambda \zeta_L$ in the Airy and Parabolic Cylinder Regions}
Before obtaining the estimates, we explain how we can arrange that
the imaginary part of the potential satisfies the assumptions~(a) or~(b)
in Proposition~\ref{prpTparabolic} and Proposition~\ref{prpTairy}.
To this end, we make use of the fact that taking the complex conjugate of the
Sturm-Liouville equation~\eqref{5ode} is again of Sturm-Liouville form,
\beq \label{5odecc}
\left( -\frac{d^2}{du^2} + \overline{V} \right) \overline{\phi} = 0 \:,
\eeq
but now with the opposite sign of~$\im V$. For the construction of the 
resolvent, we are free work either with the original equation or with the complex
conjugate equation because if the resolvent of~\eqref{5odecc} has been
constructed, the corresponding resolvent of~\eqref{5ode} is obtained simply
by complex conjugation, preserving all our estimates.
With this in mind, we can proceed as follows: In the parabolic cylinder region,
the function~$\im V$ either has a zero, or it is everywhere positive or negative.
If it has a zero, we are in case~(b) of Proposition~\ref{prpTparabolic}.
If it is everywhere positive, we are in case~(a) of Proposition~\ref{prpTparabolic}.
If it is everywhere negative, we work with the complex conjugate equation and are
again in case~(a).
If we are in the Airy case, Lemma~\ref{lemmaairy123} gives three possible cases.
In case~(i), we may apply Proposition~\ref{prpTairy} in case~(a).
In case~(ii), we work with the complex conjugate equation and again
apply case~(a) in Proposition~\ref{prpTairy}.
Finally, in case~(iii) the assumption~(b) in Proposition~\ref{prpTairy}
are satisfied for~$\delta<\frac{1}{2}$.
We conclude that with this procedure, the assumptions~(a) or~(b)
in Proposition~\ref{prpTparabolic} and Proposition~\eqref{prpTairy}
can always be satisfied. In what follows, we can take them for granted.

We consider the interval~$[\ur, v]$, where (cf.~\eqref{regions})
\beq \label{vdef} v = \left\{ \begin{array}{cl} u_+ & \text{in the Airy case} \\
\umax & \text{in the parabolic cylinder case} \\
\umax & \text{in the WKB case} \:. \end{array} \right.
\eeq
The following lemma is trivial in the WKB case because
in this case the interval~$[\ur, v]$ reduces to a single point (cf.~\eqref{urdef} and~\eqref{vdef}).
But in the Airy and parabolic cylinder cases, the next lemma gives control of the region near
the zero of~$\re V$.

\begin{Lemma} \label{lemma97} For any~$u \in [\ur, v]$,
\begin{align*}
\big| &\partial_\lambda \zeta_L(u) - \partial_\lambda \zeta_L(\ur) \big| \\
&\leq \frac{2 \Const_2^2}{|\phi(\ur)|^2} \left( \big| \partial_\lambda \log \phi(\ur)  \big|\: (u-\ur)
+ \frac{\Const_2^2}{2} \:\big| \partial_\lambda y(\ur) \big| \:(u-\ur)^2 + \frac{\Const_2^4}{6}\, (u-\ur)^3  \right) \:.
\end{align*}
\end{Lemma}
\Proof 
We first integrate the differential equation~\eqref{yleq} from~$\ur$ to~$u$ to obtain
\[ \phi^2 \partial_\lambda y \big|_{\ur}^u = \int_{\ur}^u \phi^2 \:. \]

In the parabolic cylinder case, we now apply Proposition~\ref{prpTparabolic}. Similarly, in the Airy case,
we apply Proposition~\ref{prpTairy}. This gives
\beq \label{philamy}
\big| \phi^2(u) \partial_\lambda y(u) \big|
\leq \big| \phi^2(\ur) \partial_\lambda y(\ur) \big| + \Const_2^2\, |\phi(\ur)|^2\: (u-\ur)
\eeq

Next, we want to estimate~$\partial_\lambda \log \phi$.
Differentiating the relation~$\partial_u \log \phi = y$ with respect to~$\lambda$, we obtain
\[ \frac{\partial}{\partial u} \frac{\partial}{\partial \lambda}\: \log \phi = \partial_\lambda y \:. \]
Integrating from~$\ur$ to~$u$ gives
\[ \partial_\lambda \log \phi \big|_{\ur}^u = \int_{\ur}^u \partial_\lambda y\:. \]
Now we can apply~\eqref{philamy} and again Proposition~\ref{prpTparabolic}, respectively
Proposition~\ref{prpTairy} to obtain
\begin{align*}
\big| & \partial_\lambda \log \phi(u) \big| - \big| \partial_\lambda \log \phi(\ur)  \big|
\leq \frac{\Const_2^2}{|\phi(\ur)|^2} \int_{\ur}^u \big| \phi^2 \partial_\lambda y \big| \\
&\leq \frac{\Const_2^2}{|\phi(\ur)|^2} \left( 
\big| \phi^2(\ur) \partial_\lambda y(\ur) \big| \:(u-\ur) + \frac{\Const_2^2}{2}\, |\phi(\ur)|^2\: (u-\ur)^2 \right) \\
&= \Const_2^2 \:\big| \partial_\lambda y(\ur) \big| \:(u-\ur) + \frac{\Const_2^4}{2}\, (u-\ur)^2 \:.
\end{align*}

Finally, we compute the $\lambda$-derivative of~$\zeta_L$,
\begin{align*}
\partial_\lambda \zeta_L \big|_{\ur}^u
&= \int_{\ur}^u \frac{1}{\phi^2} = -2 \int_{\ur}^u \frac{1}{\phi^2}\: \partial_\lambda \log \phi \\
\end{align*}
Using the above estimate for~$\partial_\lambda \log \phi(u)$, we obtain
\begin{align*}
\big| &\partial_\lambda \zeta_L(u) - \partial_\lambda \zeta_L(\ur) \big|
\leq \frac{2 \Const_2^2}{|\phi(\ur)|^2} \int_{\ur}^u \big| \partial_\lambda \log \phi \big| \\
&\leq \frac{2 \Const_2^2}{|\phi(\ur)|^2} \int_{u^-}^u \left( \big| \partial_\lambda \log \phi(\ur)  \big| 
+ \Const_2^2\: \big| \partial_\lambda y(\ur) \big| \:(u-\ur) + \frac{\Const_2^4}{2}\, (u-\ur)^2  \right) \:.
\end{align*}
Carrying out the integral gives the result.
\QED

\subsection{Proof of the Lower Bound for~$|\zeta_L|$}
According to the estimates near the poles (Lemmas~\ref{lemmaremu} and~\ref{lemmaremu2}),
by suitably increasing~$\Const_5$ we can arrange that
\beq \label{zetaules}
\big| \zeta_L(\ul) \big| < \delta\:.
\eeq
As is obvious from the expansion of the potential in~\eqref{V0ex} and~\eqref{Vn0ex},
by further increasing~$\Const_5$ we can arrange that on~$[\ul, u_0]$,
\[ -\re V \geq \frac{\re \lambda}{2} \geq \frac{\Const_5}{2}\: |\Omega| \:. \]
Using furthermore that~$\im V$ is bounded by~$\lesssim |\Omega|$
(see again~\eqref{relamright} and~\eqref{inter}), possibly by again increasing~$c_3$
we can again arrange that the real part of the potential
dominates the imaginary part. Hence
\beq \label{reVes}
\int_{\ul}^{u_0} \re \sqrt{V} \leq 2 \int_{\ul}^{u_0} \frac{\im V}{\sqrt{|\re V|}}
\lesssim (u_0-\ul)\: \frac{|\Omega|}{\sqrt{\Const_5\, |\Omega|}} \lesssim \frac{u_0}{\sqrt{\Const_5}}\:
|\Omega|^{\frac{1}{2}}\:.
\eeq
In view of the value of~$u_0$ as given in~\eqref{u00def} and~\eqref{u0Ldef}, one
concludes that by further increasing~$\Const_5$, we can make the left side in~\eqref{reVes}
as small as we like. In view of the formula for the radius of the osculating circle in the
WKB region~\eqref{pRapprox}, this means that that~$R$ is constant on the interval~$[\ul, u_0]$,
up to an arbitrarily small error. Since~$R(u_0)=\frac{1}{2}$ (as is again obvious from~\eqref{pRapprox}),
we can thus arrange by choosing~$\delta$ sufficiently small that
\beq \label{Rules}
\Big| R(\ul) - \frac{1}{2} \Big| \leq \varepsilon \:.
\eeq

Combining~\eqref{zetaules} with~\eqref{Rules} and using that~$\zeta_L$ lies on the osculating
circle with center~$\p$, the triangle inequality gives
\[ \Big| |\p(\ul)| - \frac{1}{2} \Big| \leq 2 \varepsilon\:. \]
Moreover, we know from~\eqref{pRapprox} that~$\p$ is nearly constant in the WKB region.
Thus we can arrange that
\[ \Big| |\p| - \frac{1}{2} \Big| \leq \frac{5}{2}\: \varepsilon \qquad \text{on~$[\ul, \ur]$}\:. \]

Combining~\eqref{zLumax} with the estimates of Lemma~\ref{lemma89},
we obtain the following estimates,
\begin{align*}
|\zeta_L(\lambda)|\big|_\umax  &\geq \left| \zeta_L(\lambda) \big|_{u_+} - \zeta_L(\lambda_0) \big|_{u_+} \right|
-3 \varepsilon \\
&\geq \left| \zeta_L(\lambda) \big|_{\ur} - \zeta_L(\lambda_0) \big|_{\ur} \right|
- \left| \zeta_L(\lambda) \big|_{\ur}^{u_+} - \zeta_L(\lambda_0) \big|_{\ur}^{u_+} \right|
-3 \varepsilon \\
|\zeta_L(\lambda)|\big|_\umax &\leq \left| \zeta_L(\lambda) \big|_{\ur} - \zeta_L(\lambda_0) \big|_{\ur} \right|
+ \left| \zeta_L(\lambda) \big|_{\ur}^{u_+} - \zeta_L(\lambda_0) \big|_{\ur}^{u_+} \right|
+ 3 \varepsilon \:.
\end{align*}
In order to estimate~$|\zeta_L(\lambda) \big|_{\ur} - \zeta_L(\lambda_0) \big|_{\ur}$,
we go back to the osculating circle estimates of Proposition~\ref{prposclam}.
Knowing that at~$\ul$, the function~$\zeta_L$ is small~\eqref{zetaules} and the
osculating circle has a radius close to one half~\eqref{Rules}, the mean value theorem yields
\begin{align*}
\Big| \log K(\lambda)|_{\ul} - \log K(\lambda_0)|_{\ul} \Big| &= \Big| \log R(\lambda)|_{\ul} - \log R(\lambda_0)|_{\ul} \Big|
= \frac{1}{\tilde{R}}\: \Big| R(\lambda)|_{\ul} - R(\lambda_0)|_{\ul} \Big| \\
&\leq \left(\frac{1}{2} -\varepsilon \right)^{-\frac{1}{2}} \: 2 \varepsilon \leq \frac{5 \varepsilon}{2}\:.
\end{align*}
Moreover, elementary trigonometry shows that the angles satisfy the inequality
\[ \left(\frac{1}{2} - \varepsilon \right) 2 \sin \left| \frac{\vartheta(\lambda)|_{\ul} - \vartheta(\lambda_0)|_{\ul}}{2}
\right| \leq \Big| |\zeta(\lambda)|_{\ul} - \zeta(\lambda_0)|_{\ul} \Big| \leq 2 \varepsilon \]
and thus
\[ \Big| \vartheta(\lambda)|_{\ul} - \vartheta(\lambda_0)|_{\ul} \Big| \leq  
2 \arcsin \left(  \left(\frac{1}{2} - \varepsilon \right)^{-1} \varepsilon \right) \leq 5 \varepsilon \:. \]

In order to control the behavior on the interval~$[\ul, \ur]$, we apply
Proposition~\ref{prposclam} (note that the condition~\eqref{WKB2} is
satisfied in view of Proposition~\ref{prpimVsqrtV}).
It follows that one of the following two inequalities holds:
\begin{gather*}
30\, \varepsilon \leq \Big| \vartheta(\lambda) \big|_{\ur} - \vartheta(\lambda_0) \big|_{\ur} \Big|
\leq 200\, \varepsilon \\
60\, \varepsilon \leq \Big| \log K(\lambda) \big|_{\ur} - \log K(\lambda_0) \big|_{\ur} \Big|
\leq 400\, \varepsilon\:.
\end{gather*}
This in turn implies that the change of~$|\zeta_L|$ can be estimated from below and above by
\begin{align*}
\Big| \zeta_L(\lambda)|_{\ur} - \zeta_L(\lambda_0)|_{\ur} \Big| & \geq 10\, \varepsilon\:R(\lambda_0)|_{\ur}
- \Big| \p(\lambda)|_{\ur} - \p(\lambda_0)|_{\ur} \Big|
\geq 4 \varepsilon \\
\Big| \zeta_L(\lambda)|_{\ur} - \zeta_L(\lambda_0)|_{\ur} \Big| & \leq 400\, \varepsilon\:R(\lambda_0)|_{\ur}
+ \Big| \p(\lambda)|_{\ur} - \p(\lambda_0)|_{\ur} \Big|
\leq 500\, \varepsilon \:.
\end{align*}

Finally, the estimate of Lemma~\ref{lemma97} shows that the change of~$\zeta_L$
in the Airy region or the parabolic cylinder region is much smaller than the change of~$\zeta_L$
in the WKB region. In particular, by choosing~$|\Omega|$ sufficiently large
(and noting that the size of the parabolic cylinder and Airy regions tends to zero
as~$|\Omega| \rightarrow \infty$), we can arrange that
\[ \left| \zeta_L(\lambda) \big|_{\ur}^{u_+} - \zeta_L(\lambda_0) \big|_{\ur}^{u_+} \right| \leq \varepsilon \:. \]
This concludes the proof of Proposition~\ref{prp94}.

\section{The Green's Function for a Double-Well Potential} \label{secdwp}
The goal of this section is to derive pointwise estimates of the Green's function.
For the statement of the result, we need the parameter~$\hat{u}$ defined as follows.
If~$\re V(\umax) \leq 0$, we simply set~$\hat{u}=\umax$
(this case includes the WBK case and part of the parabolic cylinder case).
If conversely~$\re V(\umax)>0$, we denote the zeros of the real part of the potential
in the parabolic cylinder or Airy regions by~$v_{L\!/\!R}$,
\[ \re V(v_L) = 0 = \re V(v_R) \qquad \text{and} \qquad v_L < \umax < v_R \:. \]
Then~$\hat{u}$ is defined by the requirement that
\beq \label{hatudef}
\int_{v_L}^{\hat{u}} \re \sqrt{V} =  \int_{\hat{u}}^{v_R} \re \sqrt{V} \:.
\eeq
Since our potential is almost symmetric around~$\piot$, the points~$\hat{u}$ and~$\umax$
are all close to~$\piot$. Working with our definitions has the advantage that we
do not need to quantify how close these points are.

Here is the main result of this section.
\begin{Prp} \label{prpresnew}
Assume that for suitable constants~$C, \delta>0$, the following inequalities hold,
\begin{gather}
\displaystyle \big| \zeta_L({\hat{u}})\: \phi^2_L({\hat{u}}) \big|,\;
\displaystyle \big| \zeta_R({\hat{u}})\: \phi^2_R({\hat{u}}) \big| \geq \displaystyle
\frac{4}{|y_L({\hat{u}}) - y_R({\hat{u}})|} \label{zetalowernew} \\
|\zeta_L| \leq C \quad \text{on~$(0,{\hat{u}}]$} \:,\qquad
|\zeta_R| \leq C \quad \text{on~$[{\hat{u}},\pi)$} \label{zetaC} \\
\big| \zeta_L({\hat{u}}) \big|, \big| \zeta_R({\hat{u}}) \big| \geq \delta \:. \label{zetad}
\end{gather}
Then, setting
\beq \label{Sigmadef}
\Sigma = \min \Big( \big| \phi_L({\hat{u}}) \big|^2, \big| \phi_R({\hat{u}}) \big|^2 \Big) \:
\big|\big(y_L - y_R\big)({\hat{u}}) \big| \:,
\eeq
the kernel of the Green's function~$s_\lambda(u,u')$ for~$0 < u \leq u' < \pi$ is bounded by
\begin{align*}
\big| & s_\lambda(u,u') \big| \lesssim \\
& \left\{
\begin{array}{cc}
\displaystyle \frac{1}{\delta^2}\: \frac{\big| \phiD_L(u)\, \phiD_R(u') \big|}{\Sigma} &
\hspace*{-1cm} \text{if~$u<{\hat{u}}$ and~$u'>{\hat{u}}$} \\[1em]
\displaystyle \bigg( \frac{C}{\delta} + \frac{C}{\delta^2 \Sigma} \bigg)
\big| \phi_R(u)\, \phiD_R(u') \big| & \text{if~$u,u' > u_+^R$} \\[1em]
\displaystyle \bigg( \frac{C}{\delta} + \frac{C}{\delta^2 \Sigma} \bigg)
\big| \phiD_L(u)\, \phi_L(u') \big| & \text{if~$u,u' < u_+^L$} \\[0.8em]
\displaystyle \bigg\{ \frac{1}{\delta \Sigma}
+ \Big( \frac{1}{\delta} + \frac{1}{\delta^2 \Sigma} \Big) \int_{\hat{u}}^u \frac{1}{|\phi_R|^2}
\bigg\}\:
\big| \phi_R(u)\, \phiD_R(u') \big|
& \text{if~${\hat{u}} \leq u \leq u_+^R$} \\[1em]
\displaystyle \bigg\{ \frac{1}{\delta \Sigma}
+ \Big( \frac{1}{\delta} + \frac{1}{\delta^2 \Sigma} \Big) \int_{u'}^{\hat{u}} \frac{1}{|\phi_L|^2}
\bigg\}\: \big| \phiD_L(u)\, \phi_L(u') \big|
& \text{if~$u_+^L \leq u' \leq {\hat{u}}$} \:.
\end{array} \right.
\end{align*}
\end{Prp}

We begin by estimating the Wronskian from below.
\begin{Lemma} \label{lemmanoeigen} Assume that~\eqref{zetalowernew} holds. Then
\[  \big| w(\phiD_L, \phiD_R) \big| \geq
\frac{1}{2}\: \Big| \zeta_L({\hat{u}})\, \zeta_R({\hat{u}})\: w(\phi_L, \phi_R) \Big| \:. \]
\end{Lemma}
\Proof We evaluate~\eqref{wronskirel} at~$u={\hat{u}}$, take the absolute value and use the inequalities~\eqref{zetalowernew}. This gives
\[ \big| w(\phiD_L, \phiD_R) \big| \geq \frac{1}{2}\: \big|
\phi_L\,\zeta_L \: \phi_R \,\zeta_R \:(y_L  - y_R) \big| ({\hat{u}}) 
= \frac{1}{2}\: \big| \zeta_L \,\zeta_R \:w(\phi_L, \phi_R) \big| ({\hat{u}}) \:, \]
concluding the proof.
\QED

In order to estimate the Green's function, we need to control both~$\phiD_L$ and~$\phiD_R$
on the whole interval~$(0, \pi)$. The estimates so far, however, only
give us control of~$\phiD_L$ on the interval~$(0, \frac{\pi}{2})$
(and similarly of~$\phiD_R$ on the interval~$[\frac{\pi}{2}, \pi]$).
The following lemma gives a formula for~$\phiD_L$ on the remaining interval~$[\frac{\pi}{2}, \pi]$.

\begin{Lemma} \label{lemmaright} For any~$u \in [\frac{\pi}{2}, \pi]$,
\beq
\phiD_L(u) = \phi_R(u) \left( \zeta_L({\hat{u}}) \:\frac{\phi_L({\hat{u}})}{\phi_R({\hat{u}})} 
+\zeta_R \Big|_{{\hat{u}}}^u\: \lim_{v \searrow 0} \frac{\phi_R(v)}{\phi_L(v)} \right) \label{phiDL} \:,
\eeq
where
\beq
\lim_{v \searrow 0} \frac{\phi_R(v)}{\phi_L(v)} = \frac{\phi_R({\hat{u}})}{\phi_L({\hat{u}})} + w(\phi_L, \phi_R)\:
\zeta_L({\hat{u}}) \label{phiR0} \:.
\eeq
\end{Lemma}
\Proof First,
\[ \phiD_L(u) = \phi_R(u)\: \frac{\phiD_L(u)}{\phi_R(u)} 
=  \phi_R(u)\: \frac{\phiD_L({\hat{u}})}{\phi_R({\hat{u}})}  + \phi_R(u)
\int_{{\hat{u}}}^u \Big(\frac{\phiD_L}{\phi_R} \Big)^\prime(\tau)\: d\tau \:. \]
Computing the derivative on the right gives
\[ \Big(\frac{\phiD_L}{\phi_R} \Big)^\prime = \frac{(\phiD_L)' \phi_R - \phiD_L (\phi_R)'}{\phi_R^2} 
= \frac{w(\phiD_L, \phi_R)}{\phi_R^2} \:. \]
Integrating this equation from~$\hat{u}$ to~$u$, we obtain
\[ \phiD_L(u) = \phi_R(u)\: \frac{\phiD_L({\hat{u}})}{\phi_R({\hat{u}})}  + w(\phiD_L, \phi_R)\:
\phi_R(u) \int_{{\hat{u}}}^u \frac{1}{\phi_R^2} \:. \]
The Wronskian appearing here is most conveniently computed asymptotically at the origin,
\[ w(\phiD_L, \phi_R) = \lim_{v \searrow 0} \big( (\phiD_L)'(v)\, \phi_R(v) - \phiD_L(v)\, (\phi_R)'(v) \big) =
\lim_{v \searrow 0} \frac{\phi_R(v)}{\phi_L(v)} \:, \]
where in the last step we differentiated~\eqref{phiDdef} and used
the asymptotics of the fundamental solutions as stated in~\eqref{genasy}, \eqref{ex1} and~\eqref{ex2}.
Applying~\eqref{phiDzeta} gives~\eqref{phiDL}.

In order to prove~\eqref{phiR0}, we begin with the computation
\begin{align*}
\phi_R(v) &= \phi_L(v)\: \frac{\phi_R(v)}{\phi_L(v)}
=\phi_L(v)\: \frac{\phi_R({\hat{u}})}{\phi_L({\hat{u}})} - \phi_L(v) \int_v^{\hat{u}} \Big(\frac{\phi_R}{\phi_L} \Big)' \\
&=\phi_L(v)\: \frac{\phi_R({\hat{u}})}{\phi_L({\hat{u}})} + \phi_L(v) \int_v^{\hat{u}} \frac{w(\phi_L, \phi_R)}{\phi_L^2}\:.
\end{align*}
We now divide by~$\phi_L(v)$ and take the limit~$v \searrow 0$. This concludes the proof.
\QED

\Proof[Proof of Proposition~\ref{prpresnew}.]
Applying Lemma~\ref{lemmanoeigen}, our task is to estimate the absolute value of the expression
\beq \label{Edef}
E := \frac{\phiD_L(u)\, \phiD_R(u')}{\zeta_L({\hat{u}}) \, \zeta_R({\hat{u}})\, w(\phi_L, \phi_R)}
\qquad \text{for all~$0 < u \leq u' \leq \pi$}\:.
\eeq
We consider the different cases after each other.
In the case~$u<{\hat{u}}$ and~$u'>{\hat{u}}$, we apply~\eqref{zetad}
to~\eqref{Edef} to obtain
\[ |E| \leq \frac{1}{\delta^2}\: \frac{\big|\phiD_L(u)\, \phiD_R(u')\big|}
{w(\phi_L, \phi_R)} \:. \]
Using that
\[ \big| w(\phi_L, \phi_R)\big| = \big|\phi_L({\hat{u}})\, \phi_R({\hat{u}})\: (y_L-y_R)({\hat{u}}) \big|
\geq \Sigma\:, \]
we obtain the desired estimate.

Using the symmetry under reflections at~$\piot$, it remains to consider the case~$u,u' \geq {\hat{u}}$.
Applying Lemma~\ref{lemmaright}, a straightforward computation yields
\begin{align}
&E = -\frac{\zeta_R({\hat{u}}) - \zeta_R(u)}{\zeta_R({\hat{u}})} \: \phi_R(u)\, \phi_R(u')\: \zeta_R(u') \label{E1} \\
&\;- \frac{1}{y_L({\hat{u}}) - y_R({\hat{u}})}\: \frac{\zeta_R({\hat{u}}) - \zeta_R(u)}{\zeta_L({\hat{u}}) \phi_L({\hat{u}})^2\:
\zeta_R({\hat{u}})}\: \phi_R(u)\, \phi_R(u')\: \zeta_R(u') \label{E2} \\
&\;- \frac{1}{y_L({\hat{u}}) - y_R({\hat{u}})}\: \frac{1}{\zeta_R({\hat{u}})\,\phi_R({\hat{u}})^2}\:
\phi_R(u)\, \phi_R(u')\: \zeta_R(u') \label{E3} \:.
\end{align}
In the case~$u>u_+^R$, we estimate these terms by
\begin{align*}
|\eqref{E1}| &\leq \frac{C}{\delta}\: \big|\phi_R(u)\, \phiD_R(u')\big| \\
|\eqref{E2}| &\leq \frac{C}{\delta^2 \Sigma}\: \big|\phi_R(u)\, \phiD_R(u')\big| \\
|\eqref{E3}| &\leq \frac{1}{\delta \Sigma}\: \big|\phi_R(u)\, \phiD_R(u')\big| \,
\end{align*}
giving the desired estimate.

In the remaining case~${\hat{u}} \leq u \leq u_+^R$, we use the identity
\[ \big| \zeta_R({\hat{u}}) - \zeta_R(u) \big| = \bigg| \int_{{\hat{u}}}^u \frac{1}{\phi_R^2} \bigg| \]
to obtain
\begin{align*}
|\eqref{E1}| &\leq \frac{1}{\delta} \, \big|\phi_R(u)\, \phiD_R(u')\big| \: \bigg| \int_{{\hat{u}}}^u \frac{1}{\phi_R^2} \bigg| \\
|\eqref{E2}| &\leq \frac{1}{\delta^2 \Sigma} \, \big|\phi_R(u)\, \phiD_R(u')\big| \: \bigg| \int_{{\hat{u}}}^u
\frac{1}{\phi_R^2} \bigg| \\
|\eqref{E3}| &\leq \frac{1}{\delta \Sigma}\: \big|\phi_R(u)\, \phiD_R(u')\big| \:.
\end{align*}
This gives the result.
\QED

\section{Deforming the Potential} \label{secdeform}
In the following estimates, we distinguish the cases that the real part of the potential
is positive and large near~$u=\umax$ or that it is negative or small there.
We refer to these cases as the {\bf{double-well case}} and the {\bf{single-well case}},
respectively. Qualitatively speaking, in the double-well case the potential looks like
in the Airy case, whereas the single-well case comprises the WKB and parabolic cylinder
cases. However, these regions are not exactly the same, making it necessary to use
a new notation. The important difference is that the single- and double-well cases are
defined without referring to the constants~$\Const_1, \Const_4, \ldots$.
Instead, we work with the integrals~\eqref{hatudef} and introduce a new constant~${\mathscr{K}}>0$
(which will be specified Section~\ref{secdoublewell} below).
More precisely, if~$\re V(\umax)=0$, we are by definition in the single-well case.
If~$\re V(\umax)>0$ (so that~$\hat{u}$ is defined by~\eqref{hatudef}), we distinguish the
\beq \label{singledouble}
\left\{ \begin{array}{rl}
\text{single-well case}: \qquad & \displaystyle \int_{v_L}^{v_R} \re \sqrt{V} \leq {\mathscr{K}} \\[1.2em]
\text{double-well case}: \qquad & \displaystyle \int_{v_L}^{v_R} \re \sqrt{V} > {\mathscr{K}} \:.
\end{array} \right.
\eeq

\subsection{Estimates in the Double-Well Case} \label{secdoublewell}
\begin{Lemma} \label{lemmaKdef}
By choosing~${\mathscr{K}}$ sufficiently large, we can arrange that in the double-well case
the following estimates hold:
\begin{align}
\big| \phi_L(\hat{u}) \big| ,\; \big| \phi_R(\hat{u}) \big| &\gtrsim \frac{1}{\sqrt[4]{|V|}}\: e^{\int_{v_L}^{\hat{u}}
\re \sqrt{V}} \label{doublel1} \\
\big| y_L(\hat{u}) - y_R(\hat{u}) \big| &\geq \sqrt{|V(\hat{u})|} \:. \label{doublel2}
\end{align}
Moreover, the function~$\Sigma$ defined by~\eqref{Sigmadef} is bounded by
\beq \label{Sigmaes}
\Sigma \gtrsim \exp \left(2 \int_{v_L}^{\hat{u}} \re \sqrt{V} \right) \:.
\eeq
\end{Lemma}
\Proof If~${\mathscr{K}}$ is chosen sufficiently large, we know from the estimates
in Section~\ref{secupumax} that the function~$|\phi_L|$ 
(and similarly~$|\phi_R|$) can be approximated by the WKB wave function~\eqref{WKBgen}. Moreover,
as stated after~\eqref{WKBgen}, the coefficient of the exponentially increasing fundamental solution
is non-zero. This implies~\eqref{doublel1}. Moreover, differentiating~\eqref{WKBgen}, one finds
that~$y_L(\hat{u}) \approx \sqrt{V(\hat{u})}$. Similarly, $y_R(\hat{u}) \approx -\sqrt{V(\hat{u})}$,
proving~\eqref{doublel2}.

Finally, the estimate~\eqref{Sigmaes} follows immediately by using~\eqref{doublel1} and~\eqref{doublel2}
in~\eqref{Sigmadef}.
\QED
From now on, we choose~${\mathscr{K}}$ so large that the the statement of this lemma applies.

\begin{Lemma} \label{lemmasmallzeta}
For any~$\varepsilon>0$, by increasing~${\mathscr{K}}$ we can arrange that for every eigenvalue~$\lambda_0$,
\beq \label{alternate}
\big|\zeta_L(\hat{u}) \big| < \frac{\varepsilon}{2} \qquad \text{or} \qquad
\big|\zeta_R(\hat{u}) \big| < \frac{\varepsilon}{2}\:.
\eeq
\end{Lemma}
\Proof Let~$\lambda_0$ be an eigenvalue. Then the eigenvalue condition~\eqref{evalcond3} is satisfied
at~$u=\hat{u}$, implying that
\[ \frac{1}{|\phi_L^2 \zeta_L|} +  \frac{1}{|\phi_R^2 \zeta_R|} \geq |y_L-y_R| \qquad \text{at~$\hat{u}$}\:. \]
Applying Lemma~\ref{lemmaKdef}, we obtain
\[ \frac{1}{|\zeta_L(\hat{u})|} +  \frac{1}{|\zeta_R(\hat{u})|} \gtrsim 
\exp \left(2 \int_{v_L}^{\hat{u}} \re \sqrt{V} \right)  \geq e^{2{\mathscr{K}}} \:. \]

Applying again the argument after~\eqref{expsmall}, by increasing~$\Const_1$
we can make the exponential factors as small as we like. Hence, for any given~$\varepsilon>0$
we can arrange that
\[ \frac{1}{|\zeta_L(\hat{u})|} + \frac{1}{|\zeta_R(\hat{u})|} \geq \frac{2}{\varepsilon} \:. \]
This gives the claim.
\QED

We now apply Proposition~\ref{prp94}. In order to combine estimates for~$\zeta_L$ and~$\zeta_R$,
we modify~\eqref{Dellam} according to
\beq \label{Dellamnew}
\Delta \lambda := 75\,\varepsilon \min \left\{ 
\bigg( \int_{\ul^L}^{\ur^L} \frac{1}{\sqrt{|V|}} \bigg|_{\lambda_0} \bigg)^{-1} ,
\int_{\ur^R}^{\ul^R} \frac{1}{\sqrt{|V|}} \bigg|_{\lambda_0} \bigg)^{-1} \right\} \:.
\eeq
Thus we choose~$\Delta \lambda$ so small that we can work with the same~$\Delta \lambda$
both for~$\zeta_L$ and~$\zeta_R$. Decreasing~$\Delta \lambda$ in this way
can be described equivalently by making the parameter~$\varepsilon$ in Proposition~\eqref{prp94}
smaller. The reason why this procedure is unproblematic is that the parameter~$\varepsilon$
is changed at most by a uniform constant:

\begin{Lemma} \label{lemmashift}
\[ \int_{\ul^L}^{\ur^L} \frac{1}{\sqrt{|V|}} \bigg|_{\lambda_0} \simeq
\int_{\ur^R}^{\ul^R} \frac{1}{\sqrt{|V|}} \bigg|_{\lambda_0} \:. \]
\end{Lemma}
\Proof Similar as in the proof of Lemma~\ref{lemma1110} and Section~\ref{secreflect},
we make use of the fact that the potential is approximately reflection symmetric around~$u=\piot$.
We thus obtain
\[ \left| \int_{\ul^L}^{\ur^L} \frac{1}{\sqrt{|V|}} \bigg|_{\lambda_0} - \int_{\ur^R}^{\ul^R} \frac{1}{\sqrt{|V|}} \bigg|_{\lambda_0} 
\right| \lesssim \int_{\ul^L}^{\ur^L} \frac{\big|V(u) - V(\pi-u) \big|}{|V|^\frac{3}{2}}
\lesssim \frac{|\Omega|}{\Const_1\, |\Omega|} \:\int_{\ul^L}^{\ur^L} \frac{1}{\sqrt{|V|}} \:. \]
This gives the result.
\QED

\begin{figure}%
\psscalebox{1.0 1.0} 
{
\begin{pspicture}(0,-1.48)(9.235,1.48)
\definecolor{colour0}{rgb}{0.8,0.8,0.8}
\psframe[linecolor=black, linewidth=0.02, fillstyle=solid,fillcolor=colour0, dimen=outer](7.86,1.48)(5.4,-0.98)
\psframe[linecolor=black, linewidth=0.02, fillstyle=solid,fillcolor=colour0, dimen=outer](8.16,0.98)(5.7,-1.48)
\psframe[linecolor=black, linewidth=0.02, fillstyle=solid, dimen=outer](2.81,0.69)(2.45,0.33)
\psframe[linecolor=black, linewidth=0.02, fillstyle=solid,fillcolor=colour0, dimen=outer](3.76,1.48)(1.3,-0.98)
\psframe[linecolor=black, linewidth=0.02, fillstyle=solid, dimen=outer](2.86,0.58)(2.2,-0.08)
\rput[bl](0.0,1.015){\normalsize{$A_L(\lambda_0)$}}
\psbezier[linecolor=black, linewidth=0.04](2.995,0.14480114)(3.0052397,0.58550054)(2.9470189,0.68051547)(2.6155553,0.675)(2.284092,0.6694845)(2.087253,0.727148)(2.08,0.36455607)(2.0727468,0.0019641726)(2.1255355,-0.16993138)(2.3426447,-0.205)(2.559754,-0.24006861)(2.9847603,-0.29589826)(2.995,0.14480114)
\rput[bl](3.075,0.54){\normalsize{$\Gamma$}}
\psframe[linecolor=black, linewidth=0.02, fillstyle=solid, dimen=outer](6.96,0.58)(6.3,-0.08)
\psframe[linecolor=black, linewidth=0.02, fillstyle=solid, dimen=outer](7.26,0.18)(6.6,-0.48)
\psframe[linecolor=black, linewidth=0.02, dimen=outer](7.86,1.48)(5.4,-0.98)
\psframe[linecolor=black, linewidth=0.02, dimen=outer](6.96,0.58)(6.3,-0.08)
\psbezier[linecolor=black, linewidth=0.04](6.58,0.645)(6.195,0.635)(6.23,0.71)(6.23,0.285)(6.23,-0.14)(6.32,-0.3)(6.37,-0.37)(6.42,-0.44)(6.435,-0.585)(6.93,-0.565)(7.425,-0.545)(7.33,-0.64)(7.33,-0.215)(7.33,0.21)(7.38,0.21)(7.155,0.445)(6.93,0.68)(6.965,0.655)(6.58,0.645)
\rput[bl](7.29,0.395){\normalsize{$\Gamma$}}
\rput[bl](4.105,1.01){\normalsize{$A_L(\lambda_0)$}}
\rput[bl](8.275,-1.38){\normalsize{$A_R(\nu)$}}
\end{pspicture}
}
\caption{Choice of the closed contour~$\Gamma$.}
\label{figannulus2}
\end{figure}
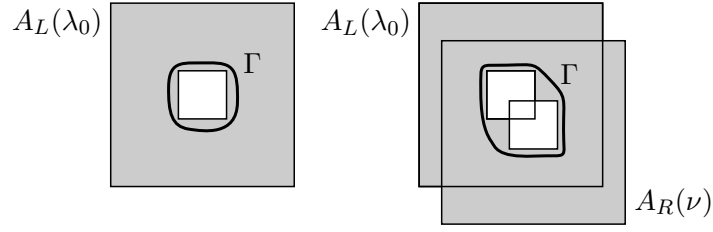

\begin{Lemma} \label{lemmazetalower}
Let~$\lambda_0$ be an eigenvalue. Then, under the assumptions of
the above Lemmas~\ref{lemmaKdef} and~\ref{lemmasmallzeta},
there is a closed contour~$\Gamma$
of length at most~$20 \,\Delta \lambda$ such that
\[ |\zeta_L|, |\zeta_R| \geq \varepsilon \qquad \text{along~$\Gamma$}\:. \]
\end{Lemma}
\Proof Let~$\lambda_0$ be an eigenvalue. Using Lemma~\ref{lemmasmallzeta},
we can arrange that~\eqref{alternate} holds.
Without loss of generality we may assume that
\[ \big|\zeta_L(\hat{u}) \big| < \frac{\varepsilon}{2} \:, \]
because otherwise we repeat the proof with~$L$ and~$R$ interchanged.
Applying Proposition~\ref{prp94}, we know that~$|\zeta_L(\hat{u})|>\varepsilon$ inside
the annular region~$A_L(\lambda_0)$.
We choose a contour~$\Gamma$ inside~$A_L(\lambda_0)$
close to the inner boundary (see the left of Figure~\ref{figannulus2}).
If the inequality~$|\zeta_R(\hat{u})| > \varepsilon$ holds along the contour~$\Gamma$, there is nothing left to prove.
Otherwise, there is a point~$\nu$ on~$\Gamma$ such that~$|\zeta_R(\hat{u})| \leq \varepsilon$.
Again applying Proposition~\ref{prp94}, it follows that~$|\zeta_R(\hat{u})>\varepsilon$ inside
the annular region~$A_R(\nu)$. This makes it possible to choose the contour~$\Gamma$ as shown
on the right of Figure~\ref{figannulus2}.
\QED

\begin{Prp} \label{prpresdouble}
The resolvent~$s_\lambda$ exists along~$\Gamma$ and is bounded by
\[ \big| s_\lambda(u,u') \big| \lesssim
\left\{
\begin{array}{cl}
\displaystyle e^{-\int_{v_L}^{v_R} \re \sqrt{V}}\: \big| \phiD_L(u)\, \phiD_R(u') \big| &
\text{if~$u<\hat{u}$ and~$u'>\hat{u}$} \\[1em]
\displaystyle \big| \phi_R(u)\, \phiD_R(u') \big| & \text{if~$u,u' > u_+^R$} \\[1em]
\displaystyle 
\big| \phiD_L(u)\, \phi_L(u') \big| & \text{if~$u,u' < u_+^L$} \\[0.8em]
\displaystyle e^{-2 \int_{u}^{v_R} \re \sqrt{V}}\: 
\big| \phi_R(u)\, \phiD_R(u') \big|
& \text{if~$\hat{u} \leq u \leq u_+^R$} \\[1em]
\displaystyle e^{-2 \int_{v_L}^{u'} \re \sqrt{V}}\: 
\big| \phiD_L(u)\, \phi_L(u') \big|
& \text{if~$u_+^L \leq u' \leq \hat{u}$} \:.
\end{array} \right. \]
\end{Prp}
\Proof We apply Proposition~\ref{prpresnew}. By symmetry, in the estimates
it suffices to consider the index~$L$.

We begin by estimating~$|\zeta_L|$.
In the WKB region, the total variation of the radius~$R$ of the osculating circle
is under uniform control by Proposition~\ref{prpimVsqrtV}. In the region~$(0,\ul)$ near the pole,
on the other hand, this total variation is even small by Lemma~\ref{lemmazul}.
In the Airy region, Proposition~\ref{prpTairy} provides uniform estimates.
Finally, Lemma~\ref{lemma89} controls the total variation
of~$\zeta_L$ on the interval~$[u_+, \hat{u}]$. Combining these results, we conclude that
\[ |\zeta_L| \lesssim 1 \qquad \text{on~$(0, \hat{u}]$}\:. \]
This gives~\eqref{zetaC}. Moreover, we know from Lemma~\ref{lemmazetalower}
that~\eqref{zetad} holds for~$\delta=\varepsilon$.
We conclude that the assumptions~\eqref{zetaC} and~\eqref{zetad} in Proposition~\ref{prpresnew}
are satisfied.

Let us verify the assumption~\eqref{zetalowernew}. From~\eqref{doublel1} and~\eqref{doublel2}
we know that
\[ \big| \zeta_L(\hat{u}) \phi_L^2(\hat{u}) \big| \gtrsim \frac{1}{\sqrt{|V|}}\:
\exp \left(2 \int_{v_L}^{\hat{u}} \re \sqrt{V} \right)
\gtrsim \frac{e^{2{\mathscr{K}}}}{|y_L(\hat{u}) - y_R(\hat{u})|} \:. \]
Thus by further increasing~${\mathscr{K}}$ we can arrange~\eqref{zetalowernew}.
Therefore, Proposition~\ref{prpresnew} applies.

We estimate the integrals over~$1/\phi_R^2$ by
\begin{align*}
\int_{\hat{u}}^u \frac{1}{|\phi_R|^2} &\lesssim \int_{\hat{u}}^u
\re \sqrt{V} e^{-2 \int_v^{u_0^R} \re \sqrt{V}}\: dv
=  \frac{1}{2} \int_{\hat{u}}^u \frac{d}{dv} e^{-2 \int_v^{u_0^R} \re \sqrt{V}} \\
&\lesssim \left( e^{-2 \int_u^{u_0^R} \re \sqrt{V}} - 
e^{-2 \int_{\hat{u}}^{u_0^R} \re \sqrt{V}} \right)
\lesssim e^{-2 \int_u^{v_R} \re \sqrt{V}} \:.
\end{align*}
Using the lower bound for~$\Sigma$ in~\eqref{Sigmaes} as well as~\eqref{hatudef}
gives the desired estimate.
\QED

\subsection{Estimates in the Single-Well Case}
In the previous section we chose the parameter~${\mathscr{K}}$.
We now proceed with estimates in the single-well case for this fixed value of~${\mathscr{K}}$.

\begin{Prp} \label{prpressingle}
There is a suitable choice of~$\varepsilon<1/(1000)$ such that for
sufficiently large~$\Const_1$ the following statement holds.
If~$\lambda_0$ is eigenvalue in the single-well case, then the resolvent
exists inside the annulus~$A_L(\lambda_0)$ (defined by~\eqref{ALdef}) and is bounded by
\[ \big| s_\lambda(u,u') \big| \lesssim \const\, \big| \phiD_L(u) \,\phiD_R(u') \big| 
\qquad \text{if~$u \leq u'$} \]
(where the constant~$\const$ may depend on~${\mathscr{K}}$ and~$\varepsilon$).
\end{Prp}
\Proof
Let~$\lambda_0$ be an eigenvalue. Then the eigenvalue condition~\eqref{evalcond3}
holds for any~$u \in (0, \pi)$. We now choose a specific value~$\breve{u}$ where
the analysis of the expression in~\eqref{evalcond3} is particularly simple. To this end, we consider
the osculating circle corresponding to~$\phi_L$ on the interval~$(0, \ur)$.
According to Lemma~\ref{lemmazul} , we can make~$|\zeta(\ul)|$ arbitrarily small.
Thus for any~$\delta>0$ by increasing~$\Const_1$ we can arrange that
\[ \big| R(\ul) -|\p(\ul)| \big|< \delta \:. \]
Next, in the WKB region~$(\ul, \ur)$, we know from Proposition~\ref{prppRapprox}
that the center~$\p$ of the osculating circle is approximately fixed, whereas the change of the radius~$R$
is given explicitly in terms of the integral of~$\re \sqrt{V}$. Using the estimate~\eqref{resqrtapprox}
together with Proposition~\ref{prpdelta}, for any given~$\delta>0$ we can arrange that that
\[ \big| R(\ur) -|\p(\ur)| \big|< \delta \:. \]
If we start at~$u=\ur$ and decrease~$u$, the function~$\zeta(u)$ will move along the osculating circle
with an angular velocity as given by~\eqref{thetaprel}. We choose~$\breve{u}$ as the
largest value of~$u$ where the angle~$\vartheta$ is such that~$\zeta$ is close to the origin, i.e.
\beq \label{eps1}
\big| \zeta_L(\breve{u}) \big|< 2 \delta \:.
\eeq

In what follows, we evaluate all functions at the point~$\breve{u}$.
For ease in notation, we shall omit the arguments of these functions.
From the estimates in Section~\ref{secesregion} we know that
\beq \label{phiLR}
|\phi_L|^2, |\phi_R|^2 \simeq \frac{1}{\sqrt{|V|}}
\eeq
(where the constants in the upper and lower bounds may depend on~${\mathscr{K}}$).
Moreover,
\beq \label{yLR}
\big| y_L - y_R \big| \lesssim \sqrt{|V|}\:.
\eeq
Therefore, the eigenvalue condition~\eqref{evalcond3} gives rise to the estimate
\[ \frac{1}{|\zeta_R|} \gtrsim \frac{1}{|\zeta_L|} - \frac{\big| y_L - y_R \big|}
{\sqrt{|V|}} \gtrsim \frac{1}{\delta} \:, \]
implying that
\beq \label{eps2}
\big| \zeta_R \big| \lesssim \delta \:.
\eeq
Combining~\eqref{eps1} and~\eqref{eps2} we conclude that for any~$\tilde{\delta}>0$, we can arrange
by increasing~$\Const_1$ that at~$\breve{u}$,
\[ \big| \zeta_L \big|, \big| \zeta_R \big| < \tilde{\delta}\:. \]

We now vary the spectral parameter~$\lambda$ (for fixed~$\breve{u}$ and~${\mathscr{K}}$).
Our goal is to analyze how the left side of~\eqref{evalcond3} depends on~$\lambda$.
To this end, it is most convenient to include the phase of the factor~$\phi^2$
into the corresponding factor~$\zeta$. Thus, using the notation in~\eqref{phiarg}, we set
\[ z = e^{2 i \vartheta} \zeta \qquad \text{so that} \qquad
\phi^2 \zeta = |\phi|^2\, z\:. \]
Then, according to~\eqref{zetarel},
\[ z = \p\, e^{2 i \vartheta} + \frac{i}{K} \:. \]
Hence~\eqref{wronskirel} can be written as
\begin{align*}
\frac{w(\phiD_L, \phiD_R)}{\phiD_L \phiD_R} &= 
y_{L} - y_{R} + \frac{1}{|\phi_{L}|^2 z_{L}} - \frac{1}{|\phi_{R}|^2 z_{R}}
= y_{L} - y_{R} + \frac{|\phi_{R}|^2 z_{R} - |\phi_{L}|^2 z_{L}}{|\phi_{L}|^2 z_{L} \, |\phi_{R}|^2 z_{R}} \:.
\end{align*}
It follows that
\[ \bigg| \frac{w(\phiD_L, \phiD_R)}{\phiD_L \phiD_R} \bigg| \geq 
\bigg| \frac{|\phi_{R}|^2 z_{R} - |\phi_{L}|^2 z_{L}}{|\phi_{L}|^2 z_{L} \, |\phi_{R}|^2 z_{R}}
\bigg| - \big| y_{L} - y_{R}\big| \:. \]
Now we can vary the angle~$\vartheta$ and the radius~$R=|K|^{-1}$ of the osculating circles
using the formulas in Proposition~\ref{prp85}. Keeping in mind that the variation of~$\vartheta_R$
and~$K_R$ involves a minus sign (because we consider the differential equations backwards in~$u$),
one sees that the variations of~$\re z_L$ and~$\re z_R$ (and similarly of~$\im z_L$ and~$\im z_R$) have
opposite signs.

We now choose~$\lambda$ in the annulus~$A_L(\lambda_0)$ as defined by~\eqref{ALdef}.
The resulting variations of~$z_L$ and~$z_R$ are obtained by integrating the
infinitesimal variations, exactly as explained in the proof of Proposition~\ref{prp94}.
Given~$\varepsilon>0$, by choosing~$\tilde{\delta}$ sufficiently small we can arrange
the variation is much larger than~$z$ at~$\lambda=\lambda_0$
(see the formulas of Proposition~\ref{prposclam}). Thus the above consideration for the sign of
infinitesimal variations implies that~$\re z_L$ and~$\re z_R$
(and similarly~$\im z_L$ and~$\im z_R$) have opposite signs.
This gives rise to the inequality
\[ \big| |\phi_{R}|^2 z_{R} - |\phi_{L}|^2 z_{L} \big| \geq \big| \phi_{L}^2 z_{L} \big| \:. \]
It follows that inside the annulus~$A_L(\lambda_0)$,
\[ \bigg| \frac{w(\phiD_L, \phiD_R)}{\phiD_L \phiD_R} \bigg| \geq 
\frac{1}{\big|\phi_{R}^2 \zeta_{R}\big|}- \big| y_{L} - y_{R}\big| \:. \]
Using~\eqref{phiLR} and~\eqref{yLR} together with the fact that~$|\zeta_L|$ and~$|\zeta_R|$ are
uniformly bounded from above,
we conclude after choosing~$\varepsilon$ sufficiently small, 
the first summand majorizes the second summand, i.e.
\[ \bigg| \frac{w(\phiD_L, \phiD_R)}{\phiD_L \phiD_R} \bigg| \gtrsim \frac{1}{\big|\phi_{R}^2 \zeta_{R}\big|} \:. \]
Multiplying this inequality by~$|\phiD_L|=|\zeta_L \phi_L$ and~$|\phiD_R|$, we obtain
\[ \big| w(\phiD_L, \phiD_R) \big| \gtrsim \frac{\big| \phi_L \zeta_L \big|}{|\phi_{R} |} \:. \]
Using again that~$|\zeta_L|=|z_L|$ is bounded from below, we conclude that
\[ \big| w(\phiD_L, \phiD_R) \big| \gtrsim 1 \:. \]
Finally, we use this estimate in~\eqref{sldef}. This concludes the proof.
\QED

\subsection{Tracking the Eigenvalues} \label{sectrack}
We are now in the position to complete the proof of Theorem~\ref{thmmain}.
In order to locate the eigenvalues, we use the following deformation argument. We consider
for~$\tau \in [0,1]$ the homotopy
\[ V_\tau = \re V + \tau \im V \]
with~$V$ as in~\eqref{5ode}.
Then at~$\tau=0$, the potential is real. As a consequence, the
Sturm-Liouville operator is self-adjoint. It has a purely discrete spectrum with real eigenvalues.
If~$\tau>0$, on the other hand, the potential is complex. Our method is to track each eigenvalue
as~$\tau$ is increased.

At~$\tau=0$, we choose the contours as shown in Figure~\ref{figcircles},
with~$N$ chosen as follows. First, we choose~$N$ at least as large as
in Proposition~\ref{prplamlower} for~$c_4$ so large that the eigenvalue~$\lambda_N$
satisfies the inequality~\eqref{relamleft}. The next lemma makes it possible to arrange
a spectral gap.
\begin{Lemma} By increasing~$N$ at most by four, we can arrange that for a suitable
constant~$\const>0$ and large~$|\Omega|$,
\beq \label{gapN}
\lambda_{N} - \lambda_{N-1} \geq \const\, |\Omega|\:.
\eeq
\end{Lemma}
\Proof In preparation, we want to show that there
is a constant~$\const_1>0$ such that for sufficiently large~$|\Omega|$,
the eigenvalues of the Hamiltonian~\eqref{Hamiltonreal} satisfy the inequality
\beq
\lambda_{N+4} - \lambda_N \geq \const_1 \:|\Omega| \:. \label{lamN2}
\eeq
To this end, we first note that the analysis in the proof of Proposition~\ref{prplamlower}
shows that the solution~$\phiD$ has no zeros
on the interval~$(0, u_-^L)$ near the pole at~$u=0$. Similarly, the solution~$\phiD$ has no
zeros on the interval~$(u_-^R, \pi)$ near the pole at~$u=\pi$.
Moreover, on the interval~$(u_+^L, u_+^R)$ where the potential is non-negative, 
we know from Lemma~\ref{lemmaatmostone} that~$\phiD$ has at most one zero.
Therefore, counting the number of zeros of~$\phiD$ on the intervals~$(u_-^L, u_+^L)$ and~$(u_-^L, u_+^L)$,
this number differs at~$\lambda_{N+4}$ and~$\lambda_N$ at least by three.
As a consequence, on one of the intervals~$(u_-^L, u_+^L)$ or~$(u_-^L, u_+^L)$,
the number of zeros of~$\phiD$ differs at least by two.
By symmetry, we may assume without loss of generality that this is the case on the interval~$(u_-^L, u_+^L)$.
It follows from Lemma~\ref{lemmaZes} that
\[ \int_{u_-^L}^{u_+^L} \Big( \im y|_{\lambda_{N+4}} - \im y|_{\lambda_{N}} \Big) \geq \pi \:. \]
Applying the mean-value theorem in the parameter~$\lambda$, we obtain
\[ \big( \lambda_{N+4} - \lambda_N \big)
\sup_{\lambda \in [\lambda_N, \lambda_{N+4}]}
\int_{u_-^L}^{u_+^L} \big| \partial_\lambda y \big| \geq \pi \:. \]
In order to estimate~$\partial_\lambda y$, we use the formula~\eqref{ylam}.
Inserting the asymptotic expansions near the poles (see~\eqref{K0} and~\eqref{phias2})
and using the rigorous estimates in~\cite[Section~5 and~8]{tinvariant}, a straightforward computation
gives the inequality~\eqref{lamN2}.

The inequality~\eqref{lamN2} shows that for any~$\Omega$, one can choose~$N$ such
that~\eqref{gapN} holds~(for~$\const=\const_1/4$). In order to show that~$N$
can be chosen uniformly in~$\Omega$ (for large~$|\Omega|$), we make
use of the fact that the leading powers in~$\Omega$ in the potential~\eqref{Vdef}
are symmetric around~$u=\piot$. This implies the small eigenvalues have the same asymptotics
as~$\re \Omega \rightarrow \pm \infty$ (with the corresponding eigenfunctions transforming as~$\phi_n(u)
\rightarrow \phi_n(\pi-u)$). As a consequence, we can choose~$N$ such that~\eqref{lamN2}
holds for all~$\Omega$ with~$|\Omega|$ sufficiently large.
\QED

For the chosen~$N$, we choose the contour~$\Gamma_0$ such that it encloses the lowest~$N$ eigenvalues,
where~$N$ is chosen at least as large as in Proposition~\ref{prplamlower}.
The other contours enclose one eigenvalue (see Figure~\ref{figcircles}).
Then all eigenvalues {\em{not}} enclosed by~$\Gamma_0$ satisfy the inequality~\eqref{relamleft}.

Now we continuously change the parameter~$\tau$ and follow each of the eigenvalues.
We also continuously deform the contours~$\Gamma_0$, $\Gamma_1$, \ldots
such that they enclose the corresponding points in the spectrum.
If~$\lambda_0$ is an eigenvalue in the single-well case, according to Proposition~\ref{prpressingle}
the resolvent is well-defined and bounded
along a closed contour~$\Gamma$ in the annular region~$A_L(\lambda_0)$
(which can be chosen for example as on the left of Figure~\ref{figannulus2}).
For an eigenvalue~$\lambda_0$ in the double-well case, on the other hand,
there are two possible subcases. Either the eigenvalue is isolated in the sense
that we can again choose the contour as on the left of Figure~\ref{figannulus}.
Or else the eigenvalue can be close to another point in the spectrum, in which case we
choose the contour as on the right of Figure~\ref{figannulus2}.
In both subcases, the resolvent estimate of Proposition~\ref{prpresdouble} applies.
Finally, we enclose the remaining~$N$ lowest spectral points again by a contour~$\Gamma_0$.
Defining the corresponding operators~$Q_n$ by
\[ Q_n := -\frac{1}{2 \pi i} \ointctrclockwise_{\Gamma_n} s_\lambda\: d\lambda
\:,\qquad n \in \N_0 \:, \]
we obtain a family of operators.

For clarity, we point out that spectral points may move from the single-well case to the
double-well case and vice versa. Moreover, the contours~$\Gamma_n$ do not need to
be chosen continuously in~$\tau$. Indeed, if two eigenvalues come close together,
the corresponding contours must be changed discontinuously because the two contours
enclosing the two eigenvalues (as on the left of Figure~\ref{figannulus2})
must be joined to form one contour (as on the right of Figure~\ref{figannulus2}).
As a consequence, the operators~$Q_n$ will in general not depend continuously
in~$\tau$.

It remains to verify a-posteriori that the inequality~\eqref{relamleft} holds for all~$\tau \in [0,1]$ and for all
spectral points {\em{not}} enclosed by~$\Gamma_0$.
To this end, we first point out that the eigenvalues may change considerably compared
to the size of the gaps between neighboring eigenvalues (this is why the theory of slightly self-adjoint perturbations
does not apply here). But the previous methods tell us how the eigenvalues change in~$\tau$
(in particular see Section~\ref{secosc}, Section~\ref{seclocate} and Section~\ref{secWKBrep}).
In simple terms, these results show that the eigenvalues must satisfy the complex Bohr-Sommerfeld
condition~\eqref{bohrc1} with well-defined errors. Combining this formula with the mean value theorem
\beq \label{mvt}
\Big| \sqrt{V_1} - \sqrt{V_0} \Big| \leq \sup_{\tau \in [0,1]} \frac{|\im V|}{\sqrt{|V_\tau|}}
\eeq
and using that on the interval~$(\ul, \ur)$ the absolute value of the potential is larger than~$\Const_1\,|\Omega|$,
whereas~$|\im V| \lesssim |\Omega|$, one sees that the eigenvalues change in such a way that~\eqref{relamleft}
remains valid.

\subsection{Uniform Boundedness of the~$Q_n$ and Completeness} \label{seccomplete}
It remains to verify that the constructed operators~$Q_0, Q_1, \ldots$ at~$\tau=1$ have all the
required properties. The idempotence and mutual orthogonality of these operators follows immediately
from Lemma~\ref{lemmacontourproduct}.
We now proceed by estimating the $\sup$-norm of the operators~$Q_n$.
For the operator~$Q_0$ we again apply the theory of slightly non-selfadjoint perturbations:
\begin{Prp} \label{prpQ0} There is a constant~$\const_2$ such that for all sufficiently large~$|\Omega|$,
\[ \| Q_0 \| \leq \const_2 \:. \]
\end{Prp}
\Proof In view of the gap estimate~\eqref{gapN}, by increasing~$|\Omega|$ we can arrange
that $|\im V|$ is much smaller than the gap. This makes it possible to find a contour~$\Gamma_0$
enclosing the first~$N$ spectral point whose distance to the spectrum is much larger than~$|\im V|$.
This makes it possible to estimate the corresponding contour integral in~\eqref{cint} for~$n=0$
by estimating the Neumann series~\eqref{neumann}. This gives the result.
\QED

We next estimate the operators~$Q_\ell$ with~$\ell \geq 1$.
We begin with a preparatory lemma.
\begin{Lemma} \label{lemmaHS}
Along the contours~$\Gamma_1, \Gamma_2, \ldots$, the 
Hilbert-Schmidt norm of the resolvent is bounded by
\[ \|s_\lambda\|_{\text{\rm{\tiny{HS}}}} \lesssim \int_{\ul^L}^{\ur^L} \frac{1}{\sqrt{|V|}}\bigg|_{\lambda_0} 
+ \int_{\ur^R}^{\ul^R} \frac{1}{\sqrt{|V|}}\bigg|_{\lambda_0} \:. \]
(with~$\lambda_0$ as in~\eqref{Dellamnew}).
\end{Lemma}
\Proof The Hilbert-Schmidt norm of the resolvent can be expressed in terms of its kernel by
\[ \|s_\lambda\|_{\text{\rm{\tiny{HS}}}}^2 = \int_0^\pi du \int_0^\pi du'\: \big| s_\lambda(u,u') \big|^2 \:. \]

We begin with the single-well case. Proposition~\ref{prpressingle} gives the estimate
\begin{align}
\|s_\lambda\|_{\text{\rm{\tiny{HS}}}}^2 &\leq 2 \int_0^\pi du \int_u^\pi du'\: \big| \phiD_L(u)\, \phiD_R(u') \big|^2 \notag \\
&= 2 \int_0^\pi du \int_u^\pi du'\: \big| \zeta_L(u)\, \phi_L(u)\;\zeta_R(u') \phiD_R(u') \big|^2 \:. \label{doubleint}
\end{align}
Near the pole at~$u=0$, the functions~$\phi_L$ and~$\phi_R$ may have a pole
(see~\eqref{phias2}). On the other hand, the function~$\zeta_L(u)$ vanishes at~$u=0$.
As a result, the integrand in~\eqref{doubleint} is bounded near the poles, as the following argument shows.
We introduce the functions~$\rho_L$ and~$\rho_R$ by
\[ \rho_L(u) = \sup_{(0,u]} |\zeta_L| \qquad \text{and} \qquad \rho_R(u) = \sup_{[u,\pi)} |\zeta_R| \:. \]
Then the integrand in~\eqref{doubleint} can be bounded by
\[ \big| \zeta_L(u)\, \phi_L(u)\;\zeta_R(u') \phiD_R(u') \big|^2 \leq \rho_L(u)\, |\phi_L(u)|^2 \rho_R(u)
\;  \rho_L(u)\, |\phi_R(u')|^2 \rho_R(u) \:, \]
giving rise to the estimate
\beq \label{intdouble}
\|s_\lambda\|_{\text{\rm{\tiny{HS}}}}^2 \lesssim
\prod_{c=L,R} \int_0^\pi \rho_L(u)\, |\phi_c(u)|^2\, \rho_R(u)\:.
\eeq
In order to estimate the obtained integrand near~$u=0$, we first note that~$\zeta_R$ is uniformly
bounded. Considering the asymptotics near~$u=0$ (see~\eqref{phi0asy} and~\eqref{phias2}),
one sees that the function~$|\phi_c|^2$ may have a pole at~$u=0$.
But in this case, inserting the asymptotics into~\eqref{zetadef}, one finds that~$|\zeta_L|$
(and therefore also~$\rho_L$) tends to zero at the inverse rate.
We thus conclude that the integrand in~\eqref{intdouble} is indeed bounded near the poles.

Away from the poles, we can use the asymptotics~\eqref{phiLR} to obtain the result. This
concludes the proof in the single-well case.

In the double-well case, we work similarly with the resolvent estimate of Proposition~\ref{prpresdouble}.
The behavior near the poles is estimated just as in the single-well case.
The only additional issue is the behavior on the interval~$[u_+^L, u_+^R]$ in the Airy case.
In this case, combining the exponential factor~$\exp(-2 \int_{u}^{v_R} \re \sqrt{V})$
with the WKB asymptotics gives rise to the estimate
\[ e^{-2 \int_{u}^{v_R} \re \sqrt{V}}\: 
\big| \phiD_R(u)\, \phi_R(u') \big| \lesssim \frac{1}{\sqrt[4]{|V(u)|\: |V(u')|}}\:
e^{-2 \int_{u}^{\min(u', v_R)} \re \sqrt{V}} \:. \]
A straightforward computation using the exponentially decaying factor on the right
gives the result.
\QED

\begin{Prp} \label{prpQnbound}
There is a constant~$c_2$ such that the
operators~$Q_1, Q_2, \ldots$ are bounded by
\[ \|Q_n \| \leq c_2 \:, \]
uniformly in~$\Omega$.
\end{Prp}
\Proof We estimate the contour integrals by
\[ \|Q_n\| \leq L(\Gamma_n)\: \sup_{\lambda \in \Gamma_n} \|s_\lambda\| \:, \]
where~$L(\Gamma_n)$ denotes the length of the contour.
The length of the contour is bounded by~$20\, \Delta \lambda$
with~$\Delta \lambda$ as given by~\eqref{Dellamnew}
(this is obvious in the single-well case, whereas in the double-well case
it was proven in Lemma~\ref{lemmazetalower}).
Applying Lemma~\ref{lemmaHS} and Lemma~\ref{lemmashift}
and using that the Hilbert-Schmidt norm
majorizes the $\sup$-norm, the result follows.
\QED

It remains to prove completeness in the sense of~\eqref{strongcomplete}
with strong convergence of the series.
Since completeness is a statement for fixed~$\Omega$, we can rely
on the theory of slightly non-selfadjoint perturbations.
Namely, Proposition~\ref{prpbounded} yields that there is~$\tilde{N}$
(which might be much larger that the parameter~$N$ in
the statement of Theorem~\ref{thmmain})
a family of operators~$\tilde{Q}_0, \tilde{Q}_1, \ldots$ with the
completeness property
\[ \sum_{\tilde{n}=0}^\infty \tilde{Q}_{\tilde{n}} = \1 \quad \text{with strong convergence}\:. \]
The spectral decompositions~$(Q_n)$ and~$(\tilde{Q}_{\tilde{n}})$
are related to each other simply by forming finite sums, i.e.
\[ \sum_{n \in \Lambda} Q_n = \sum_{\tilde{n} \in \tilde{\Lambda}} \tilde{Q}_{\tilde{n}} 
\qquad \text{(with~$\Lambda, \tilde{\Lambda} \subset \N_0$)} \]
whenever the operators on the left and right describe the same spectral points.
In particular, for large~$n$, the operators~$Q_n$ are sums of one or two
of the operators~$\tilde{Q}_{\tilde{n}}$. This implies that the series
in~\eqref{strongcomplete} also converges strongly. Since every the spectral point
is taken into account in exactly one of the operators~$Q_n$, it follows that
\[ \sum_{n=0}^\infty Q_n = \sum_{\tilde{n}=0}^\infty \tilde{Q}_{\tilde{n}}  \quad \text{with strong convergence}\:. \]
This concludes the proof of Theorem~\ref{thmmain}.

\Thanks {{\em{Acknowledgments:}}
We would like to thank Ole Christensen for helpful comments on Riesz bases.
We are grateful to the Vielberth Foundation, Regensburg, for generous support.

\providecommand{\bysame}{\leavevmode\hbox to3em{\hrulefill}\thinspace}
\providecommand{\MR}{\relax\ifhmode\unskip\space\fi MR }
\providecommand{\MRhref}[2]{%
  \href{http://www.ams.org/mathscinet-getitem?mr=#1}{#2}
}
\providecommand{\href}[2]{#2}

\end{document}